\documentclass[12pt]{article}

\usepackage{newtxtext,newtxmath}

\usepackage{graphicx}

\usepackage[letterpaper,margin=1in]{geometry}

\renewenvironment{abstract}
	{\quotation}
	{\endquotation}

\date{}

\makeatletter
\renewcommand{\fnum@figure}{\textbf{Figure \thefigure}}
\renewcommand{\fnum@table}{\textbf{Table \thetable}}
\makeatother

\usepackage{scicite}

\usepackage{url}

\usepackage{xcolor}%
\definecolor{gender_blue}{RGB}{116,210,226}
\definecolor{gender_red}{RGB}{234,119,118}

\definecolor{median_blue}{RGB}{39,82,161}
\definecolor{mean_red}{RGB}{198,79,130}

\definecolor{training_cyan}{RGB}{76,178,207}
\definecolor{training_orange}{RGB}{222,69,48}

\definecolor{WTF_yellow}{RGB}{252,185,0}
\definecolor{Follower_green}{RGB}{25,132,70}

\newtheorem{customlist}{List}

\def\scititle{
	Recommender system in X inadvertently profiles ideological positions of users
}
\title{\bfseries \boldmath \scititle}

\author{
	Paul~Bouchaud$^{1,2\dagger}$,
	Pedro~Ramaciotti$^{1,2,3\ast\dagger}$\and
	\small$^{1}$Complex Systems Institute of Paris Ile-de-France CNRS, Paris, France.\and\
	\small$^{2}$médialab, Sciences Po, Paris, France.\and
	\small$^{3}$Learning Planet Institute, CY Cergy Paris University, Paris, France.\and
	\small$^\ast$Corresponding author. Email: pedro.ramaciotti-morales@cnrs.fr\and
	\small$^\dagger$These authors contributed equally to this work.
}

\begin{document} 

\maketitle

\begin{abstract} \bfseries \boldmath
Several data protection laws restrict processing that reveals political opinions, irrespective of the controller's intent. Whether recommender systems do so as a by-product of optimizing relevance has not been measured. From 2.5 million ``Who to Follow'' recommendations shown to 682 volunteers in France, we reconstructed an approximation of the embedding used by X's recommender for 26,509 accounts, computing survey-calibrated ideology scores. One direction in this embedding orders users by Left-Right position (Pearson $\rho$ = 0.887), distinct from directions tracking age, gender or popularity. We show this scale exists and affects the recommendations computed from the embedding. Removing it diversified recommendations at a limited cost in accuracy. We document a consequential form of emergent political representation that current definitions of profiling do not clearly address.
\end{abstract}

\noindent
Recommender systems on social media are specifically trained to predict engagement, not to infer political opinions. Yet the behavioral traces they learn from, such as the content they share, the users they follow and reply to, are among the strongest known predictors of ideology~\cite{Kosinski2013}. It is thus expected that these systems may therefore come to internally represent political positions \cite{faverjon2026recommender}. In diverse national contexts, including South Korea, Swirzerland and Brazil, regulations have been introduced to constrain the exploitation of sensitive personal data such as political opinions. The European GDPR prohibits processing personal data ``revealing [...] political opinions'' (among other sensitive personal attributes) without explicit consent, a prohibition the Court of Justice of the EU has held applies whether or not the controller aims to obtain such information (Court of Justice of the European Union, CJEU, ruling C-252/21, para. 69), and even when sensitive information is only revealed indirectly (CJEU, ruling, C-184/20). Whether deployed recommenders form such representations for ordinary users, and whether these can be measured with instruments used in political science, is of relevance for AI and industrial policy, governance, regulation.

Recent studies have shown that machine learning procedures can autonomously learn sensitive attributes \cite{amini2019uncovering}, and that social media data can be leveraged in estimating individual political preferences \cite{fagni2022fine}, ideological positions \cite{barbera2015birds}, and stances on issues \cite{ramaciotti2022inferring}.
This body of work has nurtured the expectation that any recommender system capable of serving relevant content on social platforms must, at least to some degree, capture and leverage political preferences among several other attributes on which personal preferences may hinge.
At the same time, beyond the question of privacy, the massification of social media and its entanglement with politics have motivated a growing body of work focused on political polarization and segregation, not least because of the role and potential consequences of AI systems computing recommendations on these platforms \cite{lorenz2023systematic}.
A better understanding of the mechanisms through which AI systems learn, represent, and leverage ideological orientations on online platforms presents significant implications for developing regulatory instruments and compliance tools, but also for studying social systems interacting with AI ones and even for AI design.

Most works studying recommender systems and politics investigate recommended items (e.g., content or contacts) either analyzing recommendations by measuring properties such as diversity and biases \cite{bakshy2015exposure, Bouchaud2023, milli2024, Duskin2024}, or assessing the causal impact of recommendations through controlled experiments with different recommendation policies \cite{nyhan2023like,huszar2022algorithmic,Jia2024}.
Several studies have addressed the question of machine representations of ideology from data during training, but mainly of contents (as opposed to individual actors), focusing on a few actors \cite{rheault2020word} (e.g., political elites such as Members of Parliament, comparing machine representations with measures such as ideology scaling of parliamentary voting), or on limited experimental settings, as opposed to large real-world systems with large user bases \cite{milbauer2021aligning}.
Fewer studies have addressed how recommender systems in social platforms learn, represent, and process political information on large numbers of regular users and contents when computing recommendations \cite{faverjon2023discovering}.
The main challenges in gaining further insights into how these AI systems work are the difficulty in using AI explainability methods in these real-world settings, and in collecting exposure data \cite{sandvig2014auditing, bakshy2015exposure,huszar2022algorithmic,Bouchaud2023,nyhan2023like}.

In this work, we develop and validate a method to infer an approximation of the embedding that X's (previously Twitter) AI recommender system leverages in computing friend and content recommendations.
X is a relevant platform for such a study because political figures actively use it \cite{van2020twitter} (making it a pervasive tool in elections \cite{jungherr2016twitter}), and because of its entanglement with the media sphere \cite{kwak2010twitter}.
First, we collected more than 2.5 million real ``Who to Follow" friend recommendations shown on the platform via a browser plug-in installed by {682} users during a data donation campaign, between January 2023 to May 2024 in France (with nearly 20 millions monthly active users as April 2024), involving more than 26 thousand both volunteer and recommended users.
Second, we used publicly available knowledge about the platform's AI system architecture to infer an approximation of the embedding computed by X's recommender system pipeline and that drives recommendations on the platform \cite{TwittersRecSys}.
At its core, this AI system embeds users, posts, and advertisements in a high-dimensional space where inner product similarities are ranked to recommend these items to users \cite{el2022twhin}.
Using this explicit formulation of inner product and ranking operations, we propose and validate a method to infer the most probable embedding positions for all users in our study.

To examine the political attributes of users in this space, we rely on ideological scaling \cite{clinton2004statistical} of the follower network connecting them to political figures \cite{barbera2015birds}, which we calibrated using political survey data \cite{ramaciotti2022inferring}.
We compute additional user attributes to inspect the embedding space, including estimated age and gender from users' profile text and photos \cite{wang2019demographic}, and their interests in various domains (e.g., news, sports, politics, or science) based on the contents of their posts.
Using these attributes we inspect the embedding space of the recommender in the platform to explain computation of recommendations in terms of these attributes.
Specifically, we ask three questions. First, is there a direction in the reconstructed embedding along which users' positions correlate with their Left-Right positions on a survey-calibrated scale? Second, is this direction distinct from directions associated with other attributes relevant to recommendation, such as age, gender, popularity and topical interests? Third, does removing this direction change the recommendations computed from the embedding? 
We do not ask whether users' internally held ideological positions \textit{cause} the positions in the embedding, which would require exogenous variation in users' internal ideological positions and would address need to address the long time-frame over which political opinion forms and the mutual dependency between opinions, digital environments and recommender systems.
Our question are about the machine representation are descriptive and relevant to privacy regulations worldwide.

If such a representations exist, it raises a question that existing work has not addressed. Research has shown that ideology positions may be estimated from behavioral data \cite{fagni2022fine,barbera2015birds,barbera2015tweeting,ramaciotti2022inferring}. Legal scholars have discussed whether platforms' inferences about users can constitute special-category data  \cite{xenidis2023beyond,ramaciotti2024depolarizing,grozdanovski2021search,wachter2019right}. What has not been shown is that the internal representation of a deployed recommender contains a scale that maps onto established political measurement instruments for large numbers of ordinary users. We provide the first demonstration of this, for X's friend recommender in France: a scale that arises from optimizing for relevance.

Our study does not answer whether this form inference constitutes profiling in the legal sense, and, on the contrary, highlights a growing inconsistency in the application of several data privacy regulations with no direct or immediate solution for current AI technologies.
{Our work suggests that the distinction between active and passive profiling may come into question when} AI systems {process data produced by large numbers of regular individuals} and {when AI explainability methods such as ours may make visible (to platforms, regulators, and regular users) how this processing maps to the treatment of uncontroversial definitions of political opinion data.}
{Our work also has implications for research on AI mediation of digital socio-information systems.}
First, it provides a way to further investigate the dynamics of human-AI systems with regard to political opinion processes and information exposure.
Second, it enables new recommendation policies and algorithm design tools resulting from explicitly modifying the machine representations of political attributes.
{We explore the latter by investigating the potential consequences of complying with data privacy regulation, constraining the ideological leanings in the spatial representation of individuals in the recommender of X, measuring changes in content relevance and diversity.}
\section*{Results}

\subsection*{Reconstructing the embedding space of the recommender in X}
\label{subsec:main_reconstructing_embedding}

The ``Who to follow" (WTF) recommendations \cite{DiscoveringWTF} reportedly drive over one-eighth of all follower connections \cite{goel2015follow}, and uses several algorithms to surface relevant accounts that are then ranked before recommending them \cite{TwitterGithubMain}. See Extended Figure~1 of the Supplementary Information document for an overview of the architecture of the service.
{X's recommender system leverages explicit user signals—who they follow, repost (i.e., share), reply to, or click on—resulting in their representation in a multidimensional embedding space.}
Positions in this embedding space are then leveraged in computing similarities via inner product operations, producing candidates for recommendation which are then ranked and served to users. See Extended Figure~2 for an illustration of the data representation of entities that are embedded, according to X's scientific publications \cite{el2022twhin}.
While the architecture and algorithms involved in the computation of this embedding are publicly known to some extent, the data used during training and the positions of entities in the embedding of the recommender are not.
This poses a double problem for explainability: first because of the opacity of the model (i.e., the so-called ``black box'' problem), but also the lack of access to the model itself, which constitutes a major barrier for most independent scientific research on algorithmic recommendation.

\begin{figure}
    \centering
    \includegraphics[width=13cm]{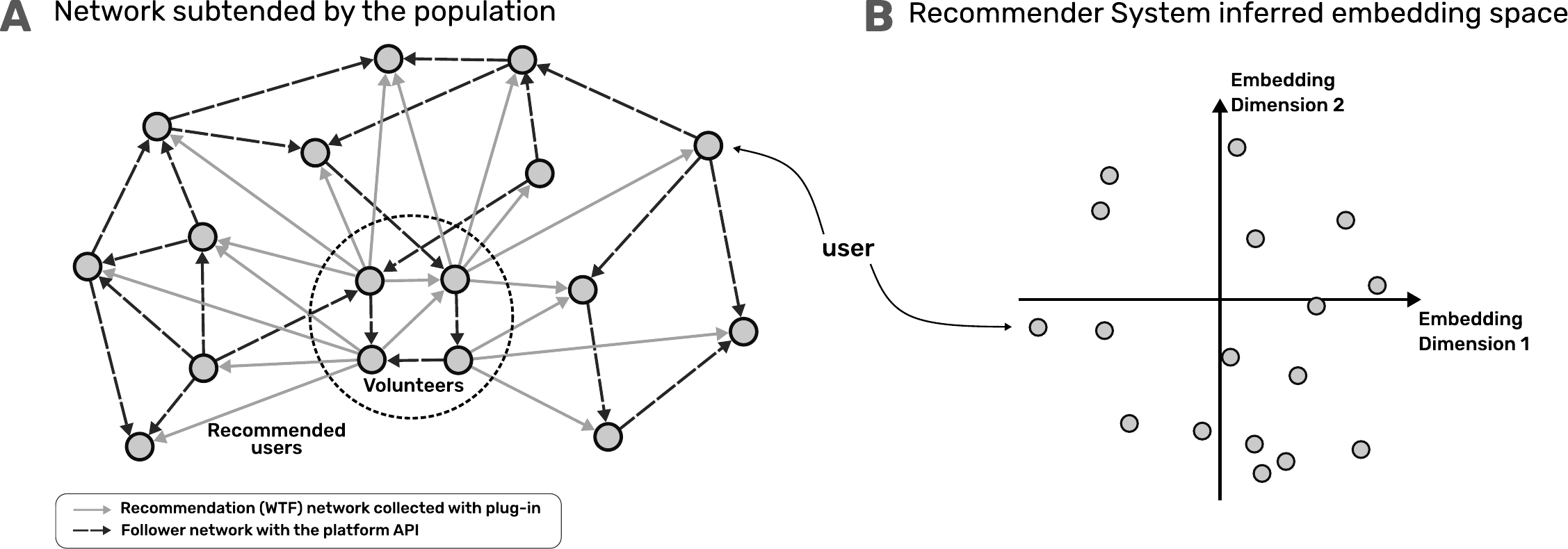}
    \caption{Schematic representation of the population of X users for the study and the inference of their position in the embedding space of X's recommender system. (A) Population of the study (data volunteers $\mathcal{V}$ and recommended users $\mathcal{U}$) subtending networks for friend recommendation (in {orange arrows}) and following relations (in {purple arrows}), collected respectively with a data donation program using a browser plug-in and X's API. (B) Schematic representation of the embedding space with every user $i$ in the population $\mathcal{E}$ having a position $\phi_i$ in the 256-dimensional embedding.}
    \label{fig:summary_embedding}
\end{figure}

To collect data traces allowing us to infer their positions in the embedding space of the WTF service, we created a data donation program in which users could install a browser plug-in that relayed to us recommendations of friends and contents serviced to them by the platform.
We advertised the browser plug-in on X and on local media outlets in the fall of 2022, offering volunteers who installed it a personalized analysis of their feed. No monetary incentive was offered to volunteers.
From January 1st, 2023, to May 1st, 2024, we collected 2,549,008 personalized friend recommendations seen on the platform by 682 fully-compliant volunteers (set $\mathcal{V}$) {i.e., individuals who opted into the data collection program, used the browser extension for at least one month, visited X to see at least 30 contact recommendations, and did not opt out of the study by uninstalling the plug-in or by later requesting their data to be deleted from our database}.
From over 105K accounts recommended during the observation period, we retain 26,338 accounts that were recommended to two or more volunteers (set $\mathcal{U}$) to define the population for our study as $\mathcal{E}=\mathcal{U}\cup\mathcal{V}$ ($|\mathcal{E}|=26,509$), for which we have data related to recommendations.

Using the platform's API, we also collected the follower network subtended by population $\mathcal{E}$, as well as profiles and posts produced by them.
We cast the problem of approximating the positions of users in $\mathcal{E}$ in the embedding of the recommender as an optimization problem for the prediction of observed recommendations, constrained to the known architecture of the embedding used by the WTF service \cite{el2022twhin}.
Following X's own design, we consider a 256-dimensional embedding space \cite{twhinbert} and we solved the optimization problem (namely, estimating the embeddings that best explained observed data) with a gradient descent algorithm.
We {assess the quality of} the inferred positions in the embedding by evaluating the accuracy with which the resulting recommender can predict observed WTF recommendations in our study.
Holding out 10\% of our data (including both recommended and not recommended users), our trained model yields an AUC-ROC value of 0.700. For comparison, three baseline approaches—recommending random second neighbors, second neighbors with the highest number of followers, or second neighbors followed by the most first neighbors—all perform below chance level (AUC of 0.467, 0.425, and 0.423 respectively). {The probability of a random baseline---predicting edges either uniformly or weighted by frequencies observed in the training dataset--- to equal or outperform our trained model is less than $10^{-5}$.}

When evaluating only positive test recommendations (i.e., friend recommendations observed in the study) against a pool of 100 randomly chosen non-recommended second-order friends, the model achieves a Precision@1 of $0.725$ and a Precision@3 (the number of recommendations shown on the right-side panel on X) of $0.691$.

We tested the robustness and reliability of the inferred positions in the embedding for alternative training parameters and sub-sampling strategies simulating potential sources of biases, including: {i) the possibility that volunteers saw recommendations on devices outside our study, ii) age, gender and ideological skews in our population, iii) changes in the activity and population of the platform during the period of observation (including changes around May 2023 related to change in platform ownership), iv) changes in the structure of the follower graph during the period of observation, and v) alternative parameters of the optimization problem.}
See the Methods section for a detailed description of the constrained optimization problem, its implementation, parametrization, validation and experiments demonstrating robustness and reliability.

\subsection*{Inferring political and demographic attributes of users}

\begin{figure}
    \centering
    \includegraphics[width=\textwidth]{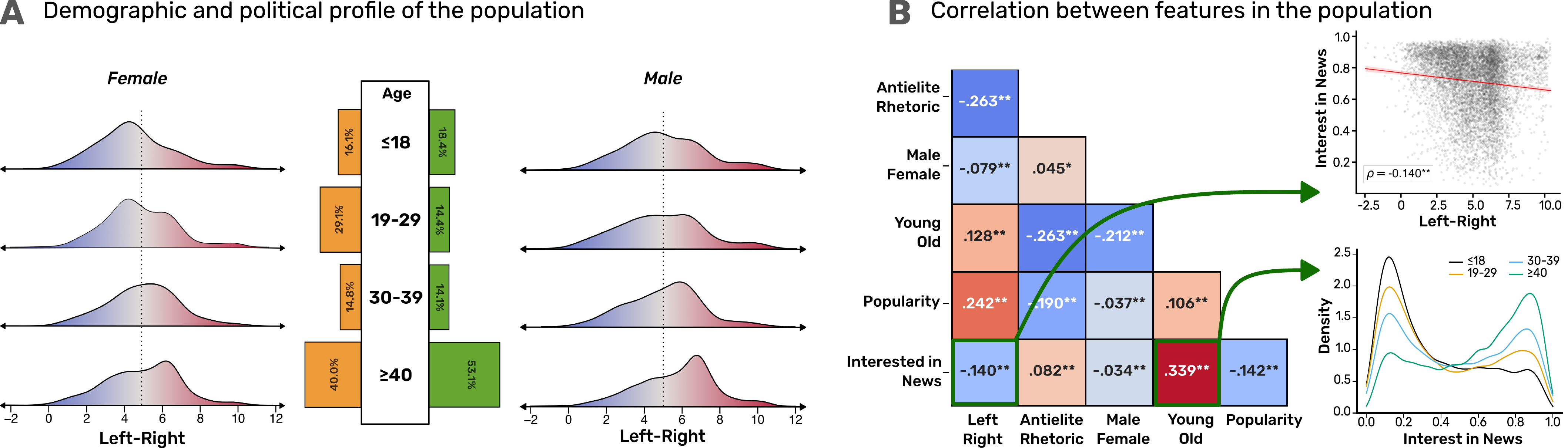}
    \caption{Characterization of the population in the study. (A) Distribution of Left-Right ideological leaning of X users segmented by age and gender. (B) Pairwise Spearman correlations between the inferred attributes of users (* indicates p-value $<0.01$, ** indicates p-value $<0.001$). The correlation between Left-Right leaning and degree of interest in news is -0.14**. The density plots for the degree of interest in news of users in different age groups illustrates a comparatively higher correlation between this attribute and age.}
    \label{fig:fig1_characterization_population}
\end{figure}

To consider political attributes of users in our study, we rely on a dataset of users in X with inferred political positions on a Left-Right scale in 2023 \cite{ramaciottimorales:hal-04807916}.
The authors of this dataset use multidimensional ideology scaling of follower networks on X \cite{barbera2015birds,barbera2015tweeting} and political survey data \cite{Jolly2022} for calibrating positions of large populations of users pooled from followers of Members of Parliament \cite{ramaciotti2022inferring}.
The dataset includes a Left-Right scale calibrated to have three reference points: 0 (the leftmost position for political parties), 10 (the rightmost positions), and 5 standing as the political center.
The dataset includes a second political dimension called anti-elite rhetoric, measuring attitudes towards elites and institutions, deemed by authors of the dataset as a second relevant dimension in the country of our study (France). 
8,249 users in this dataset are also in population $\mathcal{E}$ (i.e., 31.1\%{}; see Figure~\ref{fig:fig1_characterization_population}.A).
Extended Figures~14~\&{}~15 show the positions of users in the dataset from \cite{ramaciottimorales:hal-04807916} along the two dimensions and metrics assessing the quality of these calibrated positions.
These positions are estimated from users' follow relations to Members of Parliament, which are also used by the recommender system. Section ~\ref{subsubsec:robustness_training_parameters} reproduces the inference of the embedding space without follower relations involving MPs to remove the possibility of artifact Left-Right directions in the embedding.

We include additional user attributes in our study, both to control for potential covariates that might explain Left-Right positions of users encoded in the recommender system, and to further examine how these attributes are processed in the embedding space. To infer these attributes, we collected their profile bios (text and image), their followed accounts, and their last 200 published posts (either original or reposted), using the API of the platform.
We infer the age and gender of users with the \texttt{M3} model developed by Wang et al. \cite{wang2019demographic}, analyzing profile pictures, screen names, and biographies. 
The \texttt{M3} model provides complementary probabilities (i.e., adding to 1) of being either ``male'' or  ``female'', allowing us to consider a Male-Female dimension ranging from the highest probability of being classified as male to the highest probability of being classified as female.
The \texttt{M3} model also provides complementary probabilities of being in age groups: 0-18, 19-29, 30-39, 40 and older (Extended Figures~11~\&{}~12 show agreement of inferences using \texttt{M3} by comparison with those achieved with other models {and with attributes self-reported by volunteers}).

We also consider the popularity of accounts, measured as the percentile rank of number of followers to account for long-tailed distributions (results are unchanged when considering the logarithm of the number of followers), and attributes capturing the interest that users have in different topics by computing a topic modeling classification of their last 200 posts, leveraging the model developed by Antypas et al. \cite{cardiffNLP}, specifically designed for posts on (then) Twitter.
Among these available topics, we specifically consider interest in news as a covariate that might explain political positions encoded by a recommender system.

Attributes in our population exhibit low absolute Spearman correlations (Figure~\ref{fig:fig1_characterization_population}.B), with the two highest being between age and interest in news ($\rho=0.339$; i.e., older users tend to exhibit more interest in news content) and between age and negative attitudes towards elites and institutions ($\rho=-0.263$).
Left-Right leaning (our main variable of interest) exhibits a lower absolute correlation with age ($\rho=0.128$) and gender ($\rho=-0.079$). All the above-cited correlations have p-values $<0.001$.

\subsection*{Measuring attributes encoded by the recommender system}

Next, we proceed to the identification and measurement of information on {attributes of} users that might be encoded in the embedding space.
The main hypothesis behind this identification and measurement procedure is that these AI systems might develop representations that match relevant constructs, such as Left-Right ideological leanings.
This expectation is motivated by the use of inner product similarities in ranking, based on positions on the embedding, on the part of the recommender.

To identify and measure the information of attributes of users encoded in the embedding, we leverage a search of spatial directions in this space and that induce high correlations with attributes, using Canonical Correlation Analysis (CCA), treating user embeddings as a multivariate dataset and inferred attributes as univariate ones.
For each user attribute, a CCA identifies the spatial direction in the embedding space that maximizes correlation between the projected positions of users along the direction and the selected attribute.
Because our goal to provide an explanation for the embedding on the semantic provided by attributes, we compute a CCA regression independently for each feature for interpretability.
The CCA directions determined in this way, achieve Pearson correlations of different qualities for the selected attributes (see Figure~\ref{fig:fig2_results}.A; all correlations have p-value $<0.0001$): Left-Right leaning ($\rho=0.887$), anti-elite rhetoric  ($\rho=0.863$), interest in news ($\rho=0.848$), popularity ($\rho=0.730$), age ($\rho=0.562$), and gender ($\rho=0.384$).
Extended Table~2 shows all Pearson correlations achieved by directions identified with the CCA for political and demographic attributes, as well as for attributes regarding interests in topics, and platform activity metrics.
%
%
Each CCA direction gives the best linear fit between an attribute and the reconstructed embedding. These correlations are computed in-sample; cross-validated values are given in Extended Table~\ref{tab:embedding_cca_correlations} in Material and Methods.

The embedding encodes several attributes not explicitly available to the recommender, such as interest in news, popularity and age. Left-Right leaning differs in its legal status only: political opinions are a special category of personal data under GDPR Article 9, while the other attributes are not. We focus on Left-Right leaning for this reason. The choice does not imply that the system treats it differently from other attributes.

To further validate that these directions produce a reliable ordering of users by the value of their attributes, we assessed the quality of their positions projected onto these directions, for Left-Right leaning ($\mathbf{w}_{LR}$), for anti-elite rhetoric ($\mathbf{w}_{AntiElite}$), for age ($\mathbf{w}_{age}$), for gender ($\mathbf{w}_{gender}$), for interest in news ($\mathbf{w}_{News}$), and for popularity ($\mathbf{w}_{pop}$).
We assess the spatial ordering of users that follow Members of Parliament of parties with clear ideological leaning (using a Mann-Whitney U test): the center-leaning party \textit{Rennaissance} (RE), the Right-leaning party \textit{Rassemblement National} (RN), and the Left-leaning party \textit{La France Insoumise} (FI) (obtaining respectively 27.85\%{}, 23.15\%{} and 21.95\%{} of the vote in the last presidential elections in 2022, followed by a fourth candidate obtaining 7.07\%{} of the vote).
Direction $\mathbf{w}_{LR}$ induces an order (Extended Figure~19) of users such that followers of MPs from FI are to the left of those of the followers of MPs from RE (p $< 0.0001$), which are to the left of those of the followers of MPs from RN (p $< 0.0001$).
Direction $\mathbf{w}_{AntiElite}$ also induces an order of users (Extended Figure~20) such that followers of MPs from FI and RN (widely regarded as anti-elite parties \cite{ivaldi:halshs-01889832}) are positioned to the right (i.e., displaying stronger anti-elite rhetoric) of followers of MPs from RE (p $< 0.0001$). 

Measured by a Mann-Whitney U test, $\mathbf{w}_{gender}$ orders users classified male or female (p $< 0.0001$), and $\mathbf{w}_{age}$ produces a relative pair-wise ordering of age groups (p $< 0.0001$). Classification of users into age and gender groups is performed at a 0.5 probability threshold on the results of the \texttt{M3} model.
To further validate positions along these CCA directions, we computed the accounts followed, and the words and hashtags used by users at their extremes, and showed in the Methods section how they relate to corresponding ideological leaning, demographic group, or topic of interest.

\subsection*{Measuring the alignment between attributes encoded in the embedding space}

Is the Left-Right spatial representation of users learned by the recommender system different from the representation of age or gender?
To assess the similarity and alignment between positions of users along the inferred attribute directions in the embedding space, we inspect two metrics: 1) the cosine similarity (ranging from -1 to 1, with 0 meaning orthogonality) between each pair of CCA directions, and 2) the correlation between the positions of users along each pair of CCA directions capturing user attributes in the embedding space of the recommender.

\begin{figure}
    \centering
    \includegraphics[width=\textwidth]{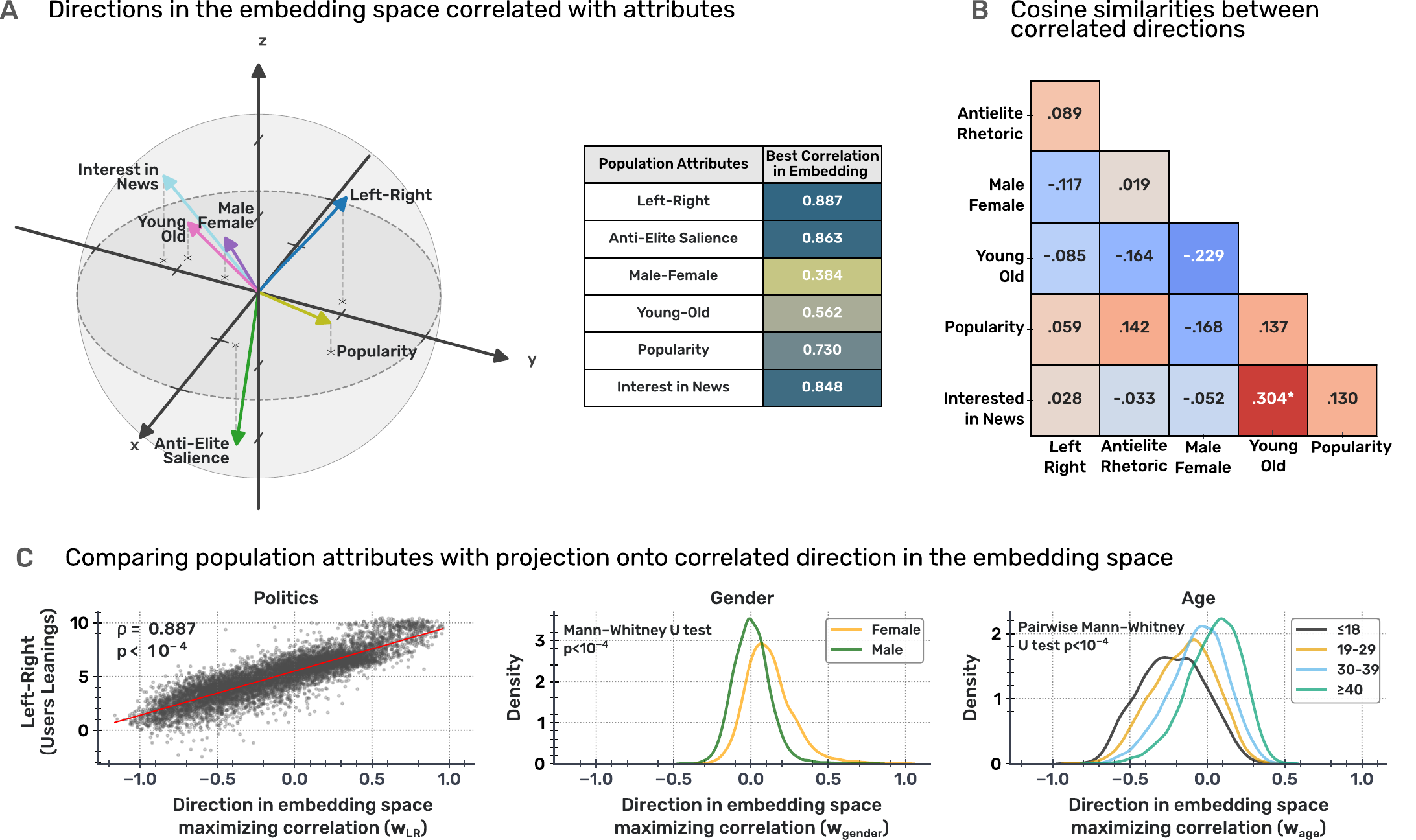}
    \caption{(A) Three-dimensional representation of directions in the 256-dimensional embedding space of the recommender system that maximize the correlation with attributes of users, identified using Canonical Correlation Analysis (CCA). The three-dimensional visualization was obtained through a Locally Preserving Projection \cite{LPP}. Vector norms of directions associated with attributes correspond to absolute correlations with these attributes, reported in the adjacent table (all p-values $< 0.0001$). (B) Absolute cosine similarity computed in the 256-dimensional embedding space between directions in the embedding space identified with CCA and that maximize correlations with attributes of users (* indicates p-value $<0.01$). (C) Comparison of attributes of users along the CCA directions in the embedding that maximize correlations with three features: Left-Right leaning $\mathbf{w}_{LR}$, age $\mathbf{w}_{age}$, and gender $\mathbf{w}_{gender}$.} 
    \label{fig:fig2_results}
\end{figure}

CCA directions do not display a high degree of spatial alignment when measured via cosine similarity (see Figure~\ref{fig:fig2_results}.B). The highest (absolute) values for cosine similarity are between age and interest in news (0.304), and age and gender (-0.229), followed by the alignment between popularity and gender (-0.168).
Only one pair of CCA directions displays a statistically significant alignment (at $\alpha=0.01$) measured by the cosine similarity: age and interest in news.
We measure the p-value as the proportion of randomized cosine similarities (i.e., permuting the attributes of users) that exceed the observed similarity between two CCA directions.

A measurement of the Spearman correlation between positions of users projected onto the CCA dimensions reveals different degrees of similarity among them.
The three highest (absolute) correlations observed (all with p-value $<0.001$) are for positions projected onto CCA directions: $\mathbf{w}_{age}$ and $\mathbf{w}_{news}$ ($\rho=0.620$), $\mathbf{w}_{age}$ and $\mathbf{w}_{Anti-elite}$ ($\rho=-0.524$), and $\mathbf{w}_{Popularity}$ and $\mathbf{w}_{LR}$ ($\rho=0.404$).
Importantly, Left-Right leanings in the recommender (captured by $\mathbf{w}_{LR}$) do not induce high correlations with machine representations of age (as captured by $\mathbf{w}_{age}$; $\rho=0.172$) nor gender  (as captured by $\mathbf{w}_{age}$; $\rho=-0.275$) (see Extended Figure~30 for the spatial distribution along these pairs of directions).
Extended Figure~29 shows the values of Spearman correlation between positions along all CCA directions.

{These results show that the information linearly encoded by the recommender along spatial directions identified with CCA, are largely orthogonal}.
An attribute of special interest for both the study of online politics and data privacy regulation are Left-Right leanings.
Direction $\mathbf{w}_{LR}$ that arranges users by Left-Right ideological leaning, induces projected positions that are only partly predicted by users' positions along the other CCA directions (Pearson $r$ = 0.89, p-value $<$ 0.0001 when all of them are used jointly; Extended Table~\ref{tab:embedding_cca_correlations} in Materials and Methods).
In other words, up to linear encoding of attributes in the embedding of the recommender, information on ideological Left-Right leaning is nearly orthogonal from other attributes.
While more complex geometrical models of encoded information on attributes might possibly be more explicative (e.g., nonlinear manifolds), our results show that, even under a conservative linear encoding assumption, the recommender system stores information on the ideological leaning of users that is not reducible to demographic or thematic attributes. 
This rules out the possibility that the observed political structure is merely an artifact of the selected correlated non-sensitive traditional relevant features such as age, gender, or news consumption.

\subsection*{Removing the Left-Right direction from recommendations}
\label{subsec:main_constrained_rec}

\begin{figure*}[t!]
    \centering
    \includegraphics[width=\textwidth]{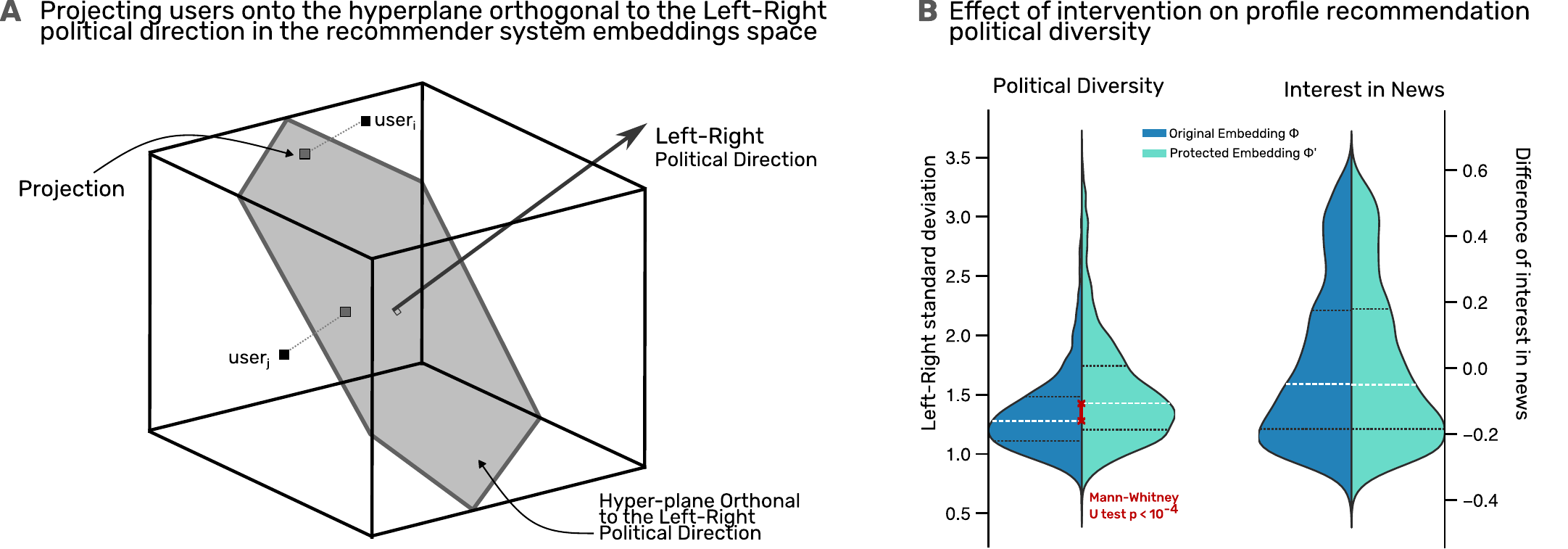}
    \caption{Constraining user recommendations from leveraging left-right political information. (A) Diagram of the intervention procedure. After identifying the direction in the latent space that maximally correlates with users' left-right political leaning, we iteratively project user representations onto the orthogonal subspace. (B) Effects on recommendations of constraining political information in the embedding of the recommender systems. For each user in the population, we compute 50 friend recommendations before (in blue) and after (in green) the intervention and compute the impact the distribution of the ideological diversity of recommendations as the standard deviation on the Left-Right scale, and the relevance as the similarity in interest in news between recipient of recommendations and recommended friends.}
    \label{fig:fig3_constrained_recommendation}
\end{figure*}

If regulators or platforms were to treat such a direction as processing of political opinions, one response would be to remove it. We test an operation aiming to remove the Left-Right information available when recommendations are computed.
The proposed procedure consists in computing the subspace in the embedding that is orthogonal to $\mathbf{w}_{LR}$ (i.e., a 255-dimensional hyper-plane in the embedding space), and projecting all recommendable entities onto it before using inner products in the resulting space to assess similarities and formulate recommendations (Figure~\ref{fig:fig3_constrained_recommendation}.A).
In the scope of our study, we propose an experiment recommending users on $\mathcal{E}$ to investigate the effects of such a procedure.
The new embedding resulting from this procedure is \textit{protected} \cite{ravfogel} from having linearly encoded information about ideological leanings of users.

This proposed procedure holds resemblance to that described as ``debiasing" of language models by Bolukbasi et al. \cite{Bolukbasi}, which allows zeroing-out the components of neutral words in the direction of a ``gender subspace" (in their work) to mitigate gender stereotypes in word embeddings.
As shown by Gonen and Goldberg \cite{gonen_goldberg} and by Ravfogel et al. \cite{ravfogel}, linearly-encoded attributes in high-dimensional spaces might be encoded in several orthogonal dimensions. This means that, even after projecting positions of users in the embedding onto the subspace that is orthogonal to $\mathbf{w}_{LR}$, in the new resulting embedding, it might be possible to find a second dimension that also encodes (with lower induced absolute correlation) ideological leanings. Therefore, as Ravfogel et al. \cite{ravfogel}, we repeat the identification and projection procedures, iteratively computing a CCA identification of directions encoding Left-Right leanings, until no subsequent identifications yield statistically-significant correlations (set at $\alpha$=0.01). This means that we iteratively remove linearly-encoded ideological information by projecting into new subspaces, until the p-value of the maximal correlation identified with CCA overflows above 0.01.

To evaluate the effect of our procedure we compare recommendations computed using X's described pipeline based on embeddings \cite{el2022twhin}, using both the original embedding and the new protected embedding void of linearly-encoded ideological leanings.
We formulate 50 friend recommendations for each user and evaluate i) the Left-Right ideological diversity of recommended profiles (measured as the standard deviation of Left-Right positions) and ii) recommendation relevance. To assess relevance, we compare, through cosine similarity, the topic distributions of posts shared by friends suggested using the original embeddings versus those suggested using the protected embeddings. Additionally, we specifically examine the interest in news and politics among users suggested by the original versus protected embeddings, where interest is quantified as the proportion of each user's last 200 posts classified as discussing news using the model by Antypas et al.~\cite{cardiffNLP}.
By selectively removing information that linearly encodes users' Left-Right ideological leanings, we observe an increase of the Left-Right political diversity of recommendations (Cohen's $d = 0.477$, $95\%$ CI $[0.459, 0.495]$ using the original embeddings as control. Confidence intervals were determined through user-level bootstrapping.) while maintaining relevance, as shown in Figure~\ref{fig:fig3_constrained_recommendation}.B. Specifically, the intervention preserves the topics shared by suggested friends ---the cosine similarity of topic distributions between friends suggested by protected and original embeddings is $0.948$ ($95\%$ CI $[0.751, 0.999]$)--- and maintains similar interest in news between recommendation recipients and recommended friends (Cohen's $d = -0.009$, $95\%$ CI $[-0.032, 0.014]$).

In summary, linearly-encoded ideological positions can be removed from the reconstructed embedding. Doing so preservers the topical profile of recommendations and increases their ideological diversity. It also lowers the accuracy with which the embedding reproduces observed recommendations, with AUC falling by about 0.08 from 0.70 (see Extended Figure~\ref{fig:weight_feature_delta_auc}).
{Section~G.4 of the Supplementary Information examines the effects of our procedure applied to other attributes, showing that Left-Right leanings encoded in $w_{LR}$ is the attribute that, when removed, leads to the most significant change in the set of recommended contacts.}
Recent literature reports mixed effects of diversified recommendations on political polarization, with some evidence suggesting potential exacerbation of affective polarization \cite{bail2018exposure}. These potentially adverse effects remain poorly understood across different national contexts and at varying levels of diversity intervention.
\section*{Discussion}

Our results show that an the embedding used by X's recommender contains a direction that approximately orders ordinary users by their Left-Right position (as estimated with ideology scaling models identified with survey data), and that removing this direction changes the recommendations computed from it. Nothing in the documented design of the system asks it to infer political opinions, unlike systems such as Community Notes that do so explicitly \cite{bouchaud2026community}. The scale Left-Right scale emerge from learning relevance on behavioral data traces. This possibility had been previously suggested based on synthetic scenarios \cite{faverjon2023discovering}. To our knowledge, this is its first demonstration in a deployed system for large numbers of regular users worldwide. Other directions track age, gender , popularity and topical interests, and are only weakly aligned with the Left-Right direction. This pattern is consistent with the linear representation hypothesis \cite{mikolov2013distributed,arora2016latent,elhage2022toy}.

This Left-Right dimension in the embedding of the recommender is not the result of circularity. The Left-Right scores we use are estimated from follow relations to Members of Parliament \cite{barbera2015birds, barbera2015tweeting,ramaciotti2022inferring}, which are also inputs to X's recommender and to our procedure for the reconstruction of its embedding. 
The dependence between ideology (and consequent homophilic follower networks) and the embedding is the phenomenon we measure, not a flaw of the measurement.
A recommender trained on those traces to predict whom users will follow optimizes for a signal in which that homophily is built in. 
That its internal representation reproduces the political scientists' scale is therefore expected. 
It is also what several data protection laws identify in their provisions. 
GDPR, for instance, forbids processing that ``reveals'' political opinions (Article 9(1)), independent of the intention of the entity conducting the processing (CJEU, ruling C-252/21, para. 69).
To rule out circularity between our method of embedding reconstruction and the estimation of Left-Right ideologies through ideology scaling, we tested reconstructing the embedding by first removing all follower links involving Members of Parliament (the ones used in ideology scaling), replicating our results (see Section~\ref{subsec:embeddings_robustness} in Materials and Methods).

We do not estimate whether users' ideology causes their position in the embedding, and our conclusions do not depend on it. 
Identifying such an effect would require exogenous variation in users' internal ideological positions. Such a setting also raises additional obstacles. First, follow decisions reflect partly unobserved motives, such as ideological proximity, information seeking and social ties, which correlate with both ideology and recommendations and create omitted-variable bias. Second, causation likely runs both ways. Ideology shapes behavioral signals, but recommendations may also shape these signals (e.g., making probability of engagement dependent on probability of exposure) and, over longer periods, possibly even their attitudes \cite{piccardi2025reranking,gauthier2026political}.
This feedback produces time-varying confounding that only fine-grained longitudinal data or a structural model could address. 
Third, the links between ideology, behavior and embedding position probably differ across users and contexts, so no single causal model would apply. 
Our design is therefore descriptive. It establishes that a Left-Right scale exists in the representation and affects its outputs. For the legal question we address, this is the relevant quantity, because regulations such as GDPR turn on what processing reveals rather than on why. 

We note some limitations of our study. Our exposure data comes exclusively from browser feeds, and we do not account for differences between browser and mobile app navigation. Additionally, our embedding reconstruction relies on the follower network, whereas X's recommendations incorporate further signals such as user actions on posts \cite{satuluri2020simclusters}. Yet, if a subset of input signals suffices for the recommender to learn ideological representations, the overall system retains this capability at least locally among subsets of users, still raising the same legal implications.
Similarly, post-ranking heuristics for diversification and moderation \cite{TwitterGithubMain} are not modeled in our study; however, the recommendations we observe still contain measurable traces of political preferences despite these heuristics.
Our volunteer cohort is self-selected and skews Left-leaning (see Materials and Methods). However, embeddings trained exclusively on recommendations served to Right- or far-Right-leaning volunteers still replicated our results (see Materials and Methods, Section~\ref{subsec:embeddings_robustness}), indicating that the political structure captured by the recommender is inherent to the relational signals the system processes rather than an artefact of cohort composition.
We present additional experiments in the Materials and Methods section demonstrating limited effects of these potential bias sources across alternative cohort compositions and training strategies. 

Our results bear on how data protection law treats learned representations, but they do not settle the question. GDPR Article 9(1) prohibits processing that reveals political opinions (unless exceptions apply, such as users giving consent or processing in the public's interest). This prohibition applies regardless of the controller's aim and of the accuracy of what is revealed (CJEU, C-252/21, para. 69). 
Several regulations worldwide contain similar provisions, including that of Brazil, South Korea, and Switzerland, with profiling standing globally as a relevant matter of legal scholarship.
Our results supply the empirical case these debates have lacked: a landmark political opinion scale, calibrated on an established surveys instrument, that emerges inside a deployed system serving content to hundreds of millions of users.

Our study exposes an increasing tension, brought about by the pervasiveness of computation in society, the flexibility of machine learning models capable of treating large and heterogeneous inputs, and the continuous improvement of AI explainability methods such as ours.
%
%
If the emergence of protected attributes falls outside the scope of data protection regulations, data controllers such as social platforms may essentially and legally gain access to variables that amount to the protected attributes. In the case of our study, for instance, X would have trivial access to an unnamed variable, produced without intent, that approximates and passively and massively administered Left-Right self-positioning survey, nullifying the spirit of privacy protection in data regulation laws.
It is worth noting that the direction may be identified, as we did, computing survey-identified ideology scaling models, but also through options legally available to private companies, such as identifying Left-Right positions of political figures or media outlets in the embedding.
%
%
If, on the contrary, emergent representations of protected categories fall within data protection, regulation would needs to move, from specifying what attributes are protected, to narrowly defining these attributes and indicating which statistical learning systems must actively inspect for their emergence.
For if emergent representations are to be regulated, it is conceivable that a wide range of statistical learning systems might develop them, from simple linear regressions used across several industries, to large recommendation systems in social platforms and AI systems.
Protected personal attributes include, for instance in EU's GDPR, racial or ethnic origin, political opinions, religious or philosophical beliefs and sexual orientation, among other, which would require operational definitions enabling active detection efforts.
Both alternatives raise important concerns and our study is, to the best of our knowledge, the first to provide concrete evidence on its materialization on a real system deployed for the public serving large populations worldwide.

Our article also explores the possibility of removing political opinion from the embedding of the recommender.
%
%
Our experiments suggest that removing political information marginally affects relevance while increasing the political diversity of recommendations. The overall effects of such a procedure, in terms of accuracy, relevance and diversity, are to be tested in the future in different settings and populations.


\clearpage 

\bibliography{references} 
\bibliographystyle{sciencemag}



\section*{Acknowledgments}


\paragraph*{Funding:}

This work has been partially funded by the “European Polarisation Observatory” (EPO) of CIVICA Research (co-)funded by EU’s Horizon 2020 programme under grant agreement No 101017201, by European Union Horizon program project “Social Media for Democracy” under grant agreement No 101094752 (www.some4dem.eu), by the Very Large Research Infrastructure (TGIR) Huma-Num of CNRS, Aix-Marseille Universit´e and Campus Condorcet, and by Project Liberty Institute project “AI-Political Machines” (AIPM). P.B. acknowledges financial support from the CFM Foundation for sciences.

\paragraph*{Author contributions:}

Conceptualization: P.B, P.R.; Methodology: P.B, P.R.; Software: P.B., P.R.; Formal analysis: P.B, P.R.; Investigation: P.B, P.R.; Visualization: P.B.; Writing – original draft: P.B, P.R.; Writing – review \&{} editing: P.B, P.R.; Supervision: P.R.; Funding acquisition: P.R.

\paragraph*{Competing interests:}

The authors declare that they have no competing interests.

\paragraph*{Data and materials availability:}

The data and the code needed to reproduce our results are available at \url{https://osf.io/mz4aw/?view_only=7bc63998850d4167befbe2db5dab79d7}, and will be identified with a DOI upon acceptance. No physical materials were generated in this work.


\subsection*{Supplementary materials}
Materials and Methods


\newpage


\renewcommand{\thefigure}{S\arabic{figure}}
\renewcommand{\thetable}{S\arabic{table}}
\renewcommand{\theequation}{S\arabic{equation}}
\renewcommand{\thepage}{S\arabic{page}}
\setcounter{figure}{0}
\setcounter{table}{0}
\setcounter{equation}{0}
\setcounter{page}{1} 


\begin{center}
\section*{Supplementary Materials for\\ \scititle}

Paul~Bouchaud$^{\dagger}$,
Pedro~Ramaciotti$^{\ast\dagger}$\\ 
\small$^\ast$Corresponding author. Email: pedro.ramaciotti-morales@cnrs.fr\\
\small$^\dagger$These authors contributed equally to this work.
\end{center}

\renewcommand{\contentsname}{Materials and Methods}
{\small
\tableofcontents
}

\newpage


\section{X's personalized recommendations}
\label{sec:x_recsys}

\subsection{The ``Who to Follow" service}
\label{subsec:x_recsys_wtf}

The ``Who to Follow" service aims to enhance Twitter's user engagement and network growth by recommending accounts that users don't currently follow but may find interesting, based on shared interests and common connections \cite{DiscoveringWTF}. This feature is important for the platform to sustain platform engagement through user discovery and crucial for new accounts. 
By the platform's own report, by 2015, 35\% of new Twitter users started following friend recommendations on their first day \cite{goel2015follow}. Launched in 2010, Twitter's ``Who to Follow" service encompasses two complementary components \cite{gupta2013wtf}: 1) recommending accounts that users may be ``Interested in" based on content affinity, and 2) suggesting ``similar users" who exhibit analogous following patterns, grounded in the homophily principle \cite{McPherson2001}. With Twitter's vast user base in the hundreds of millions, this dual approach sought to surface relevant accounts from the platform's extensive network, and has been reported to drive over one-eight of the new connections on Twitter \cite{goel2015follow}.\newline

\begin{figure}[h]
\centering
\includegraphics[width=\textwidth]{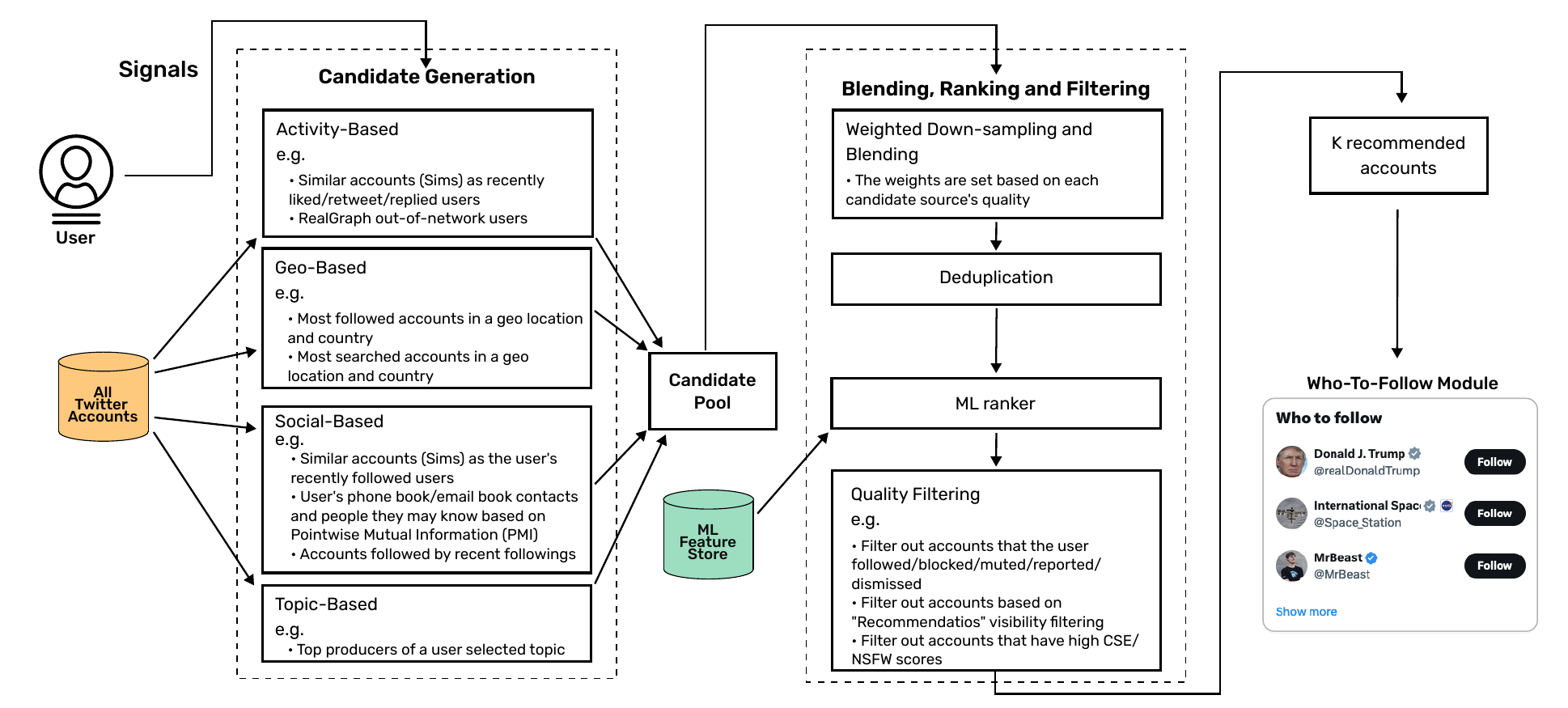}
\caption{Diagram (adapted from \cite{TwAccountRecommendations}) of Twitter/X ``Who to Follow" (WTF) account recommendation system that computes the three friend recommendations that are proposed on the right-hand lateral bar in the website of X. The WTF recommender system can be organized in three main sub-parts: 1) candidate generation (left part of the figure), 2) integration of features computed by machine learning models (at the center of the figure), and 3) candidate mixing and ranking (on the right part of the figure).}
\label{fig:diagramTwitter}
\end{figure}

To accomplish such a task, rather than operating as a monolithic system, Twitter's ``Who to Follow" service integrates recommendations from diverse algorithms ---a blend of 20 different algorithms in 2013 according to Gupta et al. \cite{gupta2013wtf}. An overview of the service is displayed in Extended Figure~\ref{fig:diagramTwitter}. The service combines, for instance, Personalized PageRank \cite{Fogaras2005} and SALSA \cite{Lempel2001} ranking algorithms with heuristics based on the assumption of social homophily, such as pooling candidates from most common second-degree neighbors.\newline

Acknowledging that not all follow relationships carry equal significance in terms the interest and preferences of users, Twitter measures the predicted ``engagement per impression'' to evaluate recommendation quality, with the objective of increasing the overall engagement on the platform \cite{gupta2013wtf}.\newline

Additional details about Twitter's more recent architecture of the ``Who to Follow" profile recommendation system were disclosed in April 2023  together with partial release of the code base of a few key components \cite{TwitterGithubMain}. The recommender system integrates 17 components contributing candidates for profile recommendation \cite{TwitterGithubWTF}, encompassing previously established approaches like SimCluster \cite{satuluri2020simclusters}, SALSA, and Personalized PageRank, alongside additional heuristics, including: considering accounts with which users have recently interacted (through explicit actions such as likes and retweets, or implicit signals like profile visits), and accounts two-random hops from users in the network of interactions \cite{TwitterGithub2hop}. The system also considers promoted (advertiser) accounts and contacts from users' address books, when synced with Twitter.\newline

Each candidate account undergoes scoring by machine learning models that compute a weighted sum between two probabilities \cite{TwitterGithubWTF}: 

$$
P\left(\texttt{user follows account}|\texttt{account recommended to user}\right),$$

\noindent{}and 
$$
P\left(\texttt{user gives positive engagement to account}|\texttt{user follows account}\right).$$\newline

The accounts for which these probabilities are estimated then pass through refinement heuristics before a ranking of the highest-scoring candidates, which are finally recommended to users.

\subsection{Heterogeneous Information Network representation for recommendation}
\label{subsec:x_recsys_hin}

\begin{figure}
    \centering
    \includegraphics[width=\linewidth]{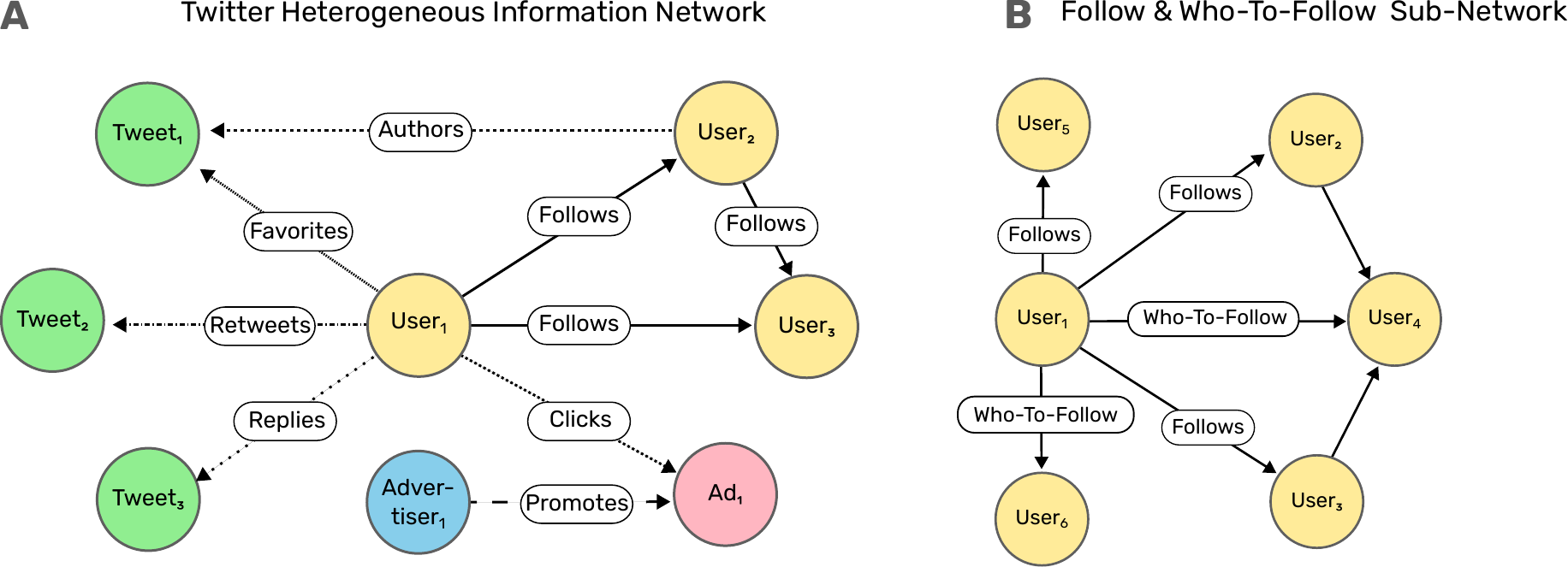}
    \caption{Diagram of the representation of the data used for training the recommender system in X, adapted from \cite{el2022twhin}  (A) X's Heterogeneous Information Network (HIN, a graph abstraction model for Knowledge Graph data) showing interconnections between entities that are relevant for recommendation (tweets, users, advertisers, advertisements) through diverse relations (authorship, follows, interactions, shares or clicks). (B) Part of the HIN representation relevant for the inference of the embedding space in our study, including ``Follow'' and  ``Who to Follow" relationships between users form a sub-network analyzed in this work.}
    \label{fig:twhin}
\end{figure}

In its candidate scoring process, the ``Who to Follow" service employs feature enrichment to capture accounts' social environment \cite{TwitterGithubFeatureHydratation}. 
To support this process Twitter researchers \cite{el2022twhin, twhinbert} implemented a Heterogeneous Information Network (HIN) for personalized recommendations, generating dense entity embeddings, lying in a high-dimensional space, that serve downstream ranking models and candidate generation.\newline
The HIN integrates diverse entity types, such as tweets, users, and advertisers, connected through varied relationships such as authorship, clicks, retweets, and follows; as illustrated on Extended Figure~\ref{fig:twhin}. Each entity and edge type is encoded as a learnable embedding vector, optimized to predict specific types of entity-entity links. Extended Figure~\ref{fig:twhin}A illustrates the HIN data representation leveraged by X's recommender system, as shown in the article by X's researchers \cite{el2022twhin}.

X's HIN-based recommender system uses observed occurrences of relations between different entities to train an embedding space for all entities and relations on which to compute rankings of potential entities to recommend.
The ranking relies on a scoring function $f$ encapsulating inner product similarity between entities in the embedding space.
Consider a scoring function $f(e) = f(\phi_s, \varphi_r, \phi_t)$ for an edge $e$ of type $r$ in the HIN connecting source entity $s$ to target entity $t$. The model optimizes entities' embeddings $\phi_s, \phi_t$ and relation's embeddings $\varphi_r$ by maximizing the log-likelihood $\log \sigma(f(e))$ (with $\sigma$ being the logistic sigmoid function) for observed edges while minimizing it for unobserved ones (i.e., relations that do not exist in the data). In \cite{el2022twhin, twhinbert}, Twitter report adopting, for model simplicity, the TransE (translating embeddings) scoring function \cite{bordes2013transE}. For an edge $e = (s, r, t)$, TransE is defined as the sum of inner products:

\begin{equation}\label{eq:transe}
f(e) = f(s, r, t) = \phi_s^T\phi_t + \varphi_r^T \phi_t.
\end{equation}

This function operates by adding the relation embedding $\varphi_r$ to the source entity embedding $\phi_s$ and computing its inner product with the target entity embedding $\phi_t$. In a successfully trained model, the inner product between a target account's embedding $\phi_t$ and, for instance, the sum of a source account's embedding and the ``Follow" relationship embedding $(\phi_s + \varphi_\text{Follow})$ should yield high values for actual follower relationships and low values otherwise.


\section{Inferring users' embeddings in the recommender system}
\label{sec:inferring_embeddings}

\subsection{Loss function defining the learning task}
\label{subsec:loss_function}

We are interested in this article in inferring users' positions in this HIN-based embedding space.
We use the detailed specification of the machine learning procedures provided by X's researchers \cite{el2022twhin} to define the task of inferring embeddings that best approximate those that were learned and leveraged by the platform's recommender system.
As displayed in Extended Figure~\ref{fig:twhin}B, we consider the sub-network solely comprised of Twitter accounts connected through two types of edges corresponding to the empirical data available for our study: ``Follow'' (following links from one user to its followees or friends) or ``Who to Follow'' (link connecting a user to those that have been recommended by the platform). Our goal is to learn user and edge type embeddings that effectively predict Twitter's ``Who to Follow" recommendations to users that we observe empirically.\newline

Integrating ``Follow" relationships in the inference of positions $\phi$ in the embedding serves two purposes: it aligns with Twitter's architecture, where user-to-user connections drive 9 of the 17 candidate sources for profile recommendations \cite{TwitterGithubMain}, and it mitigates the sparsity of ``Who to follow" observations. As shown in the training section below (Section~\ref{subsec:training}), integration ``Follow" relationships enhance the model's ability to predict ``Who to Follow" recommendations.\newline

\noindent{} To learn embeddings $\phi$ for users, we consider a convex combination of both relationships through the following loss function $\mathcal{L}$ :

\begin{equation}
\label{eq:loss_si}
\mathcal{L} = \alpha \mathcal{L}_\text{WTF} + (1-\alpha) \mathcal{L}_\text{Follow},
\end{equation}\newline

\noindent{}where each relationship-specific loss component is defined as:

\begin{equation}\label{eq:loss_type_si}
\mathcal{L}_r = \sum_{e \in \mathcal{P}_r} \log \sigma\big(f(e)\big) + \sum_{e' \in \mathcal{N}_r} \log \sigma\big(-f(e')\big)
\end{equation}\newline

In this formulation, $\mathcal{P}_r$ and $\mathcal{N}_r$ denote the sets of positive and negative edges for a relationship type $r$ with $r \in$ \{``Follow", ``Who To Follow"\}. ``Who To Follow" is hereinafter also shortened as ``WTF".

Both to generate and rank candidate accounts, Twitter's systems use account features to assess which candidates relate most strongly to a given user. Through the loss function $\mathcal{L}$, the account representations we learned from our experimental data aim to approximate Twitter's internal representations $\phi$. Two accounts will be placed closely in our representation if Twitter recommended one account to a specific user, indicating it was a top-scoring candidate according to Twitter's internal user representation system.

\subsection{Datasets used for training and approximation of the embedding}
\label{subsec:training_datasets}

Our inference of users' embeddings relies on two data sources: \ref{subsubsec:crowdsourced_dataset}) ``Who To Follow" recommendations gathered via a crowdsourcing initiative and \ref{subsubsec:follower_network_dataset}) ``Follow" relationships collected through Twitter's API.

\subsubsection{Crowdsourced (plug-in) dataset of recommendations served in the platform {and participant survey}}
\label{subsubsec:crowdsourced_dataset}

The main hurdle in the inference of embedded positions using $\mathcal{L}$ in Eq.~\eqref{eq:loss_si} stems from acquiring ``Who To Follow" recommendation data. We address this through a custom browser extension that unobtrusively collects ``Who To Follow" recommendations displayed to volunteer users that installed the extension as they scroll their feed. 

\begin{figure}
\centering
\includegraphics[width=\textwidth]{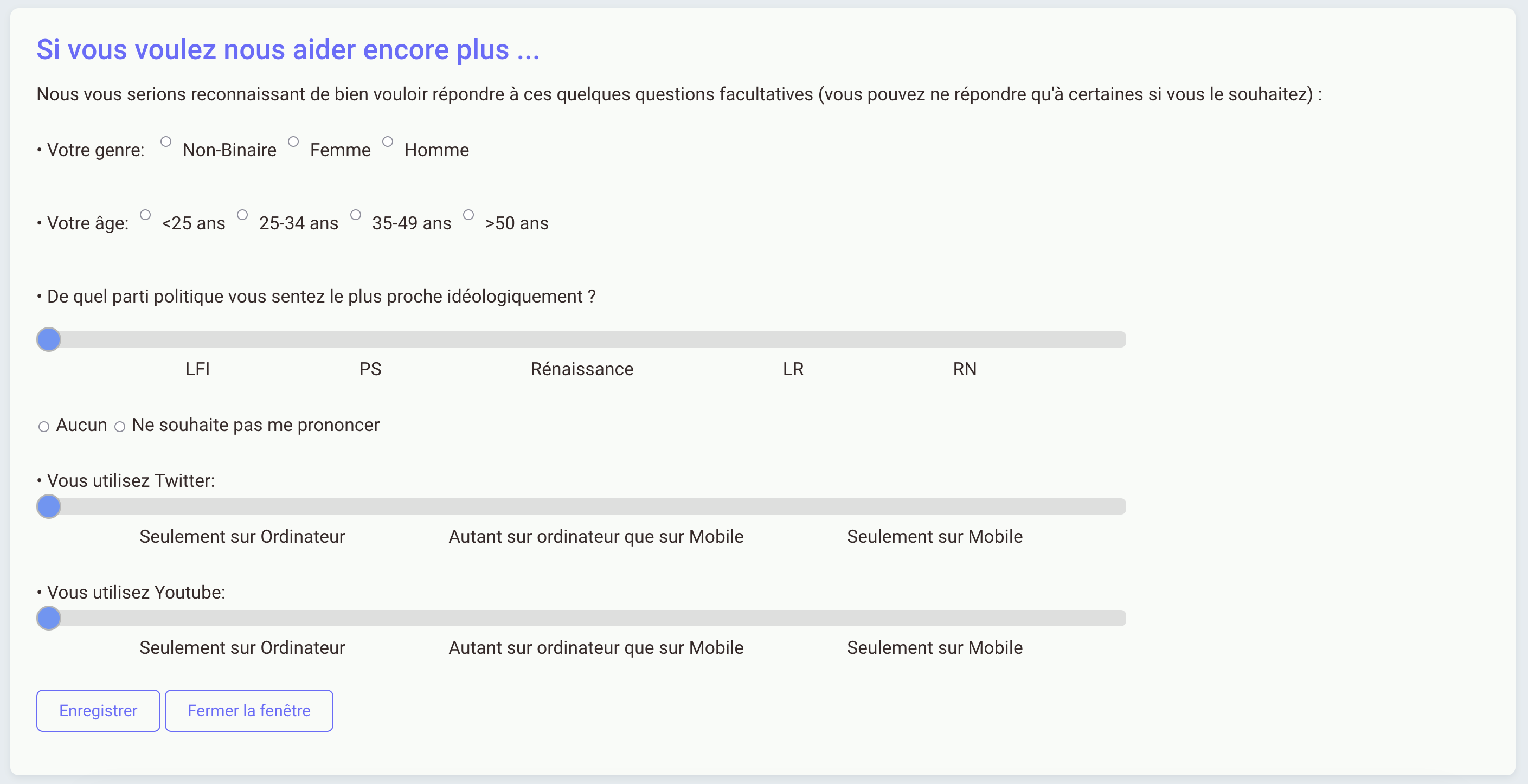}
\caption{{Screenshot of the survey presented to volunteers of the data donation program (in French), asking: their gender, age group, their ``ideological position'' on a 101-point scale with political parties arranged from Left to Right for reference and orientation, whether they connect to Twitter from their phone or their computer or both. The survey also asks the same question for YouTube (not used in this study).}}
\label{fig:horus_survey}
\end{figure}

This browser extension was installed by self-recruited participants following mainstream media coverage (newspapers and radio) and Twitter advertising in fall of 2022. The study incentivized participation by offering volunteers personalized analytics on their Twitter feed's political diversity, delivered upon sufficient data collection. No monetary incentives were offered to participants.
{Participants were required to answer a brief survey, shown in Extended Figure~\ref{fig:horus_survey}, asking their gender, age group, their ``ideological position'' on a 101-point scale (from 0 to 100) with political parties arranged from Left to Right for reference and orientation, whether they connect to Twitter from their phone or their computer or both (on a gliding 21-point scale from 0 to 100 with intervals of length 5). 
The parties used for reference are LFI (\textit{La France Insoumise}, or Indomitable France), PS (\textit{Parti Socioliste}, or Socialist Party), \textit{Rénaissance} (Macron's party, previously known as \textit{En Marche!}), LR (\textit{Les Républicains}, or The Republicans), and RN (\textit{Rassemblement National}, Le Pen's party previously known as \textit{Front National}). The ordering of parties is taken from the Chapel Hill Expert Survey \cite{Jolly2022} and are equidistant.
The survey also asks a question related to YouTube use (not used in this study). The answers to this survey are not used directly in the conclusions of our study, which relies on inferred socio-demographic and ideological attributes. We administered this survey to produce validation metrics for the inferred attributes of users. The following box provides the questions presented to the respondents, translated into English (Extended Figure~\ref{fig:horus_survey} for a screenshot).}\newline

\medskip

\noindent{}\textbf{Translation of the survey used to compute quality metrics for inferred user attributes.}\\
\noindent{}\fbox{\begin{minipage}{\textwidth}
{\it
{If you want to help us further...}\\
We would be grateful if you could answer a few optional questions (you can complete the survey partially if you want):
\begin{itemize}
 \item Your gender: $\circ$ Non-Binary\, $\circ$ Female\, $\circ$ Male
 \item Your age: $\circ$ $<$25 years\, $\circ$ 25-34 years\, $\circ$ 35-49 years\, $\circ$ $>$ 50 years
 \item To which political party do you feel the closest?\\
 (gliding scale)\\
 LFI\quad\quad PS\quad\quad Rénaissance\quad\quad LR\quad\quad RN\\
 $\circ$ None\, $\circ$ I don't whish to answer 
 \item You use Twitter:\\
 (gliding scale)\\
 Only on PC\quad\quad Equally on PC and on my mobile \quad\quad Only on my mobile
 \item You use YouTube:\\
 (gliding scale)\\
 Only on PC\quad\quad Equally on PC and on my mobile \quad\quad Only on my mobile
\end{itemize}
}
\fbox{Save}\fbox{Close}
\end{minipage}}\newline

All participants provided informed consent after receiving detailed information about the study objectives and data collection scope. Only after obtaining consent did the browser extension begin monitoring participants' Twitter interactions. The reader is referred to the Ethics and Compliance section of the main article for the reference to the Ethics Review Board and legal data deposited associated with the data used in this study.\newline

\begin{figure}
\centering
\includegraphics[width=\textwidth]{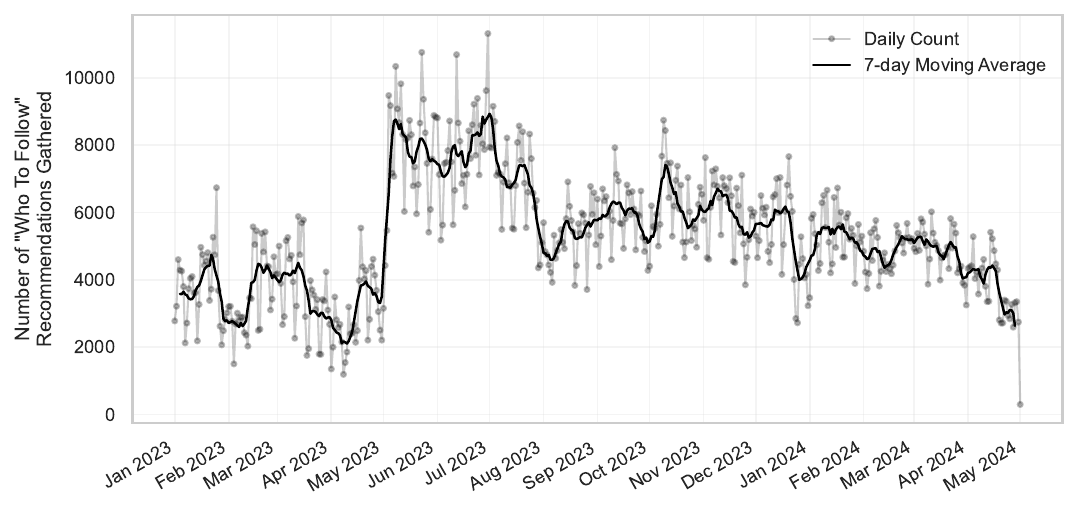}
\caption{Evolution of the number of friend recommendations (WTF) seen by volunteers in our study and collected with our browser plug-in. The observation period of the study extended from January 1st, 2023, to May 1st, 2024. The number of collected recommendations depends on the number of volunteers participating in the project, which saw an increase in May 2024 due to the visibility of changes in content moderation and data accessibility brought about by the change in ownership of the platform. We include tests of accuracy and reliability tests in the inference of the embedding space, and that use different temporal windows in sampling, to demonstrate that the inference is robust throughout the whole period of observation.}
\label{fig:horus_collect_timeline}
\end{figure}

From January 1st, 2023 to May 1st, 2024, we gathered over 2,549,008 "Who To Follow" profile recommendations displayed to 682 unique Twitter users logged-in using our browser plug-in. {To ensure the robustness of our results, we only include users having received 30 ``Who To Follow'' profile recommendations,} henceforth referred to as volunteers from the set of volunteers $\mathcal{V}$. Extended Figure~\ref{fig:horus_collect_timeline} shows the daily rate of collection of recommendations from users that used our browser plug-in.

Participation fluctuated with media coverage of the project, peaking in spring 2023 with over ten thousand ``Who To Follow" recommendations collected daily; as displayed on Extended Figure \ref{fig:horus_collect_timeline}. Users could freely withdraw from the study by removing the browser extension, leading to natural attrition in collection rates during periods of reduced public outreach. We observe an increase in participation around the time of policy changes in the platform linked with changing ownership. Below, in Section~\ref{subsubsec:robustness_training_datasets} we show specific experiments (experiment \ref{subsubsec:robustness_training_datasets}.\textit{iii}) aiming at testing the impact of, and ensuring that this exogenous factor does not affect the validity of our results.\newline

\begin{figure}
\centering
\includegraphics[width=\textwidth]{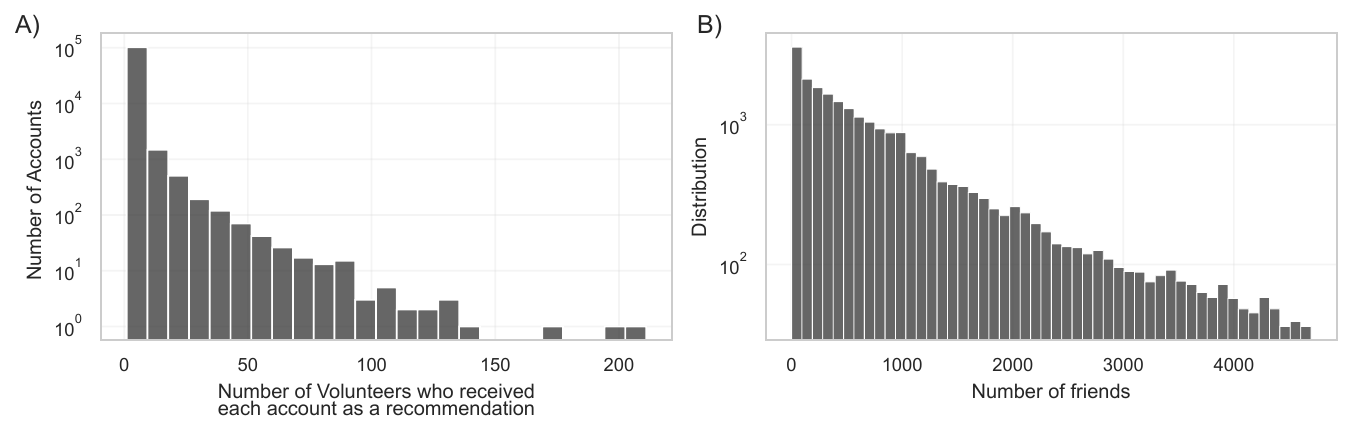}
\caption{ (A) Distribution of accounts based on how many volunteers received them as recommendations. (B) Distribution of the number of friends accounts for each user in $\mathcal{E}$.}
\label{fig:horus_distrib_rec}
\end{figure}

Within the more than 2.5 million ``Who to Follow" recommendations collected, we observed 105k unique Twitter accounts that were recommended, with a highly skewed distribution (see Extended Figure~\ref{fig:horus_distrib_rec}). For instance, while 74.5\% of accounts were recommended to only one volunteer, Elon Musk's account was recommended to 35.4\% of volunteers who did not follow him already. Extended Table~\ref{tab:twitter_recommendations} presents the five accounts recommended to the highest fraction of non-following volunteers. Besides Twitters' new owner, we find a major French news broadcast channel, a YouTube creator, followed by the French president and a former prime minister.\newline

\begin{table}
\centering
\caption{Most recommended accounts seen by our volunteers, with the rate or fraction of volunteers that saw the recommendation during the observation period.}
\label{tab:twitter_recommendations}
\begin{tabular}{|l|c|r|}
\hline
\textbf{Twitter Handle} & \textbf{Recommendation Rate} & \textbf{Description} \\
\hline
@elonmusk & 35.4\% & Owner, CTO and Exec. Chairman of Twitter \\
\hline
@BFMTV & 30.8\% &  French News Broadcast Channel \\
\hline
@MrBeast & 28.1\% & Popular YouTube Content Creator \\
\hline
@EmmanuelMacron & 22.3\% & President of France (2017-2027)\\
\hline
@GabrielAttal & 20.7\% & Former Prime Minister of France  (2024)\\
\hline
\end{tabular}

\end{table}

Overall, 26\,338 unique Twitter accounts were recommended by Twitter to at least two volunteers. We will refer hereinafter to this set of users as $\mathcal{U}$.

\subsubsection{Dataset of follower networks obtained with the platform's API}
\label{subsubsec:follower_network_dataset}

To complement the ``Who To Follow'' profile recommendations collected through crowdsourcing, we systematically collected the list of friends of each user in $\mathcal{E} = \mathcal{V} \cup \mathcal{U}$ via the Twitter API. Among 26\,509 users in $\mathcal{E}$, 24\,918 had public accounts and at least one friend. Our volunteers $v \in \mathcal{V}$ followed an average of 661 accounts, with a median of 320.
This collection allows us to integrate the follower network subtended by users in $\mathcal{E}$ in the loss function $\mathcal{L}$ for the inference of their embeddings $\phi$.

\begin{figure}
\centering
\includegraphics[width=0.5\linewidth]{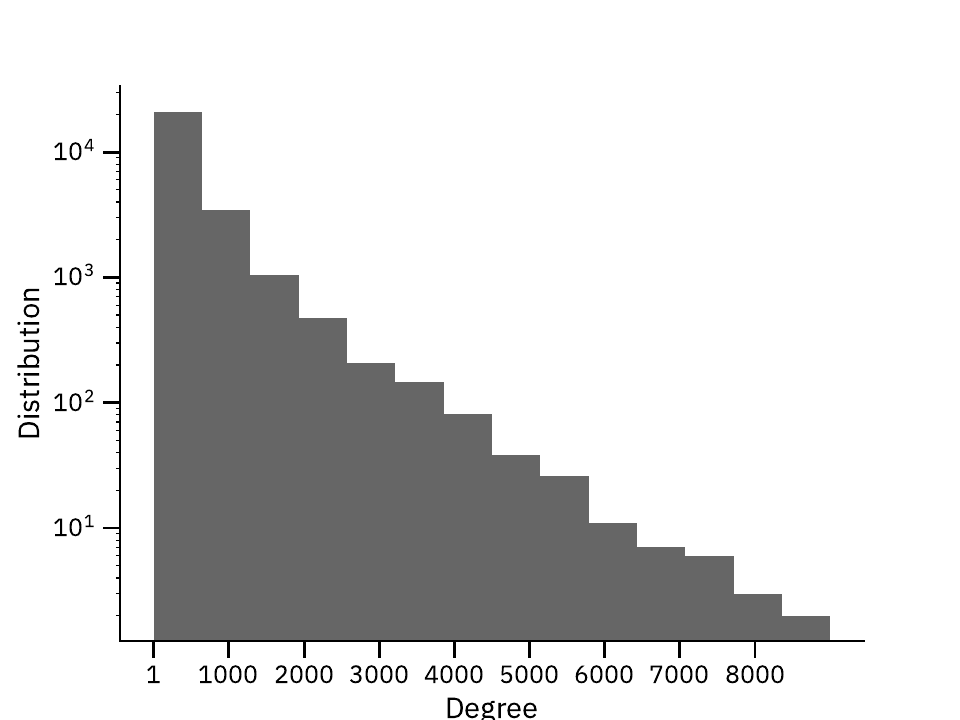}
\caption{Distribution of accounts' degree.}
\label{fig:degree_si}
\end{figure}

{We display in Figure \ref{fig:degree_si} the distribution of degrees (number of Follow and Who To Follow relationships) for each user. In particular, we observe that 99.1\% of accounts have a degree larger than 5, and 95.6\% larger than 10.}

\subsection{Negative sampling}
\label{subsec:negative_sampling}

Combining our two datasets (described in Sections \ref{subsubsec:crowdsourced_dataset} \&{} \ref{subsubsec:follower_network_dataset}), we obtain the set of positive edges for the ``Follow" and ``Who to Follow" relationships, resulting in $|\mathcal{P}_\text{Follow}| =$ 4 441 670 ``Follow'' edges and $|\mathcal{P}_\text{WTF}| =$  110 164 ``Who to Follow'' edges. As we lacked access to the exact set of candidate accounts scored by Twitter's algorithmic systems, from which the ``Who to Follow'' list is computed after ranking, we rely on negative edge sampling.
Specifically, we generated negative edges by corrupting positive ones, replacing target accounts with randomly sampled accounts. Building upon established literature on Knowledge Graph embeddings \cite{7358050} and the impact of negative sampling on model quality \cite{mikolov2013}, we extended the negative sampling strategy of PyTorch-BigGraph \cite{lerer2019pytorchbiggraph}, used by Twitter in \cite{el2022twhin}.\newline

\noindent{} Specifically, we combine three sampling methods in equal proportion:

\begin{itemize}
\item \textbf{Uniform sampling}: Target users are randomly selected from the entire pool of accounts with equal probability;
\item \textbf{Prevalence-based sampling}: Target users are sampled with probability proportional to their frequency in the positive edge set, reflecting their natural occurrence in the recommendations;
\item \textbf{Second-neighborhood sampling}: Target users are selected from friends of friends, i.e. being two hops away in the user-to-user network, capturing the local network structure.
\end{itemize}

Both uniform and prevalence-based sampling approaches present distinct limitations when used independently. Uniform sampling risks creating a model that learns to score edges primarily based on source and target node frequency rather than meaningful relationships between nodes. Conversely, prevalence-based sampling may not sufficiently penalize high scores assigned to rare entities. PyTorch-BigGraph \cite{lerer2019pytorchbiggraph}, used by Twitter in \cite{el2022twhin}, addresses these limitations by combining, in equal proportion, targets sampled using both strategies.

Informed by Twitter's own algorithms \cite{TwitterGithubMain}, we complemented the negative sampling by incorporating non-recommended second-order friends and excluding first-degree connections (i.e. direct friends). The homophilic nature of social networks \cite{McPherson2001} (and in particular Twitter networks \cite{colleoni2014echo,kang2012using,halberstam2016homophily}) motivates the search for new ties from friends of friends rather than random strangers. We observe in Twitter's algorithmic pipeline different candidate sources tied to this notion of homophily: either by computing similarity between accounts through Follow2vec, Twitter's adaptation of Word2Vec over the user-user follower network \cite{TwitterGithubSims}, embedding closely users tied to similar accounts \cite{node2vec}, or by explicitly looking for second-order connections through two-hop random walks \cite{TwitterGithub2hop} or second-order friends following the source users \cite{TwitterGithubTriangular}. 

\begin{figure}
\centering
\includegraphics[width=\linewidth]{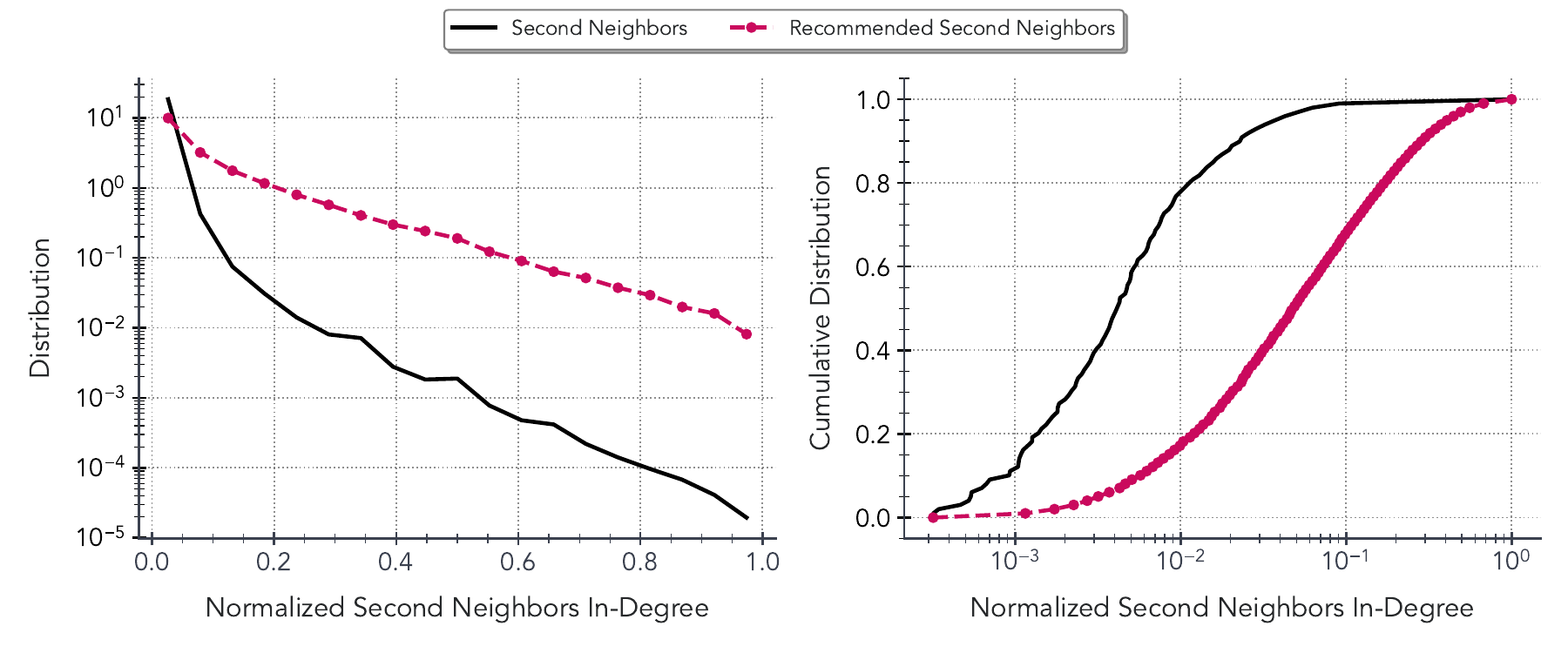}
\caption{Analysis of the degree distribution of second-degree neighbors. Distribution (left figure) and Cumulative Distribution (right figure) of the in-degree of second-degree neighbors, with respect to first-degree neighbors, normalized per volunteer by the largest in-degree. The distribution for recommended second-degree neighbors is shown as a dashed red line.}
\label{fig:wtf_distrib_degree}
\end{figure}

An analysis of the empirical ``Who to Follow" recommendations reveals that 62.4\% of recommended accounts are second-degree neighbors to the user receiving the recommendations. Among recommended second neighbors, only 15.8\% are followed by a single of  the user's friends, while half are followed by at least 8 friends. The degree distribution shown in Extended Figure \ref{fig:wtf_distrib_degree} showcases that recommended second-degree neighbors tend to have substantially more first-degree friend connections compared to non-recommended ones.

Consequently, prevalence sampling from non-recommended second-order friends provides particularly valuable negative examples, as these represent plausible but ultimately non-recommended connections ---effectively serving as ``hard negatives" for the model.\newline

Finally, we curate the negative edge sets $\mathcal{N}_\text{Follow}$ and $\mathcal{N}_\text{WTF}$ by combining uniform, prevalence-based, and second-neighborhood sampling methods in equal proportions. We adopt a negative-to-positive ratio of 3:1 for both ``Follow'' and ``Who to Follow'' relationships, ratio of the same magnitude than literature and industry standards \cite{bordes2013transE}, resulting in $|\mathcal{N}_\text{Follow}| = 3|\mathcal{P}_\text{Follow}|$ and $|\mathcal{N}_\text{WTF}| = 3|\mathcal{P}_\text{WTF}|$.

\subsection{Training an approximation of the embedding of X's recommender system}
\label{subsec:training}

\subsubsection{Implementation}

Following Twitter's implementation \cite{twhinbert}, we trained 256-dimensional users and relationship embeddings under the loss function defined in Equation \eqref{eq:loss_si}. The model parameters were Glorot initialized \cite{Glorot}, in order to prevent vanishing or exploding gradients during training, and optimized with Adagrad \cite{adagrad} in PyTorch.

We randomly segmented the curated edge set into 80\%, 10\%, and 10\% splits for training, hyperparameter tuning, and final evaluation. The splitting was performed independently for each relation type to maintain consistent Follow/WTF ratios across all sets.

\subsubsection{Tuning hyperparameter $\alpha$ in loss function $\mathcal{L}$}

To optimize prediction of ``Who to Follow" edges, we tuned the hyperparameter $\alpha \in [0,1]$ from loss function on a 10\% holdout validation dataset. The parameter $\alpha$ in Equation \ref{eq:loss_si}, balances in the loss the contributions of profile recommendations and ``Follow'' relationships. Setting $\alpha = 1$ trains the model exclusively on ``Who to Follow" data, while $\alpha = 0$ uses only ``Follow" relationships.

\begin{figure}
\centering
\includegraphics[width=\linewidth]{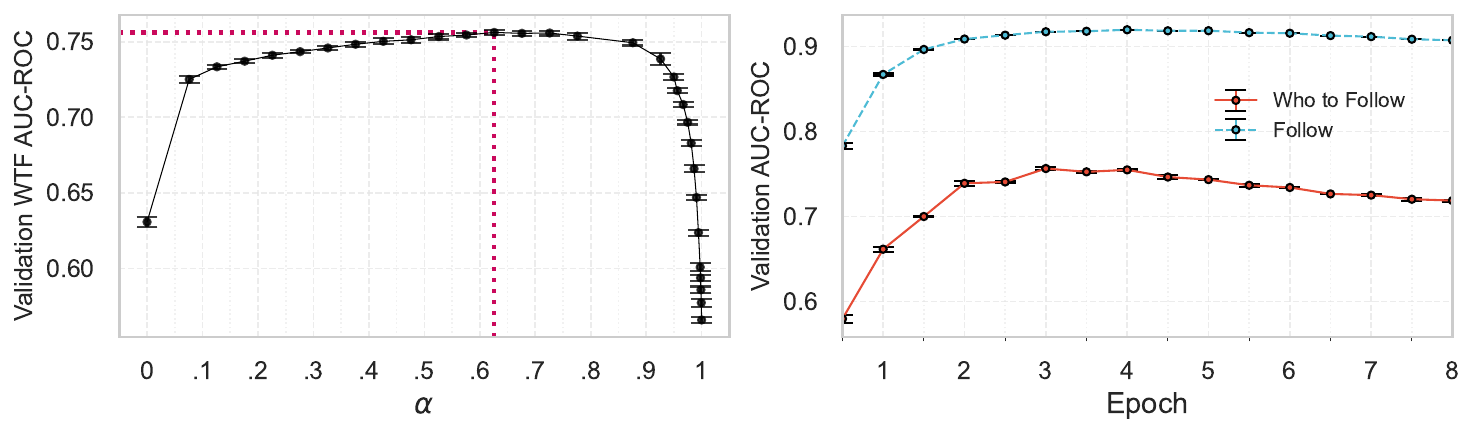}
\caption{Accuracy of the trained algorithm and its embedding space in predicting observed data for different training parameters. (Left figure) AUC-ROC scores for ``Who to Follow" predictions as a function of $\alpha$, with error bars showing standard deviation across ten replications. (Right figure) AUC-ROC values for ``Who to Follow" (in \textcolor{training_orange}{orange}) and ``Follow" (in \textcolor{training_cyan}{cyan}) predictions versus training epoch}
\label{fig:roc_hypertune_alpha}
\end{figure}

Due to the nondeterministic nature of gradient descent, we trained the model ten times for each $\alpha$ value over three epochs, evaluating the ROC AUC on the validation dataset for each replicate. We observe on Extended Figure \ref{fig:roc_hypertune_alpha} that the AUC-ROC ---measuring the probability that the model ranks a randomly chosen positive instance higher than a randomly chosen negative one--- increases with $\alpha$ until reaching a peak of $.7562 \pm .0018$ (standard deviation over ten replicates) at $\alpha = 62.6\%$, after which it declines. 

The relative scarcity of profile recommendation data, especially in comparison with follow relationships, limits predictive power when training solely on user recommendations, i.e. $\alpha = 1$ . The tuning of $\alpha$ demonstrates the benefits of heterogeneous information network embeddings, even for single-entity-single-relationship downstream tasks such as account-to-account follow recommendations, corroborating Twitter's observations \cite{el2022twhin} that guided their implementation \cite{TwitterGithubMain}.

In what follows, all training uses the optimal value $\alpha = 62.6\%$, which assigns each ``Who to Follow" training edge 1.67 times the weight of each ``Follow" edge. Also, we trained the model over eight epochs, and three independent runs, measuring the AUC-ROC on the validation dataset every half epoch. As shown in Extended Figure \ref{fig:roc_hypertune_alpha}, the optimal number of epochs was determined to be three.


\section{Validating the inferred embeddings $\Phi$ for users}

In this section, we provide different evidence to the quality of these embeddings. First, in Section~\ref{subsec:predictive_quality_embeddings} we validate our computed embeddings by showing that, if used for recommendation of friends using X's prescribed recommender system methods, they are able to predict the users that were recommended with high accuracy. Then, in Section~\ref{subsec:embeddings_robustness}, we show that these embeddings are robust under a diversity of sub-sampling strategies addressing possible biases that might have been introduced by our research design.

\subsection{Testing the performance of embeddings in predicting recommended users}
\label{subsec:predictive_quality_embeddings}

After training for three epochs on a combination (1:3) of positive and negative edges from ``Who to Follow" recommendations and follower relationships, weighted (1.67:1), we learned $\Phi\in\mathbb{R}^{|\mathcal{E}|\times 256}$, the accounts embeddings, along with ``Who to Follow" $\varphi_\text{WTF}$ and ``Follow" $\varphi_\text{Follow}$ relationships embeddings. Our evaluation approach, while constrained by our lack of definitive negative set of candidates considered by Twitter's model, demonstrates strong predictive performance.\newline

On a 10\% held-out test set, the learned embeddings achieve an AUC-ROC of $70.0\%$ for predicting ``Who to Follow" recommendations when using non-recommended second-order friends as negative edges. When evaluating positive test recommendations against a pool of randomly sampled 100 non-recommended second-order friends, the model achieves a precision@1 of $72.5\%$. Matching Twitter's ``Who to Follow" interface which typically displays slate of three recommendations, we evaluate precision@3, reaching $69.1\%$, while precision@10 equals $64.3\%$.\newline

We emphasize a key distinction in our evaluation: unlike standard recommendation systems that predict user reactions to suggestions, our model predicts Twitter's recommendation pipeline outputs. This shift in objective—predicting model behavior rather than user behavior—may explain our model's strong performance relative to literature benchmarks \cite{benchmarkGraph}.\newline

Overall, the X account embeddings $\Phi$, learned following the delineated approach, successfully predict Twitter's ``Who to Follow" recommendations, indicating that studying the inferred embedding space provides a good approximation for Twitter's internal representation of entities.

\subsection{Testing the robustness of embeddings $\Phi$ under alternative data sub-sampling and model parameter strategies}
\label{subsec:embeddings_robustness}

The learned user embeddings $\Phi$, hereinafter named \textit{reference embeddings}, depend on both the collected data and the training procedure described above.
Abusing notation, we denote by $\widetilde{\Phi}$ different embeddings that result from considering different data sub-sampling strategies and different hyper-parametrizations.
First, in Section~\ref{subsubsec:robustness_linear_transformations} we specify how we compare alternative embeddings $\widetilde{\Phi}$ with the reference embedding $\Phi$.
While the training procedure was optimized to maximize the model's predictive performance for ``Who to Follow" relationships between accounts, the stability of the resulting embeddings under various perturbations remains crucial for drawing reliable conclusions. 
Consequently, we conducted two types of robustness checks, training embeddings $\widetilde{\Phi}$ \ref{subsubsec:robustness_training_parameters}) with alternative training parameters, and \ref{subsubsec:robustness_training_datasets}) on alternative datasets.

\subsubsection{Evaluating similarity of embeddings under affine transformations}
\label{subsubsec:robustness_linear_transformations}

Considering our subsequent analysis, identifying interpretable directions in the embedding space through canonical correlation analysis (CCA; see more details below in Section~\ref{section:cca}), we assess the robustness of learned embeddings up to nonsingular affine transformations, as CCA exhibits invariance to such transformations \cite{Mathai2022}. We measure the coefficient of determination, $R^2$, between the reference embeddings $\Phi$ and each variant $\widetilde{\Phi}$ after applying the optimal nonsingular affine transformation determined through least squares optimization.

\subsubsection{Testing the robustness of embeddings for alternative training parameters}
\label{subsubsec:robustness_training_parameters}

\noindent{}\textit{\textbf{Experiment \ref{subsubsec:robustness_training_parameters}.i) Different number of epochs.}} We performed 10 training runs using the original training dataset, saving the learned representations $\widetilde{\Phi}$ after 3, 5, and 10 epochs. Reference embeddings $\Phi$ were learned at 3 epochs, as determined optimal on the validation dataset.

\noindent{}\textit{\textbf{Results of \ref{subsubsec:robustness_training_parameters}.i)}} The coefficients of determination, $R^2$, between reference embeddings $\Phi$ and linearly transformed variants $\widetilde{\Phi}$ show consistent stability across non-deterministic training runs, with highest values at 3 epochs ($R^2 = .84759 \pm .00052$). Embedding stability decreases thereafter, with $R^2$ values dropping at 5 epochs ($R^2 = .83757 \pm .00062$) and 10 epochs ($R^2 = .77506 \pm .00066$). This decline reflects a shift from learning the overall network structure to noisy over-fitting the training data, as evidenced by the validation metrics in Extended Figure \ref{fig:roc_hypertune_alpha}. This interpretation is further supported by the reduced consistency between embedding replicates ---the average pairwise $R^2$ between 10 replicates trained for 10 epochs drops to $.6410 \pm .0013$, compared to $.8496 \pm .0018$ for embeddings learned after 3 epochs. Yet, overall, reference embeddings $\Phi$ do explain a substantial fraction of variance in variant embeddings $\widetilde{\Phi}$, even after extended training.\newline

\noindent{}\textit{\textbf{Experiment \ref{subsubsec:robustness_training_parameters}.ii) Intermediate values for $\boldsymbol{\alpha}$.}} We performed five training runs on the original training dataset for 3 epochs, considering two additional $\alpha$ values. While the reference embeddings $\Phi$ were learned with $\alpha = 62.6\%$ (optimal value on the validation dataset), alternative embeddings $\widetilde{\Phi}$ were trained with $\alpha = \{42.6\%, 87.6\%\}$, achieving validation AUC-ROC scores of $.7503 \pm .0019$ and $.7494 \pm .0019$ respectively (standard deviation over ten replicates). 

\noindent{}\textit{\textbf{Results of experiment \ref{subsubsec:robustness_training_parameters}.ii)}} The average coefficient of determination, $R^2$, between reference embeddings $\Phi$ and variants $\widetilde{\Phi}$ was $.8417 \pm .0041$ for $\alpha = 42.6\%$ and $.7762 \pm .0061$ for $\alpha = 87.6\%$ (standard deviation over five replicates). The high $R^2$ values indicate that embeddings learned with different $\alpha$ values remain structurally similar, up to non-singular linear transformations, to the reference embeddings $\Phi$, despite slightly suboptimal validation performance.

\noindent{}\textit{\textbf{Experiment \ref{subsubsec:robustness_training_parameters}.iii) Training only on recommendation ($\boldsymbol{\alpha} = 1$).}} We trained for 3 epochs considering, from the original dataset, only ``Who to Follow" recommendation signal, excluding ``Follow" signal i.e. $\alpha = 1$. Beyond $R^2$, and considering the particular status of $\alpha=1$, we performed the entire subsequent analysis, finding interpretable directions in the embedding space aligned with features of interest.

\noindent{}\textit{\textbf{Results of experiment \ref{subsubsec:robustness_training_parameters}.iii)}} Through Canonical Correlation Analysis, we found in the new embedding space directions that correlated with the features of interest, yet to a lesser extent than for the optimal embedding $\Phi$. Specifically, the Pearson correlations between the optimal direction and the features are: 0.35 for Left-Right leaning, 0.34 for Anti-Elite Rhetoric, 0.15 for Male-Female, 0.26 for Age, 0.42 for Popularity (log of number of followers), and 0.32 for Interest in News. These results confirm that even when training solely on the observed recommendations, the embeddings encode users' political leaning.

\noindent{}\textit{\textbf{Experiment \ref{subsubsec:robustness_training_parameters}.iv) Alternative negative/positive edges ratio.}} Rather than the 3-1 ratio between negative and positive edges, we train embeddings with a 1-1 ratio, randomly subsampling the initial training set.

{\noindent{}\textit{\textbf{Results of experiment \ref{subsubsec:robustness_training_parameters}.iv)}} Through Canonical Correlation Analysis, we found in the new embedding space a direction that correlates with the Left-Right political alignment, the variable of interest in this article, to a comparable measure with the original embeddings. In particular, we observe a strong correlation of $\rho=0.802, p<10^{-4}$.}

{\noindent{}\textit{\textbf{Experiment \ref{subsubsec:robustness_training_parameters}.v) Alternative negative edges sampling strategy}} Rather than the sampling strategy for the negative edges used in the original setup, we experiment with a random baseline. Specifically, negative edges are randomly sampled between users, under the only constraint of not being a positive one.}

{\noindent{}\textit{\textbf{Results of experiment \ref{subsubsec:robustness_training_parameters}.v)}} Through Canonical Correlation Analysis, we found in the new embedding space a direction that correlates with the Left-Right political alignment, the variable of interest in this article, to a comparable measure with the original embeddings. In particular, we observe a strong correlation of $\rho=0.783, p<10^{-4}$.}

{\noindent{}\textit{\textbf{Experiment \ref{subsubsec:robustness_training_parameters}.vi) Alternative degree threshold}} Rather than considering accounts recommended to at least two volunteers, we increase the threshold to assess the robustness of our results and consider only those recommended to at least 10 volunteers.}

{\noindent{}\textit{\textbf{Results of experiment \ref{subsubsec:robustness_training_parameters}.vi)}} Through Canonical Correlation Analysis, we identified in the new embedding space a direction that correlates with Left-Right political alignment at a level comparable to the original embeddings, with a strong correlation of $\rho=0.807, p<10^{-4}$.}

\subsubsection{Testing the robustness of embeddings for different data sub-sampling strategies}
\label{subsubsec:robustness_training_datasets}

In this section, we consider four possible sources of biases in our collected served recommendations and address them by comparing the embeddings $\Phi$ with alternative embeddings $\widetilde{\Phi}$ computed with subsamples addressing these potential sources.
\newline

\noindent{}\textit{\textbf{Experiment \ref{subsubsec:robustness_training_datasets}.i) Users seeing recommendations outside our experiment (on other devices).}} The extension only captures recommendations displayed on Twitter's desktop interface, meaning volunteers may have been exposed to more recommendations than those collected. From the profile recommendations gathered on desktop, we observe that each account is suggested 5.11 times on average to the same volunteer (computation limited to the 435 volunteers who viewed at least 25 ``Who To Follow" triplets in their feed). Inspecting Twitter's codebase reveals no fundamental reason why mobile profile recommendations would significantly differ from desktop ones. Therefore, we postulate that most missed recommendations were redundant with desktop ones and thus already included in the training dataset. Nevertheless, as we cannot, by construction, assess this observational blind spot, we trained embeddings on corrupted training datasets to validate our findings.

Assuming volunteers spend 50\% of their Twitter consumption on desktop browsers where we collect profile recommendations, and 50\% on other devices, and assuming that the pool of recommended accounts is shared across devices, we estimate that we may miss up to 19.6\% of unique accounts recommended to volunteers. To simulate this effect, we augmented each volunteer's dataset by adding 19.6\% of \textit{artificial non-observed} positive edges, randomly sampled following the same rationale as for the previous curation of negative edges. We trained variant embeddings $\widetilde{\Phi}$ using identical hyperparameters as reference embeddings $\Phi$ (3 epochs, $\alpha = 62.6\%$) on 10 random samplings of additional ``Who to Follow" edges.\newline

\noindent{}\textit{\textbf{Results of experiment \ref{subsubsec:robustness_training_datasets}.i)}} The average coefficient of determination, $R^2$, between variant embeddings $\widetilde{\Phi}$, linearly transformed, and the reference embeddings $\Phi$ was $.8417 \pm .0033$. In a more extreme scenario where users spend 75\% of their Twitter consumption on non-observed devices ---where we would be missing potentially 39.2\% of unique account recommendations, assuming cross-device recommendation pools and accounts recommended 5.11 times on average to volunteers--- the $R^2$ between reference embeddings and 10 variants trained with 39.2\% artificial non-observed positive edges remained high at $.8177 \pm .0031$. These high $R^2$ values demonstrate the robustness of the reference embeddings $\Phi$ to missing observations. \newline

\begin{figure}
\centering
\includegraphics[width=\linewidth]{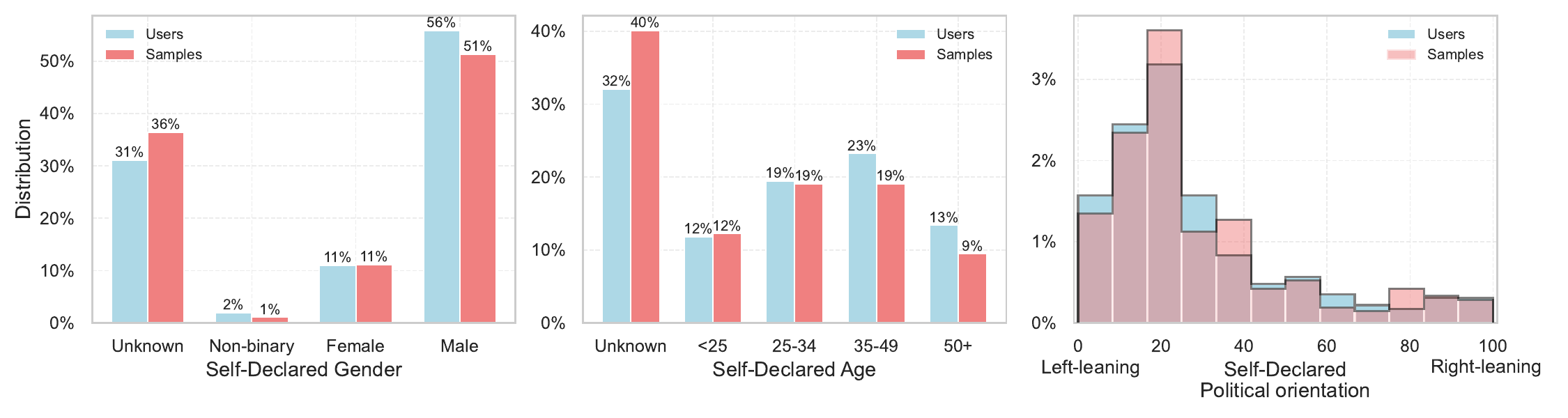}
\caption{Demographic and political characterization of volunteers in the study using surveys administered at enrollment via the browser plug-in. Distribution of volunteers' self-declared demographics (in \textcolor{gender_blue}{blue}) and weighted by ``Who to Follow" samples collected per volunteer (in \textcolor{gender_red}{red}), including gender, age, and political leaning. ``Unknown'' category reflects unfilled optional survey fields, 45.9\% of volunteers did not disclose a political preference. }
\label{fig:horus_distrib_demo}
\end{figure}

\noindent{}\textit{\textbf{Experiment \ref{subsubsec:robustness_training_datasets}.ii) Addressing population biases.}} A second limitation of our ``Who to Follow" recommendation dataset arises from volunteer self-selection, creating a sample non-representative of X's overall user base. Our data donation initiative was advertised through mainstream media, including a call for participants in LeMonde (a major French newspaper), radio coverage on France Culture (a French public radio), and X advertisements. During extension installation, volunteers could optionally indicate their age, gender, and political preference on a continuous scale from far-left to far-right. Extended Figure~\ref{fig:horus_distrib_demo} presents these survey results as both raw data and weighted by the number of ``Who to Follow" recommendations gathered per user. While our cohort spans the full spectrum of age, gender, and political leanings, respondents tended to be left-leaning, males, over 35.

To evaluate demographic skew impact, we constructed alternative training datasets of profiles suggested to subsets of volunteers who completed the survey. We applied our previous curation filters to each subset, including only accounts suggested by at least two volunteers while incorporating their follow relationships. Sample size limitations restricted our analysis to single demographic variables rather than intersections, as multi-dimensional stratification would yield insufficient subsets.

Using identical hyperparameters as for the reference embeddings $\Phi$ (3 epochs, $\alpha = 62.6\%$), we trained variant embeddings $\widetilde{\Phi}$ over the following \textcolor{blue}{four} datasets:

\begin{itemize}
    \item 284 self-identified male volunteers, yielding 49\,319 ``Who to Follow" recommendations and 11\,454 embedded accounts.
    \item 186 volunteers aged over 35, yielding 23\,868 ``Who to Follow" recommendations and 6\,542 embedded accounts.
    \item 218 Left-leaning volunteers (scoring below 40 on the political spectrum), yielding 37\,042 ``Who to Follow" recommendations and 9\,132 embedded accounts.
    \item {57 Right/far-Right leaning volunteers (scoring above 65 on the political spectrum), yielding 5\,997 ``Who to Follow" recommendations and 1\,947 embedded accounts.}
\end{itemize}

\noindent{}\textit{\textbf{Results of experiment \ref{subsubsec:robustness_training_datasets}.ii)}} Linear transformation of embeddings learned from male-volunteer induced datasets achieved a coefficient of determination, $R^2$, of $.654$ with reference embeddings $\Phi$. Left-leaning-volunteer and above-35-volunteer induced datasets yielded $R^2$ values of $.611$ and $.552$, respectively.
The $R^2$ values contrast notably with higher values obtained from variations in training parameters and observational data, highlighting the importance of demographic information encoded in the learned account representations. {Finally, even when considering a narrow, highly skewed cohort of only Right/far-Right leaning volunteers, the learned embeddings encode Left-Right political leaning. Through a CCA we find a direction whose user projection onto it correlates with left-right leaning with a Spearman correlation of $\rho=0.76, p<10^{-4}$}. Overall, we note that the reference embeddings $\Phi$ learn from the complete volunteer cohort—spanning diverse gender, age, and political orientations—captured significant structure of the variant embeddings $\widetilde{\Phi}$, maintaining $R^2$ values above $.55$. We analyze the demographic information encoded in these learned account representations in the following sections.\newline

\noindent{}\textit{\textbf{Experiment \ref{subsubsec:robustness_training_datasets}.iii) Heterogeneity of the population across the period of observation.}} Our dataset of recommendations spans a 16-month collection period, with varying collection rates as shown in Extended Figure \ref{fig:horus_collect_timeline}. A notable increase in ``Who to Follow" recommendations occurred in May 2023. To assess potential temporal effects, we split the dataset into two periods: recommendations collected before and after May 1st, 2023. We then learned embeddings using our previously described methodology.

\noindent{}\textit{\textbf{Results of experiment \ref{subsubsec:robustness_training_datasets}.iii)}} The embeddings learned from these temporal splits show strong alignment with the reference embeddings $\Phi$, achieving $R^2$ values of $.784$ and $.773$ for the pre-May and post-May 2023 periods, respectively.\newline

\noindent{}\textit{\textbf{Experiment \ref{subsubsec:robustness_training_datasets}.iv) Changing follower graph during the period of observation.}} An additional consequence of our extended observational window is that the volunteer pool of friends changed over time. In our training dataset, we aggregated all accounts followed by volunteers, irrespective of when they started following them, with one notable exception: if a ``Who to Follow" suggestion turned into a friend, we solely considered the Who to Follow observation, as the signal was scarcer for these cases. Consequently, some accounts in the training dataset labeled as ``friends" were not followed when the first Who to Follow recommendations were collected. Similarly, some accounts may no longer be friends by the experiment's end if users unfollowed them. To assess how robust our embeddings $\Phi$ are to temporal dynamics, we tracked volunteers' friend lists throughout the experiment, allowing us to derive friend addition and removal rates. This revealed that, at most, 36.9\% of friends in the training set were either a) not yet followed (26.7\%) at the experiment's start 16 months ago or b) removed from the friend pool (10.2\%) since then. We then randomly removed 36.9\% of friends from the training dataset and learned variant embeddings $\widetilde{\Phi}$, repeating this process three times.\newline

\noindent{}\textit{\textbf{Results of experiment \ref{subsubsec:robustness_training_datasets}.iv)}} Linear transformation of embeddings $\widetilde{\Phi}$ learned from those variant training dataset, achieved a coefficient of determination, $R^2$, of $.80217\pm.00028$ (standard deviation over three replicates) with reference embeddings $\Phi$.\newline

Our experiments demonstrate that the reference embeddings $\Phi$ remain stable (modulo linear transformations, as described in Section \ref{subsubsec:robustness_linear_transformations}) despite substantial changes in training parameters and dataset composition.

{\noindent{}\textit{\textbf{Experiment \ref{subsubsec:robustness_training_datasets}.v) Removing ties between users and Members of Parliament}} The ideological leaning of users is inferred from following patterns with members of parliament, and because such follow relations appear in the training dataset ---specifically 2.9\% of edges in the training dataset are follow edges between members of parliament and users whose ideology has been scaled--- we assess the robustness of our results to the removal of such ties. We explicitly remove such ties in this experiment and then learn the embeddings using our previously described methodology.}

\noindent{}\textit{\textbf{Results of experiment \ref{subsubsec:robustness_training_datasets}.v)}} A CCA between the new embeddings and the Left-Right political leaning of users results in a Spearman correlation of $\rho = 0.816, p<10^{-4}$. In comparison, as reported in the main text, we observe a correlation of $\rho = 0.887, p<10^{-4}$ when considering the full training dataset.

{\noindent{}\textit{\textbf{Experiment \ref{subsubsec:robustness_training_datasets}.vi) Removing ties between volunteers.}} Because our method of recruitment could allow snowballing effects to emerge, e.g., a volunteer recommending a friend to take part in the experiment, in this experiment we remove the ties (follow or WTF recommendation) between volunteers, removing less than 0.03\% of edges. We then learn the embeddings using our previously described methodology.
}

{\noindent{}\textit{\textbf{Results of experiment \ref{subsubsec:robustness_training_datasets}.vi)}} A CCA between the new embeddings and the Left-Right political leaning of users results in a Spearman correlation of $\rho = 0.829, p<10^{-4}$. In comparison, as reported in the main text, we observe a correlation of $\rho = 0.887, p<10^{-4}$ when considering the full training dataset.
}


\section{Inferring users' features}\label{section:usr_features}

Having learned embeddings $\Phi$ for each X user in $\mathcal{E}$, we now curate a set of user features that we will later leverage in attributing semantics to this embedding space. Specifically, we define the following families of user features: \ref{subsec:twitter_activity})  user activity metrics (e.g., popularity, number of posts and likes), \ref{subsec:demographic_features}) demographic features, \ref{subsec:topics_features}) topics of interest and linguistic features, and \ref{subsec:ideological_features}) ideological leanings of users.

\subsection{User activity metrics}
\label{subsec:twitter_activity}

Using X's API, we collect account metadata for each user in $\mathcal{E}$:

\begin{itemize}
\item \textit{Follower count}: measures social capital, ranging from 0 to 197 million with a median of 7,987 followers.
\item \textit{Follower-to-friend ratio}: assesses network reciprocity, with a median of 0.097.
\item \textit{Daily Posts count}: user total posts (status) divided by days since account creation, with a median of 2.21 posts per day.
\item \textit{Daily Likes count}: user total likes divided by days since account creation, with a median 1.40 of likes per day.
\end{itemize}

\begin{figure}
\centering
\includegraphics[width=\linewidth]{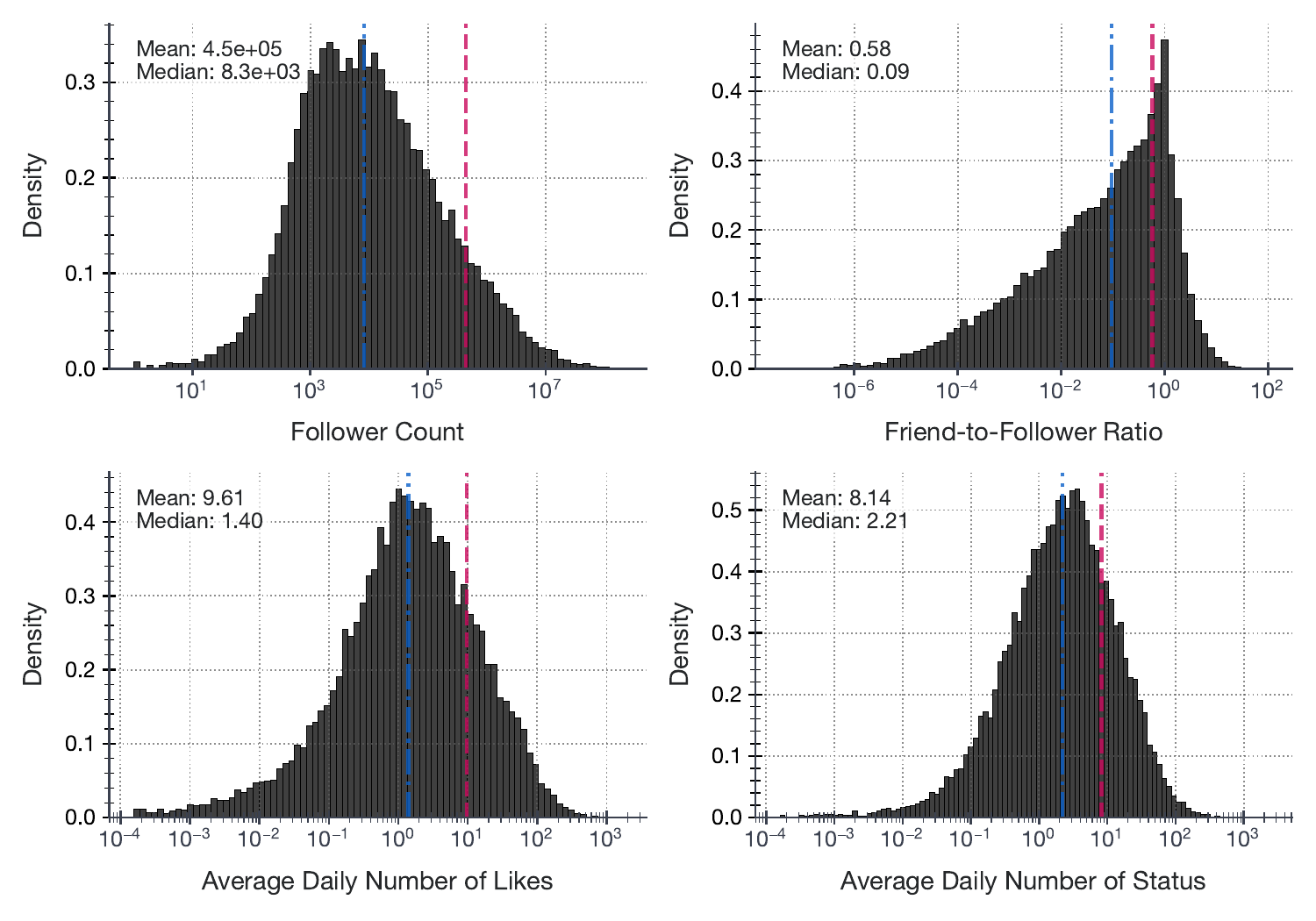}
\caption{Distribution of X users in our study by attributes related to activity on the platform, on a logarithmic scale, with mean values indicated by dashed \textcolor{mean_red}{red} lines and median by dashed \textcolor{median_blue}{blue} lines: follower count (top left), friend-to-follower ratio (top right), average daily number of likes (bottom left), and average daily number of posts/statuses (bottom right).}
\label{fig:distrib_activity_features}
\end{figure}

These metrics, whose distributions are shown in Extended Figure~\ref{fig:distrib_activity_features}, span multiple orders of magnitude. We normalized each metric as the fraction of users in $\mathcal{E}$ with strictly higher values, yielding scores in [0,1].

Leveraging an independent data collection from early 2023, we examined the Legacy verification status of 24\,214 users in $\mathcal{E}$. This status, formerly known as the blue check-mark, indicated whether an account was ``authentic, notable, and active" \cite{legacyX}, typically granted to public figures, journalists, and influential organizations before the platform's verification policy changes. Up to 30.1\% of the accounts were, as of early 2023, verified.

\subsection{Inference of demographic features}
\label{subsec:demographic_features}

\begin{figure}
\centering
\includegraphics[width=\linewidth]{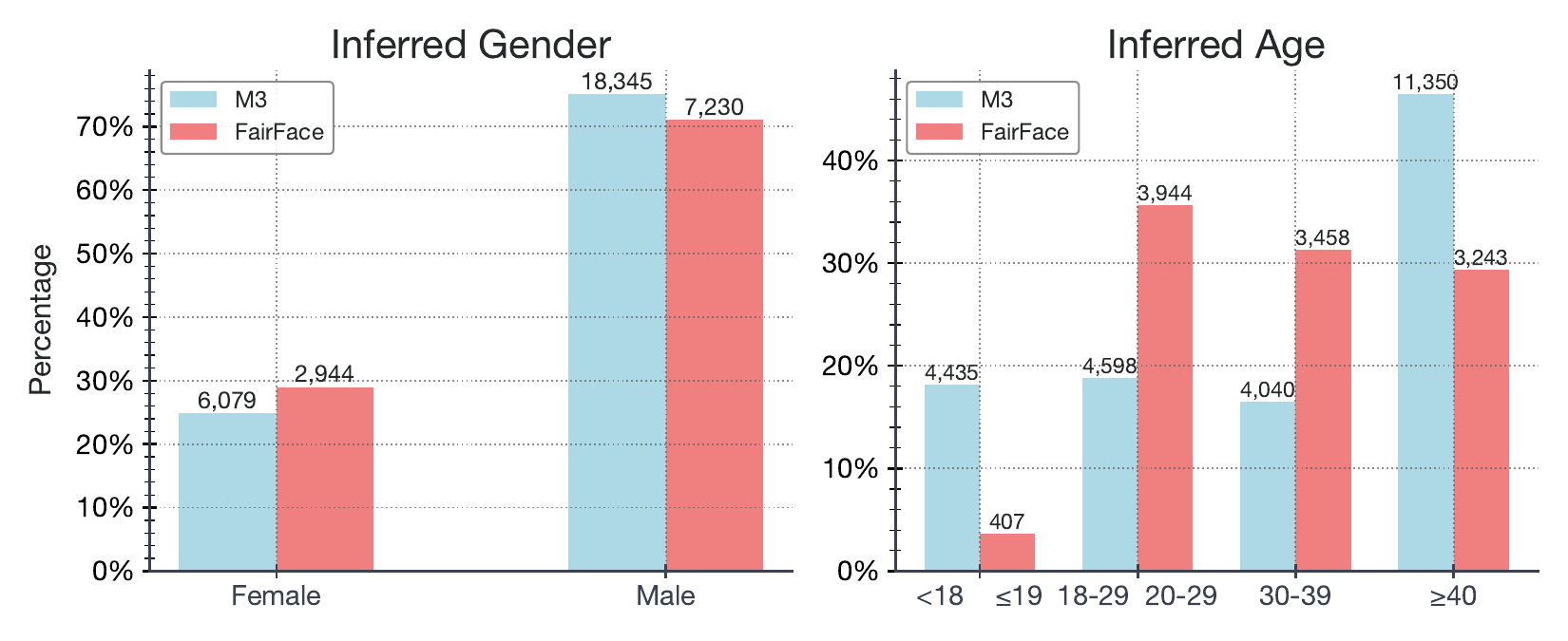}
\caption{Comparison of the inference of age and gender for users in the study (using  \texttt{M3} model by Wang et al., \cite{wang2019demographic}), with an alternative method of inference, the \texttt{FairFace} model (developed by Karkkainen and Joo \cite{Karkkainen2021}), for robustness.  Gender and age distributions predicted by \texttt{M3} (in \textcolor{gender_blue}{blue}) for 24\,425 accounts in $\mathcal{E}$ and by \texttt{FairFace} (in \textcolor{gender_red}{red}) for the subset of 10\,932 accounts with a \textit{single} visible profile face. One limitation of our study lies in the fact that users may choose profile names, descriptions, and photos that do not match their demographic groups, choosing, for instance, avatars (e.g., cartoons, historic personalities), obfuscated representations (e.g., disguises, photos where they do not appear clearly), or even misleading representations (e.g., seeking to appear to other users as being from a demographic group other than their own). We argue that this limitation does not impact the main findings of our study, as these occurrences do no affect the overall trend between features in the absence of reason to believe that their frequency should be dependent on the main variables against which we test independence to demographic features: namely, political leaning.}\label{fig:distrib_m3_deep}
\end{figure}

Leveraging account metadata collected for Twitter activity metrics, we analyze profile pictures, screen names, usernames, and biographical descriptions to infer accounts' demographic attributes. Specifically, we applied the Multimodal, Multilingual, and Multi-attribute (\texttt{M3}) model, developed by Wang et al.~\cite{wang2019demographic}, on the 24\,425 accounts in $\mathcal{E}$ with available information. We display on Extended Figure~\ref{fig:distrib_m3_deep} the gender and age distributions of profiles inferred by \texttt{M3} .

\begin{figure}
\centering
\includegraphics[width=\linewidth]{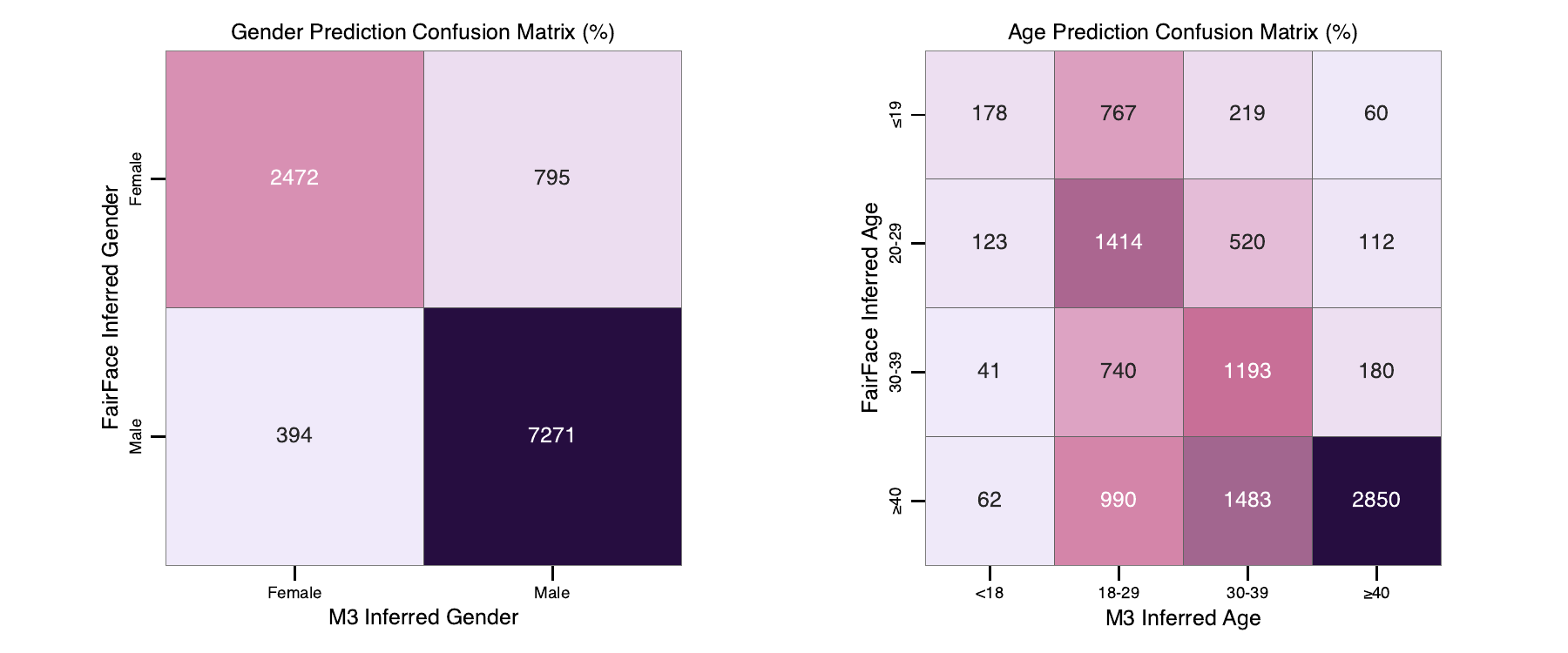}
\caption{Confusion matrices comparing the demographic inference achieved by \texttt{M3} and by \texttt{FairFace} models on users in our study, for gender (left figure) and age (right figure) for 10\,932 accounts with visible profile faces in $\mathcal{E}$.}
\label{fig:confusion_m3_FairFace}
\end{figure}

Furthermore, as to corroborate \texttt{M3} 's inference, we estimate the age and gender of 10\,932 accounts in $\mathcal{E}$ having a \textit{single} visible face in the profile pictures. These estimates were derived through the \texttt{FairFace} model, developed by Karkkainen and Joo \cite{Karkkainen2021}. This model was selected for being open-weights and offering state-of-the-art performance. Moreover, its curated training dataset enables consistent performance across race and gender groups \cite{Karkkainen2021}, contrasting with typical closed-source solutions \cite{Raji2019, GenderShades}. We show the confusion matrices for the inference of gender and age in Extended Figure \ref{fig:confusion_m3_FairFace}.\newline

{Finally, we compare the inferred age and gender with volunteers' (optional) self-declarations. We observe an overall alignment, with an F1 score of 0.84 for gender and an average F1 score of 0.82 for age buckets.}\newline

One limitation of our study lies in the fact that users may choose profile names, descriptions, and photos that do not match their demographic groups, choosing, for instance, avatars (e.g., cartoons, historic personalities), obfuscated representations (e.g., disguises, photos where they do not appear clearly), or even misleading representations (e.g., seeking to appear to other users as being from a demographic group other than their own). We argue that this limitation does not impact the main findings of our study, as these occurrences do no affect the overall trend between features in the absence of reason to believe that their frequency should be dependent on the main variables against which we test independence to demographic features: namely, political leaning.\newline

\subsubsection{Gender}
\label{subsubsec:gender_inference}

For the 10\,932 accounts where \texttt{FairFace} detected a \textit{single} face, the gender inferred by \texttt{FairFace} strongly agrees with \texttt{M3} 's inference with a Cohen's $\kappa = .731$. 

Automated gender inference systems tend to disproportionately misclassify individuals that do not identify as male or female \cite{FoschVillaronga2021}. We interpret these inferred gender as an algorithmic extension enabling large-scale aggregated trends in demographic analysis rather than definitive individual gender identification, i.e., a prediction of the most probable binary gender label that a team of annotators (akin to that involved in the training of \texttt{M3}) would provide if prompted to do so.

\subsubsection{Age}
\label{subsubsec:age_inference}

Having aligned \texttt{FairFace} age bins into the closest \texttt{M3} stratification levels ($<18$ from \texttt{M3} mapped to the $\leq 19$ from \texttt{FairFace} , 18-29 from \texttt{M3} mapped to 20-29 from \texttt{FairFace}), we observe a poor agreement between both models with Cohen's $\kappa = .326$.

This poor agreement reflects the models' distinct objectives: \texttt{FairFace} estimates age from profile pictures, while \texttt{M3} infers account owners' ages using also text description from the profiles. Manual inspection reveals accounts displaying children receive young age estimates from \texttt{FairFace} but young-parent age ranges from \texttt{M3} (average first-time parent age was 28.5 in 2015 \cite{insee2017enfant} corresponding to the second class bin in \texttt{M3} classification). Similarly, profiles classified as $\geq40$ years by \texttt{FairFace} but 18-29 by \texttt{M3} displayed celebrities like Elon Musk or Ronaldo on their profile picture. Given \texttt{M3}'s specific training for social media demographics, we utilize its predictions for subsequent analyses.

The validity of \texttt{M3}'s age categorization is supported by distinctive patterns in account characteristics, with each age group showing characteristic median account creation years ($\leq$18: 2018; 19-29: 2015; 30-39: 2013; $\geq$40: 2013) and terms used in users' biographical descriptions. We report distinctive vocabulary by extracting the 10 words, longer than 5 characters, with the highest chi-squared statistics, comparing their frequency in each age bin to their overall use in the population $\mathcal{E}$ (see Extended List~\ref{list:attributes_words_by_age}).\newline

\begin{customlist}
\label{list:attributes_words_by_age}
Distinctive vocabulary of age groups computed by extracting the 10 words, longer than 5 characters, with the highest chi-squared statistics, comparing their frequency in each age bin to their overall use in the population $\mathcal{E}$.
\begin{align*}
\leq\text{18}&: \text{YouTube, Twitch, Player, Minecraft, Streaming, Videos, Discord, Youtuber, Streamer, Gaming}\\
\text{19-29}&: \text{Student, Player, Professional, Enthusiast, Artist, Contact, Candidate, Engineer, Creator, Feminist}\\
\text{30-39}&: \text{Journalist, Doctor, Opinions, Co-Founder, Manager, Adjoint, Director, Teacher, Senior, Redactor}\\
\geq\text{40}&: \text{Official Account, President, France, MP, Association, University, Politics, Minister, District, Director}
\end{align*}

\end{customlist}

\subsection{Content-based attributes}
\label{subsec:topics_features}

Beyond activity metrics and demographic inference from Twitter profiles metadata, we characterized user content production patterns. On 27th October 2024, we collected the last 200 tweets (published or retweeted) from 25\,080 users in $\mathcal{E}$ through the Twitter API, excluding private and deleted accounts, totaling 2.3 million tweets.

This additional dataset enabled computation of content-related metrics. Specifically, we leveraged the multi-topic classification model \texttt{cardiffnlp/tweet-topic-21-multi} developed by Antypas et al \cite{cardiffNLP}, to categorize tweets into 19 generic topics including arts \& culture, news \& social concern, and science \& technology. Non-English tweets were translated using Opus-MT models \cite{OPUS2020, OPUS2023}, selected for their strong validated performance and their open-weights nature ensuring reproducibility. We calculated the proportion of tweets per topic for each user across all 19 categories considered by \cite{cardiffNLP}.

\begin{figure}
\centering
\includegraphics[width=\linewidth]{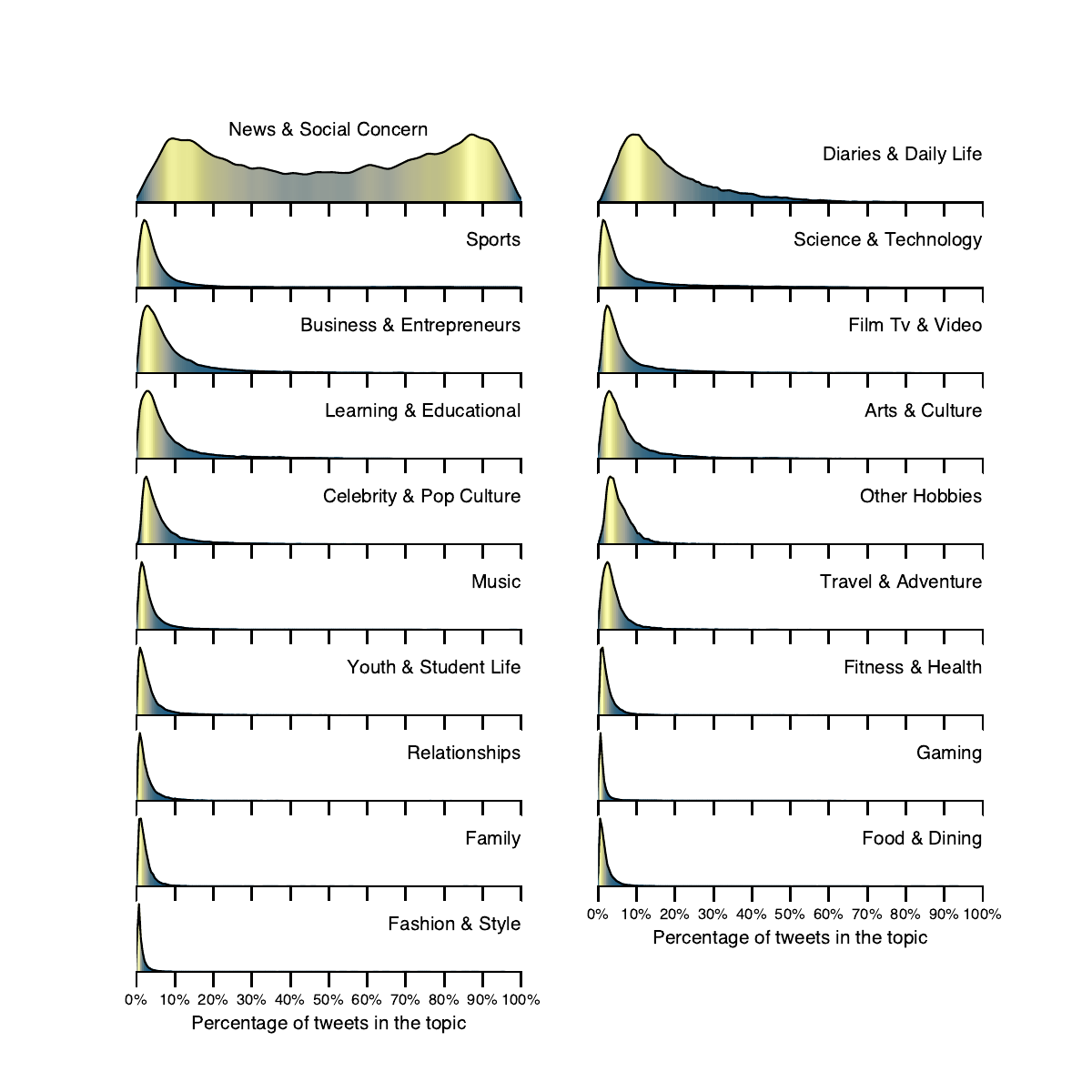}
\caption{Distribution of users by percentage of tweets that they produce on each topic, according to the topic classifier from \cite{cardiffNLP}. Most users produce a low percentage of their tweets on contents that are classifiable into topics from \cite{cardiffNLP}. The exception is the topic ``News \%{} Social Concern'': users are more evenly distributed when organized by the \%{} of tweets their produce in this topic, and slightly bimodal, concentrated around 10\%{} and 90\%{} of tweets in this topic.}
\label{fig:distrib_topic}
\end{figure}

We display on Extended Figure \ref{fig:distrib_topic} the distribution of those metrics, observing that ``News \& Social Concern" is the most discussed topic among embedded users, with some exclusively discussing this topic. The following main topics in decreasing frequency are ``Diaries \& Daily Life", ``Sports", ``Science \& Technology" and ``Business \& Entrepreneurs".

To ease topics interpretation and corroborate their validity, we report in Extended List~\ref{list:attributes_words_topics} below the 25 most overrepresented words in tweets classified under each topic, compared to their expected frequency under uniform random distribution across topics. Extended List~\ref{list:attributes_words_topics} contains only 4+ letter words (lower-cased and translated into English when required using Opus-MT model \cite{OPUS2023}).\newline

\begin{customlist}
\label{list:attributes_words_topics}
25 most overrepresented words in tweets classified under each topic, compared to their expected frequency under uniform random distribution across topics from \cite{cardiffNLP}. List contains only 4+ letter words (lower-cased and translated into English when required using Opus-MT model \cite{OPUS2023}).

\begin{itemize}
\item \textbf{Arts \& Culture}: \textit{paintings, feast, theatre, artworks, painter, tolkien, ballet, illustrations, exhibitions, retrospective, exhibit, nouveau, angoul, musicians, animations, shakespeare, arts, gogh, poem, curator, biography, treasures, photographs, exhibition, festival}
\item \textbf{Business \& Entrepreneurs}: \textit{binance, staking, paypal, multinationals, pharmaceutical, cryptocurrency, economies, ceos, sellers, subsidiary, remuneration, investors, mckinsey, consulting, profitability, outlook, investor, margin, defi, stablecoin, blackrock, eligible, founders, banking, trading}
\item \textbf{Celebrity \& Pop Culture}: \textit{lana, celine, portman, hathaway, meryl, grammy, zendaya, vmas, cotillard, starring, billie, perry, cobain, hollywood, eminem, brad, pitt, natalie, dion, thee, cooper, comedian, andrew, lipa, scarlett}
\item \textbf{Diaries \& Daily Life}: \textit{subway, stayed, shower, crush, laundry, habits, sagittarius, cashier, eats, darling, toilet, pisces, rude, reflexes, wakes, kindness, jokes, notification, fever, conditioning, kisses, asleep, sleep, boyfriend, girlfriends}
\item \textbf{Family}: \textit{dads, grandparents, mothers, cousin, eidmubarak, parent, daughters, moms, niece, mubarak, sister, grandmother, widows, grandpa, adha, daughter, father, mother, fathers, aunt, uncle, uncles, sisters, nephew, sibling}
\item \textbf{Fashion \& Style}: \textit{fashion, jewelry, dior, leather, wardrobe, clothing, outfit, makeup, necklace, nike, pants, clothes, heels, accessories, bracelet, runway, lingerie, christianlouboutin, outfits, shirts, dress, perfume, vinted, grazia, shoe}
\item \textbf{Film Tv \& Video}: \textit{batman, emmy, deadpool, succession, saga, starring, biopic, nolan, superman, amour, slayer, ahsoka, christopher, uploaded, toriyama, anatomy, pixar, loki, runner, wick, titanic, arcane, theaters, rings, venom}
\item \textbf{Fitness \& Health}: \textit{injections, pediatric, cardiovascular, endometriosis, resuscitation, addictions, parkinson, chronic, fitness, therapeutic, respiratory, cardiac, autism, cancers, surgery, vaccinated, worldmentalhealthday, dental, injection, disorders, liver, oncology, masks, transplant, practitioner}
\item \textbf{Food \& Dining}: \textit{pastry, bakery, ingredients, foods, cook, culinary, sauce, flavors, meal, juice, gourmet, burger, meals, cookies, vegetables, cooking, cuisine, kebab, chefs, eats, recipe, cheese, restaurants, cocktails, pizzas}
\item \textbf{Gaming}: \textit{warcraft, ubisoft, indiedev, warhammer, xbox, gameplay, fallout, warzone, kart, doom, skins, tabletop, pokemon, mario, nintendoswitch, gaming, overwatch, gamedev, sims, playstation, fortnite, playable, creed, zelda, gear}
\item \textbf{Learning \& Educational}: \textit{classroom, graduates, orientation, bachelor, genetet, schooling, aesh, autism, sociology, graduate, maths, vocational, fellowship, interdisciplinary, nursing, diploma, exams, principal, supervisors, college, dean, trainers, doctoral, baccalaureate, teacher}
\item \textbf{Music}: \textit{musicians, lana, electro, guitar, piano, sings, celine, vocals, tyler, eminem, billboard, grammy, concert, tiakola, tunes, singles, hiphop, eurovision, billie, music, olympia, concerto, vinyl, lipa, perform}
\item \textbf{News \& Social Concern}: \textit{burkina, blacks, judiciary, vietnam, insufficient, sudden, accountable, thailand, rage, committing, widespread, punished, relative, investigators, ppda, deprived, colonization, bastille, interruption, reservists, synagogue, vape, kirk, remuneration, golan}
\item \textbf{Other Hobbies}: \textit{jesus, christ, shall, salvation, prayer, sins, worship, blessed, resurrection, baptized, holy, righteous, lord, foryou, elder, saints, mary, remission, saves, thee, messiah, heaven, gospel, bible, mercy}
\item \textbf{Relationships}: \textit{boyfriend, romantic, married, dads, darling, girlfriends, happybirthday, girlfriend, wedding, crush, cousin, couples, marry, daddy, breakup, birthday, cherish, bride, laughs, lily, marriage, kisses, bestie, inseparable, thanksgiving}
\item \textbf{Science \& Technology}: \textit{github, methane, labs, ipad, firefox, rust, genai, artificialintelligence, mozilla, embedded, astronomy, computing, macos, ryzen, kubernetes, reactors, encryption, benchmarks, datasets, defi, browser, cyberattack, ransomware, planets, phishing}
\item \textbf{Sports}: \textit{derby, coaches, athletics, squad, vinicius, neymar, allezlesbleus, athlete, djokovic, maxime, whistle, uefa, mexicogp, flick, khelif, eurosport, bayern, cristiano, zidane, ancelotti, leon, azerbaijangp, bellingham, dauphin, chants}
\item \textbf{Travel \& Adventure}: \textit{hiking, hotels, booking, portes, trips, boarding, tourist, thailand, traveling, airports, travellers, baggage, travel, beaches, trip, travelling, thisisliving, shuttle, wuhan, bikes, infotraffic, booked, gate, pavement, rides}
\item \textbf{Youth \& Student Life}: \textit{schooling, orientation, aesh, classroom, boarding, bachelor, exam, exams, baccalaureate, college, vocational, genetet, nursery, principal, diploma, elementary, graduates, kindergarten, school, graduation, autism, primary, classes, ulis, colleges}
\end{itemize}
\end{customlist}

Finally, to assess linguistic sophistication in users' original tweets beyond topical content, we compute the average word-length; after pre-processing to remove non-linguistic elements (mentions, hashtags, emojis, URLs).
We report the 15 words with the highest chi-squared statistics, comparing their frequency in the first and last decile of users average word-length to their overall use in the population in Extended List~\ref{list:attributes_word_length}.\newline

\begin{customlist}
\label{list:attributes_word_length}
Linguistic sophistication in users' original tweets beyond topical content, shown by computing the average word-length. After pre-processing to remove non-linguistic elements (mentions, hashtags, emojis, URLs), we reporte through the 15 words with the highest chi-squared statistics, comparing their frequency in the first and last decile of users average word-length to their overall use in the population.
    
    \begin{itemize}
    \item \textbf{Low Average Word-Length}: \textit{jveux, jte, drole, jcrois, etais, wsh, jss, jpense, frr, karmine, karmine corp, oe, jvais, ptite, tqt}
    \item \textbf{High Average Word-Length}: \textit{applications, pollinators, exchange, UMR, young talents, cyberattacks, webinars, ArtificialIntelligence, RSE, white paper, innovative, access, innovative, apply, emerging}
    \end{itemize}
Remark: Most terms of the reported ``Low Average Word-Length'' are jargon or slang: e.g., \textit{jveux} (for \textit{je veux}, meaning ``I want that''), \textit{jte} (for \textit{je te}, meaning ``I <verb> you''), \textit{drole} (\textit{drôle}, translating to ``funny''), \textit{jcrois} (for \textit{je crois}, meaning ``I think so''). The term \textit{wsh} has an intricate meaning. It might come from \textit{wesh}, which is itself a term derived from the word \textit{oui} in French. It is also a word from Algerian Arab (the largest community of foreign residents), commonly used in the expression \textit{We[c/s]h rak?}, translatable to \textit{How is it going?}.
\end{customlist}

\subsection{Ideological Positions of Users}
\label{subsec:ideological_features}

One of the central objectives of our study is to identify whether the embedding space $\Phi$ of the recommender system encodes information about political leaning of users. To achieve this goal, we leverage a dataset of politically positioned X users developed by Ramaciotti et al. \cite{some4dem2023dataset}. We first present the dataset in Section~\ref{subsubsec:epo_dataset} to then analyze its intersection with the population $\mathcal{E}$ of our study in Section~\ref{subsubsec:intersection_with_epo_dataset}.

\subsubsection{Politically-positioned X population dataset from \cite{some4dem2023dataset}}
\label{subsubsec:epo_dataset}

This dataset contains 494 065 Twitter users interested in French politics, positioned along two political scales or dimensions: 1) a Left-Right dimension and 2) a dimension measuring attitudes towards elites and institutions.
The dataset comes with a detailed documentation \cite{ramaciottimorales:hal-04807916} explaining the methodology used and the quality metrics showcasing the accuracy of inferred political positions. The methodology relies on a two step approach: 1) multidimensional ideology scaling of follower networks on Twitter, and 2) calibration of political positions using political survey data.

Step (1) uses ideological scaling \cite{clinton2004statistical,imai2016fast} of follower networks. Ideology scaling of choice data has its roots in political science research in the 1980s \cite{poole1985spatial}, and was first developed for single-dimensional Liberal-Conservative ideological spatialization of Twitter users in the US by Barberà \cite{barbera2015birds}.
Ideology scaling spatialization has been subsequently developed to capture additional political dimensions \cite{ramaciotti2022inferring}.
The main motivation for this new dimension is that European settings, and in particular France for our study, cannot be reduced to a single ideological dimension (e.g., Left-Right) in empirical political spatial analyses \cite{bakker2012complexity}.
The population used for this ideological spatialization is chosen among users that follow at least three Members of Parliament (MPs) on X/Twitter (as Barberà did in his 2015 studies \cite{barbera2015birds,barbera2015tweeting,barbera2015understanding}). This allows us to select a large number of users that have an interest in national politics.
Next, the follower network subtended by these users is collected, and a political homophily process is proposed as an explanation for the observed following links connecting MPs and their users:

\begin{equation}
    P\left(\text{U}_i\rightarrow \text{MP}_j|\alpha_i,\beta_j,\gamma,\varphi_i,\varphi_j\right) = \text{logistic}\left(\alpha_i + \beta_j - \gamma \|\varphi_i - \varphi_j \|^2 \right),
    \label{eq:homophily_model_si}
\end{equation}\newline

\noindent{}where, $U_i\rightarrow MP_j$ stands for the event in which user $\text{U}_i$ follows $\text{MP}_j$, $\alpha_i$ is the level of \textit{activity} of user $\text{U}_i$ measured in number of followed accounts, $\beta_j$ is the level of \textit{popularity} of $\text{MP}_j$, measured in number of followers, $\gamma$ is a scale variable, and $\varphi_i$ and $\varphi_j$ are the multidimensional positions of $\text{U}_i$ and $\text{MP}_j$ in some unobservable multidimensional space. This data generation process is homophilic in the sense that $P\left(\text{U}_i\rightarrow \text{MP}_j\right)$ increases as the distance between unobservable positions $\|\varphi_i - \varphi_j\|$ in some latent space decreases. Step (1) involves using observed follower networks between MPs and their followers to infer the parameters of the model in equation Eq.~\eqref{eq:homophily_model_si}. The reader is referred to the work of Barberà on Twitter \cite{barbera2015tweeting} for the detailed methodology on this inference, to the work of Ramaciotti et al. \cite{ramaciotti2022inferring} for an application to French Twitter data, and to the documentation of the dataset we use \cite{ramaciottimorales:hal-04807916} for the details of the application of the methods to the data of this study.

\begin{figure}
    \centering
    \includegraphics[width=0.75\linewidth]{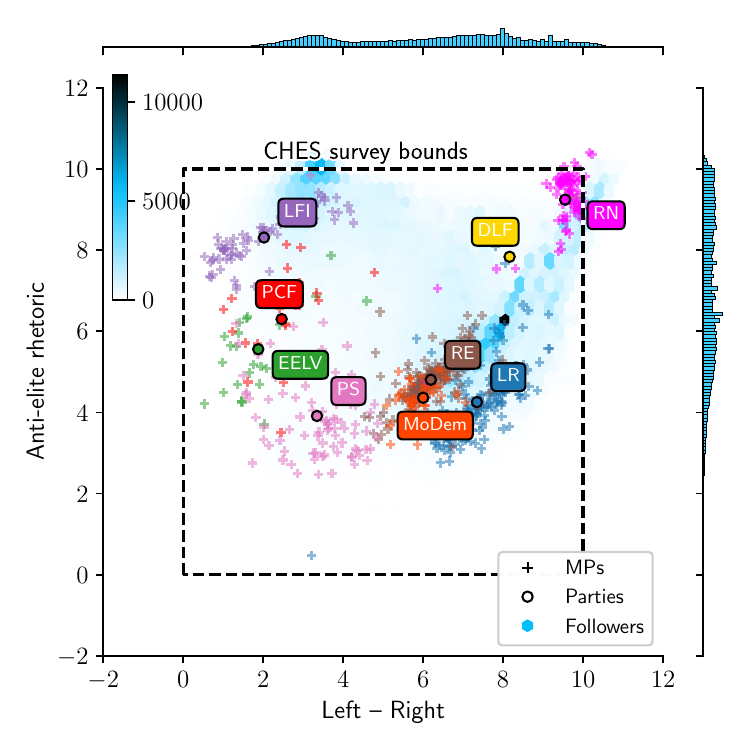}
    \caption{Figure taken from the methodology report of the dataset of politically positioned users from the Twitter French political Twittosphere and that we use in this study \cite{ramaciottimorales:hal-04807916}. Users (followers of MPs, density shown in \textcolor{cyan}{light blue}) are embedded along two political dimensions: Left-Right and an Anti-elite rhetoric dimension. Positions in the dimension are calibrated using data from the Chapel Hill Expert Survey (CHES) \cite{Jolly2022}, and that provide positions of parties as reference points in space. Positions of MPs and political parties used in the calibration are shown in colors. Along the Left-Right scale 0 is the left-most position for political parties, while 10 is the right-most, and 5 is the political center. Along the Anti-elite rhetoric scale 0 stands for no salience of anti-establishment and anti-elite rhetoric, while 10 stands for extreme salience of anti-establishment and anti-elite rhetoric.}
    \label{fig:epo_data_2D}
\end{figure}

Step (2) uses political survey data to calibrate multidimensional positions $\varphi_i$ for the population of study resulting from the Bayesian inference for the parameters of Eq.~\eqref{eq:homophily_model_si}. This step is required because, in Eq.~\eqref{eq:homophily_model_si}, the probability $P\left(\text{U}_i\rightarrow \text{MP}_j\right)$ is invariant to isometries on positions $\varphi_i$ and $\varphi_j$: i.e., for any isometry $T$, $\|\varphi_i - \varphi_j\|=\|T(\varphi_i - \varphi_j)\|$.
In particular, this identification problem means that positions $\varphi$ on any dimension cannot be taken to stand as indicators for political positions on an identifiable issue or ideology (ideology in the sense of an organized structure of attitudes \cite{campbell1960american,converse1964nature}), because isometries $T$ include flips, translations, and rotation of the latent space.
For calibration and identification, the dataset from \cite{ramaciottimorales:hal-04807916} uses party positions available for political parties, from the Chapel Hill Expert Survey (CHES).
The CHES dataset \cite{Jolly2022} is created with the responses from hundreds of experts in political science, who position political parties along dimensions such as Left-Right, attitudes towards elites and institutions, Liberal-Conservative, or attitudes towards immigration, to name a few.
Creating proxies for party positions as the centroids of MPs $\varphi_j$, the dataset from \cite{ramaciottimorales:hal-04807916} is compounded with an affine transformation going from the parameters space resulting from the Bayesian inference using Eq.~\eqref{eq:homophily_model_si}, onto the space of the dimensions of the CHES data.
The leading political dimension of the CHES data is the Left-Right dimension (ranging from 0 to 10, from Left to Right, for positions of parties). 
Because the scale used for calibration is constructed for positions of political parties, users can have positions to the left of position 0 (the Left-most position for parties) and to the right of position 10 (the Right-most position for parties).
Previous research by the authors of the dataset has shown that a second relevant dimension in France and in several countries is the dimension measuring attitudes towards elites and institutions \cite{ramaciotti2021unfolding}, which authors explain that motivates the inclusion of this dimension in the dataset. Because of its importance and availability we also leverage this second dimension in our study. Extended Figure~\ref{fig:epo_data_2D} from \cite{ramaciottimorales:hal-04807916} shows the density of users in the population along these two dimensions, and the positions of MPs and parties used for the calibration.

\begin{figure}
    \centering
    \includegraphics[width=0.45\linewidth]{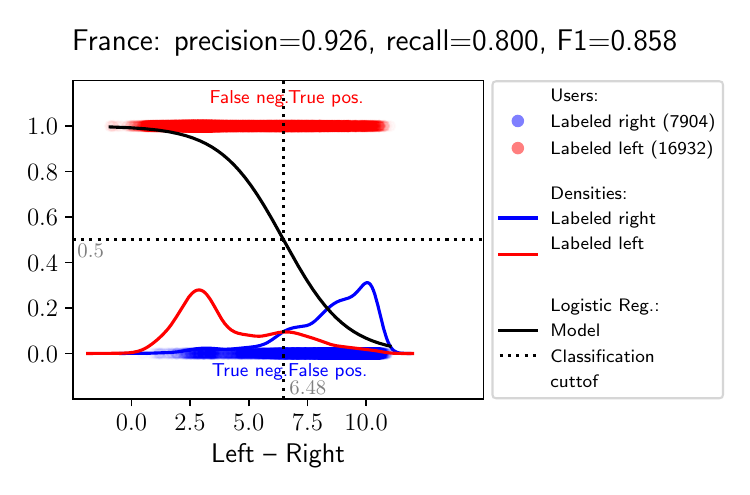}
    \includegraphics[width=0.45\linewidth]{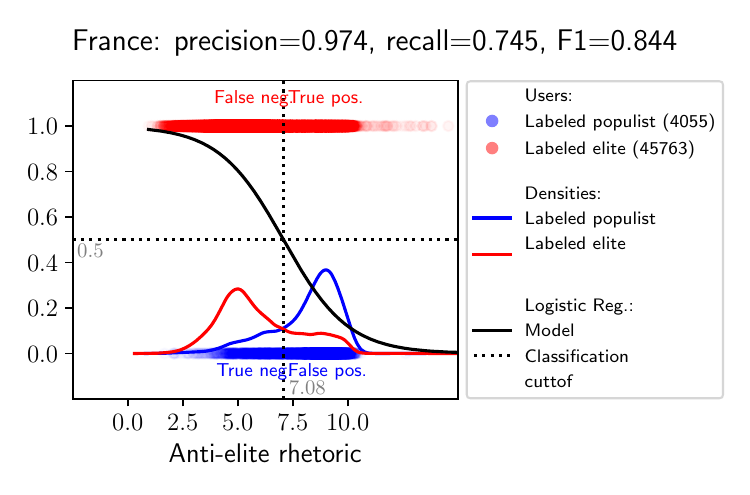}
    \caption{Figures taken from the methodological report of the dataset of politically positioned users from the Twitter French political Twittosphere \cite{ramaciottimorales:hal-04807916}, reporting the validation of spatial positioning of users along the political dimensions. Density of users labeled as identified with ``left'', ``right'', ``elite'', or ``populist'' along dimensions Left-Right (left figure) and Anti-elite rhetoric (right figure). Labeling was computed with annotations produced by the LLM Zephyr 7B $\beta$ \cite{tunstall2023zephyr}. The figure shows the goodness of fit of logistic regressions as classifiers of pairs of labels using the political dimension on which they are embedded. Individuals displaying populist discourse (e.g., subscribing to a vision of society divided into the ``people'' and ``elites'') may also be part of elite political or economic groups. The validation proposed by the authors of the dataset \cite{ramaciottimorales:hal-04807916} relies on the fact that most individuals that display high anti-elite and anti-establishment views will not be themselves part of the elite, and that if the proposed validation metrics hold, it will be \textit{despite} the fact there may be some elite individuals displaying anti-elite and/or anti-establishment rhetoric.}
    \label{fig:epo_data_validations}
\end{figure}

The authors of the dataset provide quality metrics computed using text data written by users embedded in these two political dimensions.
They collected the profile bios of users and proposed an annotation protocol based on LLM classification (using model Zephyr 7B $\beta$ \cite{tunstall2023zephyr}) to identify a subset of users that could be attributed with the labels ``left'', ``right'', ``elite'', or ``populist''. 
The quality metrics for the positions measure the degree to which the inferred political dimensions can separate these labeled users: i.e., the quality with which the Left-Right dimension can separate users labeled as ``left'' and ``right'', and the quality with which the Anti-elite rhetoric dimension can separate users labeled as ``populist'' and ``elite''.
Extended Figure~\ref{fig:epo_data_validations} from \cite{ramaciottimorales:hal-04807916} shows the quality metrics of political positions along these two dimensions. This separation is measured as the accuracy metrics (precision, recall and F1-score) resulting from using the political dimensions for classification of labeled users.
It is worth noting that individuals displaying populist discourse (e.g., subscribing to a vision of society divided into the ``people'' and ``elites'') may also be part of elite political or economic groups. The validation proposed by the authors of the dataset \cite{ramaciottimorales:hal-04807916} relies on the fact that most individuals that display high anti-elite and anti-establishment views will not be themselves part of the elite, and that if the proposed validation metrics hold, it will be \textit{despite} the fact there may be some elite individuals displaying anti-elite and/or anti-establishment rhetoric.

\subsubsection{Intersection between the politically-positioned dataset from \cite{ramaciottimorales:hal-04807916} and our population of study $\mathcal{E}$}
\label{subsubsec:intersection_with_epo_dataset}

\begin{figure}
\centering
\includegraphics[width=\linewidth]{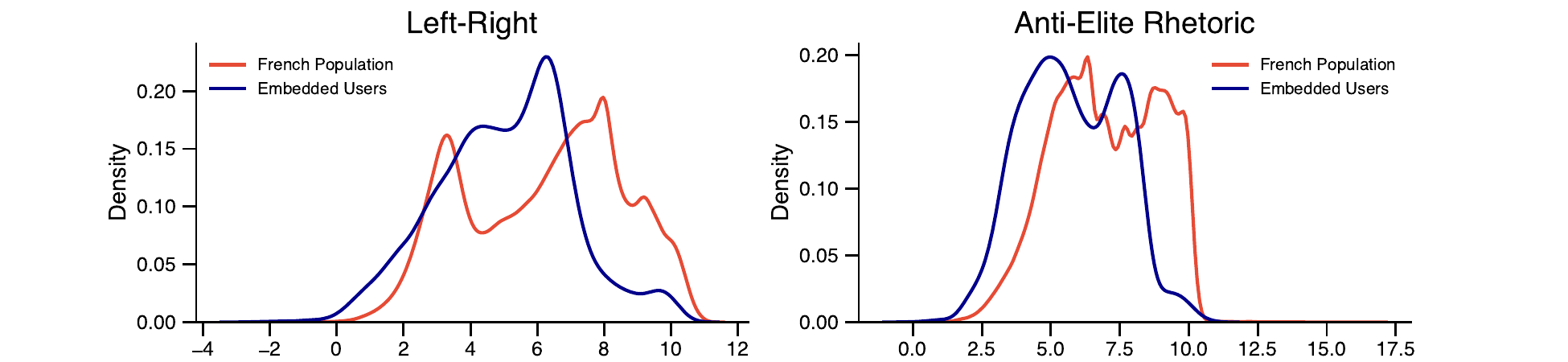}
\caption{Comparison of positions along the two political dimensions for the population available from the data from \cite{ramaciottimorales:hal-04807916}, and the intersection of 8\,249 users in the population $\mathcal{E}$ of our study (31.1\%{}). Distribution of users Left-Right (left figure) leaning and Anti-Elite rhetoric (right figure) within the embedded population $\mathcal{E}$ (in \textcolor{red}{red}) compared to the distribution within the overall French Twitter population (in \textcolor{blue}{blue}) from the dataset of politically-positioned Twitter users from \cite{ramaciottimorales:hal-04807916}.}
\label{fig:distrib_epo_vs_sample}
\end{figure}

\begin{figure}
\centering
\includegraphics[width=\linewidth]{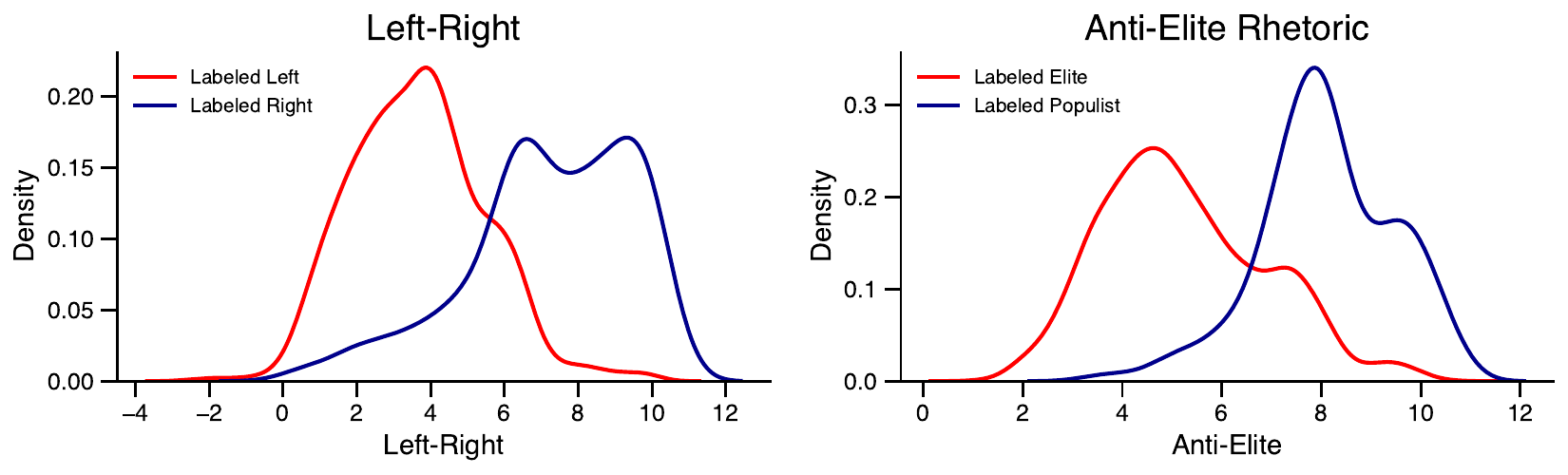}
\caption{Validation of the positions of 8\,249 users in the population $\mathcal{E}$ of our study (31.1\%{}) that are also in dataset from \cite{ramaciottimorales:hal-04807916}, by checking that labeled users are also correctly ordered by the two political dimensions. Distribution of users Left-Right leaning (left figure) and Anti-Elite rhetoric (right figure) within the embedded population $\mathcal{E}$ segmenting those labeled either as ``\textcolor{red}{Left}" or ``\textcolor{blue}{Right}", and either as ``\textcolor{red}{Elite}" or ``\textcolor{blue}{Populist}" by a LLM Zephyr 7B $\beta$ \cite{tunstall2023zephyr}.}
\label{fig:epo_valid_llm}
\end{figure}

Before using the dataset from \cite{ramaciottimorales:hal-04807916}, we compare our population $\mathcal{E}$ with that included in this dataset.
An analysis of the dataset reveals that 8\,249 users in the population $\mathcal{E}$ of our study (31.1\%{}) have also political positions in the dataset from \cite{ramaciottimorales:hal-04807916}.
Furthermore, we compare in Extended Figure~\ref{fig:distrib_epo_vs_sample} the distributions of users from $\mathcal{E}$ that have both, Left-Right and Antielite rhetoric positions, and labels in \cite{ramaciottimorales:hal-04807916}. Similarly, we specify the validation of the scales to the embedded population in Extended Figure~\ref{fig:epo_valid_llm}. The reader is referred to the documentation of the dataset we use \cite{ramaciottimorales:hal-04807916} for the details.

\begin{figure}
    \centering
    \includegraphics[width=\linewidth]{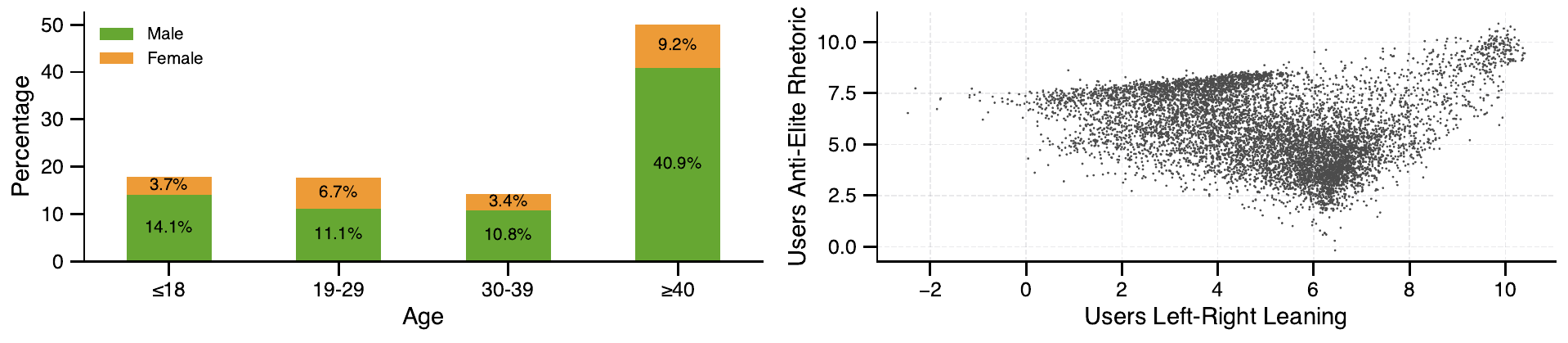}
    \caption{Summary of the main attributes of users in our population. Distribution of inferred age and binary gender for the embedded population $\mathcal{E}$ (left figure), Scatter plot of embedded users' positions on the left-right and anti-elite rhetoric scales (right figure).}
    \label{fig:summary_features_population}
\end{figure}

The core demographic and political makeup of our population of study is described in Figure~\ref{fig:fig1_characterization_population} in the main article, with Extended Figure~\ref{fig:summary_features_population} providing a more direct comparison within demographic and political variables.

\section{Inferring directions in the embedding space that reproduce user features}\label{section:cca}

Let $\Phi \in \mathbb{R}^{n \times d}$ denote the matrix of user embeddings, where each row $\phi_u^\top \in \mathbb{R}^d$ represents the embedding vector of user $u \in \{1,\ldots,n\}$ in a $d$-dimensional embedding space; in our case $d=256$ and $n = |\mathcal{E}|= 26\,509$. Given an attribute vector $\mathbf{a} \in \mathbb{R}^n$, where $a_i$ corresponds to a measured characteristic of user $i$, we aim to identify the direction in the embedding space that maximally correlate with this attribute.

For a unit vector $\mathbf{w} \in \mathbb{R}^d$, the correlation between the projection of user embeddings onto $\mathbf{w}$ and the attribute vector $\mathbf{a}$ is given by:
\begin{equation} 
\rho_{\mathbf{a}}^{\Phi}(\mathbf{w}) = \frac{(\Phi\mathbf{w})^\top\mathbf{a}}{\|\Phi\mathbf{w}\|_2\|\mathbf{a}\|_2}
\end{equation} 
where $\|\cdot\|_2$ denotes the Euclidean norm.

We seek to find $\mathbf{w}_{\mathbf{a}}$ such that the projection $\Phi\mathbf{w}_{\mathbf{a}}$ of users' embeddings $\Phi$ onto $\mathbf{w}_{\mathbf{a}}$ maximally correlates with $\mathbf{a}$:
\begin{equation} 
\mathbf{w}_{\mathbf{a}} = \arg\max_{\mathbf{w} \in \mathbb{R}^d: \|\mathbf{w}\|_2 = 1} \rho_{\mathbf{a}}^{\Phi}(\mathbf{w})
\end{equation} 

This optimization problem is a special case of Canonical Correlation Analysis (CCA), where one seeks to find a linear combination that maximizes correlation between a univariate dataset (the attribute vector $\mathbf{a}$) and a multivariate dataset (the user embeddings $\Phi$). The vector $\mathbf{w}_{\mathbf{a}}$ can then be readily obtained by solving the CCA eigenvalue problem.

To establish statistical significance, we employ a permutation test with $N=10^4$ repetitions. For each iteration, we generate $\tilde{\mathbf{a}}$ by randomly permuting the elements of $\mathbf{a}$, compute $\mathbf{w}_{\tilde{\mathbf{a}}}$ through CCA, and measure the resulting correlation $\rho_{\tilde{\mathbf{a}}}^{\Phi}(\mathbf{w}_{\tilde{\mathbf{a}}})$. This procedure yields the null distribution of correlations expected under the hypothesis of no meaningful relationship. The one-sided p-value is subsequently determined as the proportion of permuted absolute correlations that exceed the absolute observed correlation $\rho_{\mathbf{a}}^{\Phi}(\mathbf{w}_{\mathbf{a}})$.

Similarly, since the null distribution of feature directions $\mathbf{w}_{\mathbf{a}}$ depends on the distribution of user embeddings in $\mathbb{R}^d$, we use permutation tests, rather than assuming a uniform hypersphere distribution, to assess the significance of similarities between $\mathbf{w}_{\mathbf{a}}$ and $\mathbf{w}_{\mathbf{b}}$, the embedding directions associated with features $\mathbf{a}$ and $\mathbf{b}$. The one-sided p-value is computed as the proportion of randomized cosine similarities that exceed the observed similarity between $\mathbf{w}_{\mathbf{a}}$ and $\mathbf{w}_{\mathbf{b}}$, where the randomized similarities are computed between embedding directions $\mathbf{w}_{\tilde{\mathbf{a}}}$ and $\mathbf{w}_{\tilde{\mathbf{b}}}$ associated with shuffled features $\tilde{\mathbf{a}}$ and $\tilde{\mathbf{b}}$.


\section{Validating inferred embedding directions}\label{section:validating_cca}

After identifying through CCA the direction $\mathbf{w}_{\mathbf{a}}$ in the embedding space $\mathbb{R}^d$ spanned by learned account embeddings $\mathbf{\Phi}$ that maximally correlates with a given feature of interest $\mathbf{a}$, we validate its interpretation as genuinely capturing this feature.
This is needed because of two reasons. 1) Even if a direction produces a high correlation in ordering and positioning users according to attributes for users, the variance or disturbance error of an underlying linear model might still be high. 2) Secondly, not every user in $\mathcal{E}$ has known values for all attributes. Accordingly, we need to control that the CCA directions induce and ordering of users also for those that have no known attributes. The CCA direction for Left-Right leanings was computed relying on the 8\,249 users from $\mathcal{E}$ (31.1\%{}) that had a position on the Left-Right scale, and we need to control that the direction orders the rest of users in $\mathcal{E}$ correctly.

To validate the positions in the embedding independently from the data we used inferring a direction $\mathbf{w}_{\mathbf{a}}$, we analyze the friends and tweet content of users at both extremes of the direction $\mathbf{w}_{\mathbf{a}}$. Specifically, we identify overrepresented friends, words, and hashtags in the social connections and tweet content of users in the first and last deciles of $\mathbf{\Phi}\mathbf{w}_{\mathbf{a}}$ (hereafter referred to as lower and upper users over $\mathbf{w}_{\mathbf{a}}$).

We determine overrepresented entities by comparing their observed frequency in the target pool to their expected frequency under uniform random distribution along the $\mathbf{w}_{\mathbf{a}}$ direction. To protect user privacy and ensure representativity, we solely report popular accounts with over $50$k followers, followed by at least 10\% of the users in the considered decile.
More precisely, for each inferred dimension $\mathbf{w}_{\mathbf{a}}$ corresponding to a feature $\mathbf{a}$, we report the popular accounts, having more than $50$k followers, that are over-followed by the low projecting (the 10\%{} of users having the lowest positions on $\mathbf{w}_{\mathbf{a}}$) and high projecting (the 10\%{} of users having the highest positions on $\mathbf{w}_{\mathbf{a}}$) users in $\mathcal{E}$. In the case of words and hashtags, the calculation is done on the basis of those included in the tweets produced by users $\mathcal{E}$.

\subsection{Inferred Left-Right political leaning direction $\mathbf{w}_{LR}$}
\label{subsec:inferred_wlr}

\begin{figure}
\centering
\includegraphics[width=\linewidth]{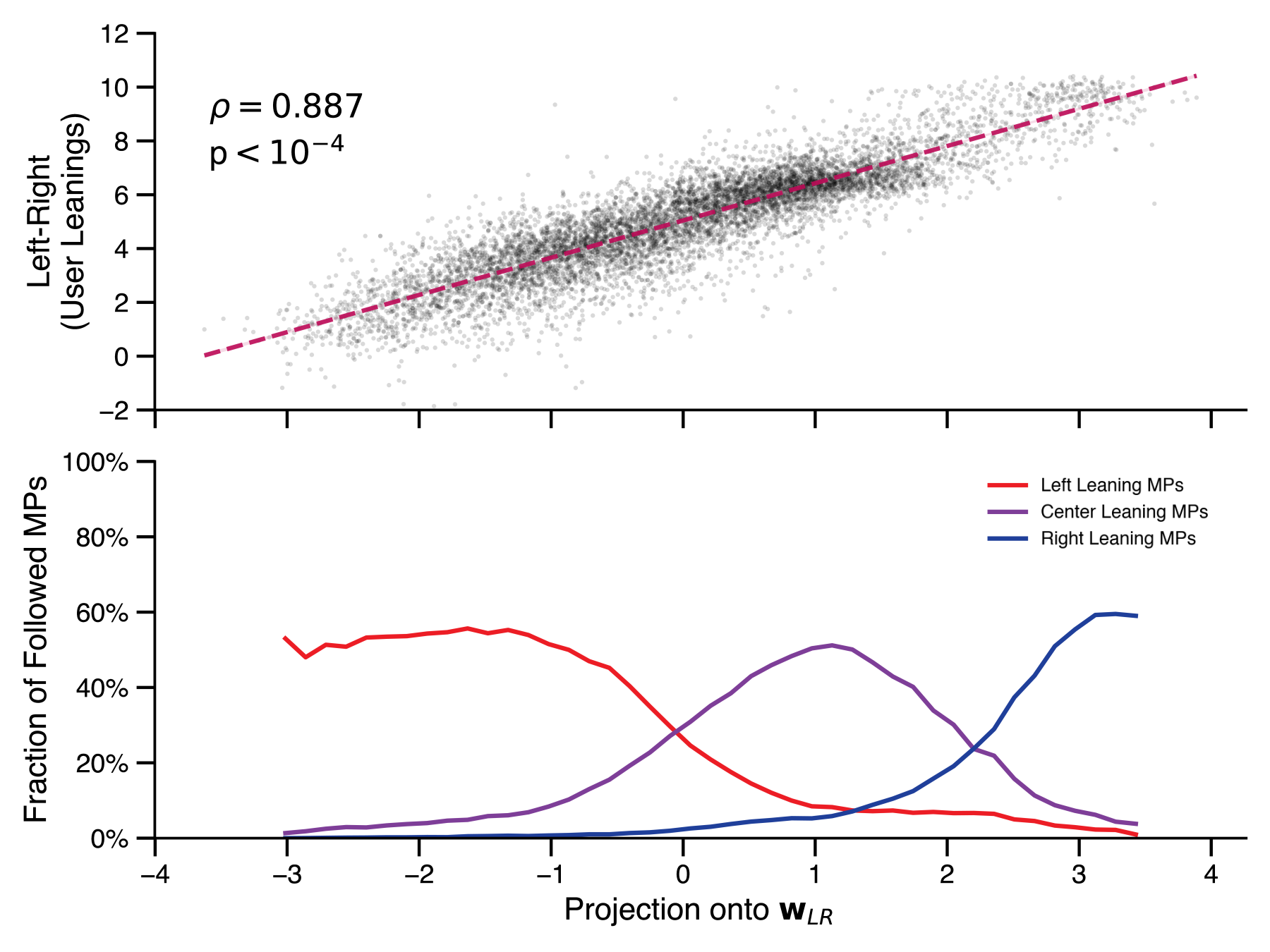}
\caption{Positioning of users by the recommender along direction $\mathbf{w}_{LR}$ and validation. (top figure) Scatter plot showing users' Left-Right political leaning against their projection onto $\mathbf{w}_{LR}$, the embedding direction maximizing feature correlation; $\rho = .887$, $p < 10^{-4}$. (bottom figure) Average fraction of followed Members of Parliament from main Left-leaning (LFI), Center  (LREM), and Right-leaning (RN) parties, as a function of users' projection onto $\mathbf{w}_{LR}$.}
\label{fig:valid_left_right}
\end{figure}

We analyze the relationship between users' political leanings and their projection onto the maximally correlated embedding direction $\mathbf{w}_{LR}$. As displayed on Extended Figure~\ref{fig:valid_left_right}, we observe a strong Pearson correlation $\rho = 0.887$, $p < 10^{-4}$. 

To validate the interpretation of this direction, we examine how users' following patterns of Members of French Parliament (MPs) vary along $\mathbf{w}_{LR}$. For each user, we compute the fraction of followed MPs from three political categories in the French National Assembly: Left-leaning (\textit{La France Insoumise}, LFI), Center  (\textit{La République en Marche}, LREM), and Right-leaning (\textit{Rassemblement National}, RN). The bottom panel of Extended Figure~\ref{fig:valid_left_right} displays these fractions as a function of users' projection onto $\mathbf{w}_{LR}$.
LREM (now rebaptized as \textit{Renaissance}), RN, and LFI are the parties having produced the first three front-runners in the first round of the last presidential election in France in 2022, obtaining correspondingly,  27.9\%{} (Emmanuel Macron), 23.2\%{} (Marine Le Pen) and 22\%{} (Jean-Luc M\'elenchon) of the vote (the next and fourth candidate obtaining 7.1\%{})

Users projecting low onto $\mathbf{w}_{LR}$ (below $-1$) predominantly follow Left-leaning MPs, with minimal following of Center MPs and virtually no following of Right-leaning MPs. Conversely, users projecting high onto $\mathbf{w}_{LR}$ exhibit an opposite pattern, with Right-leaning MPs constituting 59\% of their followed parliamentarians. Users projecting onto $\mathbf{w}_{LR}$ at around 1.1 predominantly follow MPs from the Center, who comprise 56\% of their followed parliamentarians.

Following our previously outlined methodology, we report the in Extended Lists~\ref{list:embedding_left_leaning_words}~\&{}~\ref{list:embedding_right_leaning_words} the accounts, words and hashtags overrepresented in users' following networks and content production at opposite ends of the $\mathbf{w}_{LR}$ spectrum further validate this interpretation.\newline

\begin{customlist}
\label{list:embedding_left_leaning_words}
Accounts followed, words used, and hashtag used by 10\%{} of users with the highest probability of begin Left-leaning according to the embedding space of recommender system (i.e., having low values along $\mathbf{w}_{LR}$).
    
    \begin{itemize}
    \item \textbf{Popular followed accounts:} \textit{@AvecJLM, @FiAssemblee, @PODEMOS, @europeangreens, @sebastiendelogu, @NUPES\_2022\_, @L\_insoumission, @CarolineFiat54, @LachaudB, @GabrielBoric, @alma\_dufour, @TrouveAurelie, @Ugobernalicis, @GreensEFA, @PCF}
    \item \textbf{Popular words:} \textit{austerity, ruralities, sleeping in street, perjury, rebellious, parliamentary niche, restoration of wealth tax, pension repeal, capital income, gaza west bank}
    \item \textbf{Popular hashtags:} \textit{\#{}jde2024 (Summer Days of Ecologists), \#{}noconfidencemotion, \#{}impeachmentmarch, \#{}nfp (New Popular Front), \#{}victorynfp, \#{}wevotenfp, \#{}amfis2024 (Far-left-wing Summer University), \#{}october1strike, \#{}newpopularfront, \#{}ceasefire, \#{}popularunion, \#{}macronimpeachment, \#{}sanofi (pharmaceutical company), \#{}eelv (Green Party)}
    \end{itemize}
\end{customlist}

\begin{customlist}
\label{list:embedding_right_leaning_words}
Accounts followed, words used, and hashtag used by 10\%{} with the highest probability of begin Right-leaning according to the embedding space of recommender system (i.e., having high values along $\mathbf{w}_{LR}$).
    
    \begin{itemize}
    \item \textbf{Popular followed accounts:} \textit{@gabriellecluzel, @tegnererik, @IdLibertes, @tanguy\_france, @stanislasrig, @JulienOdoul, @\_LesPatriotes, @matteosalvinimi, @AmauryBucco, @ChdOrnellas, @GregToussaint4, @Stephane\_Ravier, @kevinbossuet, @Bruno\_Attal\_, @HistoiredeFran7}
    \item \textbf{Popular words:} \textit{islamization, illegal immigrants, sovereign powers, remigration, narco, family reunification, cathedrals, allah akbar}
    \item \textbf{Popular hashtags:} \textit{\#{}deportation, \#{}hdpros (Right-Leaning TV Show), \#{}drugs, \#{}europe1 (French Radio), \#{}reconquête (Far-Right Policy Party), \#{}angryfarmers, \#{}sinwar}
    \end{itemize}
\end{customlist}

\subsection{Inferred anti-elite rhetoric direction $\mathbf{w}_{AntiElite}$}
\label{subsec:inferred_wantielite}

\begin{figure}
\centering
\includegraphics[width=\linewidth]{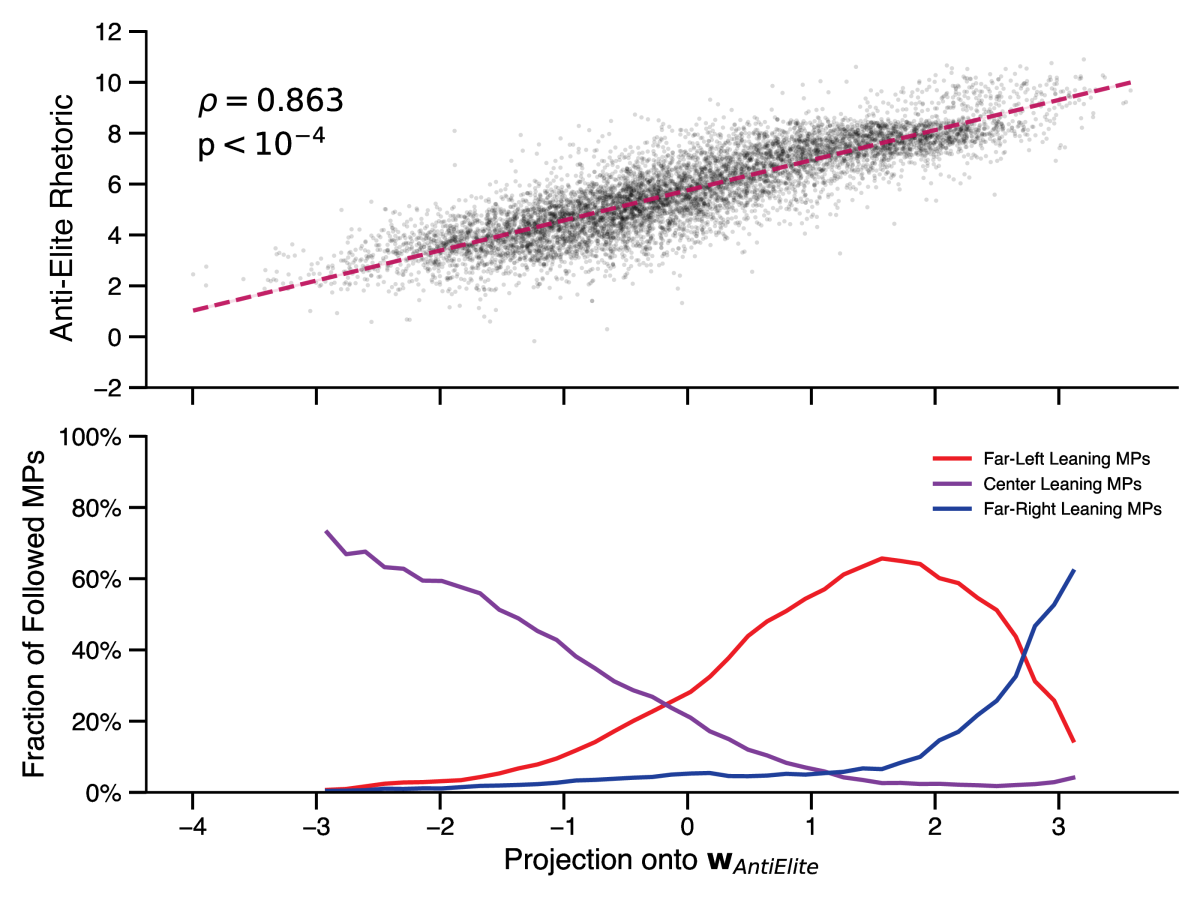}
\caption{Positioning of users by the recommender along direction $\mathbf{w}_{AntiElite}$ and validation. (Top) Scatter plot showing users' Anti-Elite Salience against their projection onto $\mathbf{w}_{AntiElite}$, the embedding direction maximizing feature correlation; $\rho = .863$, $p < 10^{-4}$. (Bottom) Average fraction of followed Members of Parliament from Left-leaning (LFI), Center (LREM), and Right-leaning (RN) parties, as a function of users' projection onto $\mathbf{w}_{LR}$.}
\label{fig:valid_antielite}
\end{figure}

Following the same approach as for Left-Right positions, we plot in Extended Figure~\ref{fig:valid_antielite} users' positions on the anti-elite rhetoric dimensions against their projection onto $\mathbf{w}_{AntiElite}$. This projection exhibits a strong Pearson correlation, $\rho = 0.863$, $p < 10^{-4}$, with users' anti-elite salience scores.

Users projecting low onto $\mathbf{w}_{AntiElite}$, do not display salient sentiments against elites and institutions, predominantly follow center-leaning MPs, who constitute 73\% of their followed parliamentarians. This proportion decreases along $\mathbf{w}_{AntiElite}$, while the fraction of far-right-leaning MPs in users' followed MPs progressively increases.

Following the previously outlined methodology, we report in Extended Lists~\ref{list:embedding_low_antielite_words}~\&{}~\ref{list:embedding_high_antielite_words} the accounts, words and hashtags overrepresented in users' following networks and content production at opposite ends of the $\mathbf{w}_{AntiElite}$ spectrum further validate this interpretation.\newline

\begin{customlist}
\label{list:embedding_low_antielite_words}
Accounts followed, words used, and hashtag used by 10\%{} with the lowest probability of displaying anti-elite rhetoric according to the embedding space of recommender system (i.e., having low values along $\mathbf{w}_{AntiElite}$).
    
    \begin{itemize}
    \item \textbf{Popular followed accounts:} \textit{@FaireEnsemblefr, @FrontRep, @DeputesEnsemble, @thalesgroup, @JeunesMacron, @RPFranceUE, @JBLemoyne, @EmmWargon, @tcabarrus, @regionbretagne, @FranceTravail, @olivierdussopt, @StanGuerini, @J\_Denormandie, @ENGIEgroup}
    \item \textbf{Popular words:} \textit{competitive, Air Force, DGA (Directorate General of Armament'), national days, brothers in arms, industrial sovereignty, civil security, circular economy}
    \item \textbf{Popular hashtags:} \textit{\#{}decarbonization, \#{}SMEs (Small/Medium Enterprises), \#{}choosefrance (investment in France initiative), \#{}france2030 (innovation investment plan), \#{}territories, \#{}cyber (cybersecurity), \#{}municipalities, \#{}cybermonth, \#{}frenchtech (French tech ecosystem), \#{}HR (Human Resources), \#{}companies, \#{}defense}
    \end{itemize}
\end{customlist}

\begin{customlist}
\label{list:embedding_high_antielite_words}

Accounts followed, words used, and hashtag used by 10\%{} with the highest probability of displaying anti-elite rhetoric according to the embedding space of recommender system (i.e., having high values along $\mathbf{w}_{AntiElite}$).
    
    \begin{itemize}
    \item \textbf{Popular followed accounts:} \textit{@louisfouch3, @BanounHelene, @StatChrisCotton, @xazalbert, @MichelJeanDomi1, @ReinfoCovidOff, @CStrateges, @VictorSinclair3, @verity\_france, @Laissonslespre1, @myriampalomba, @LaurentOzon, @gregtabibian, @CaudeHenrion, @Beatrice\_Rosen}
    \item \textbf{Popular words:} \textit{remigration, politico-media, pedophilia, Islamization, UEJF (Union of French Jewish students), Soros, scheming, family reunification}
    \item \textbf{Popular hashtags:} \textit{\#{}deportation, \#{}stopgenocide, \#{}reconquête (far-right French political party), \#{}impeachment, \#{}zelensky, \#{}angryfarmers, \#{}zemmour (French far-right politician), \#{}mercosur, \#{}macronimpeachment, \#{}impeachmentmarch, \#{}amfis2024 (Far-left-wing summer university)}
    \end{itemize}
\end{customlist}

\subsection{Inferred gender direction $\mathbf{w}_{gender}$}
\label{subsec:inferred_wgender}

\begin{figure}
\centering
\includegraphics[width=\linewidth]{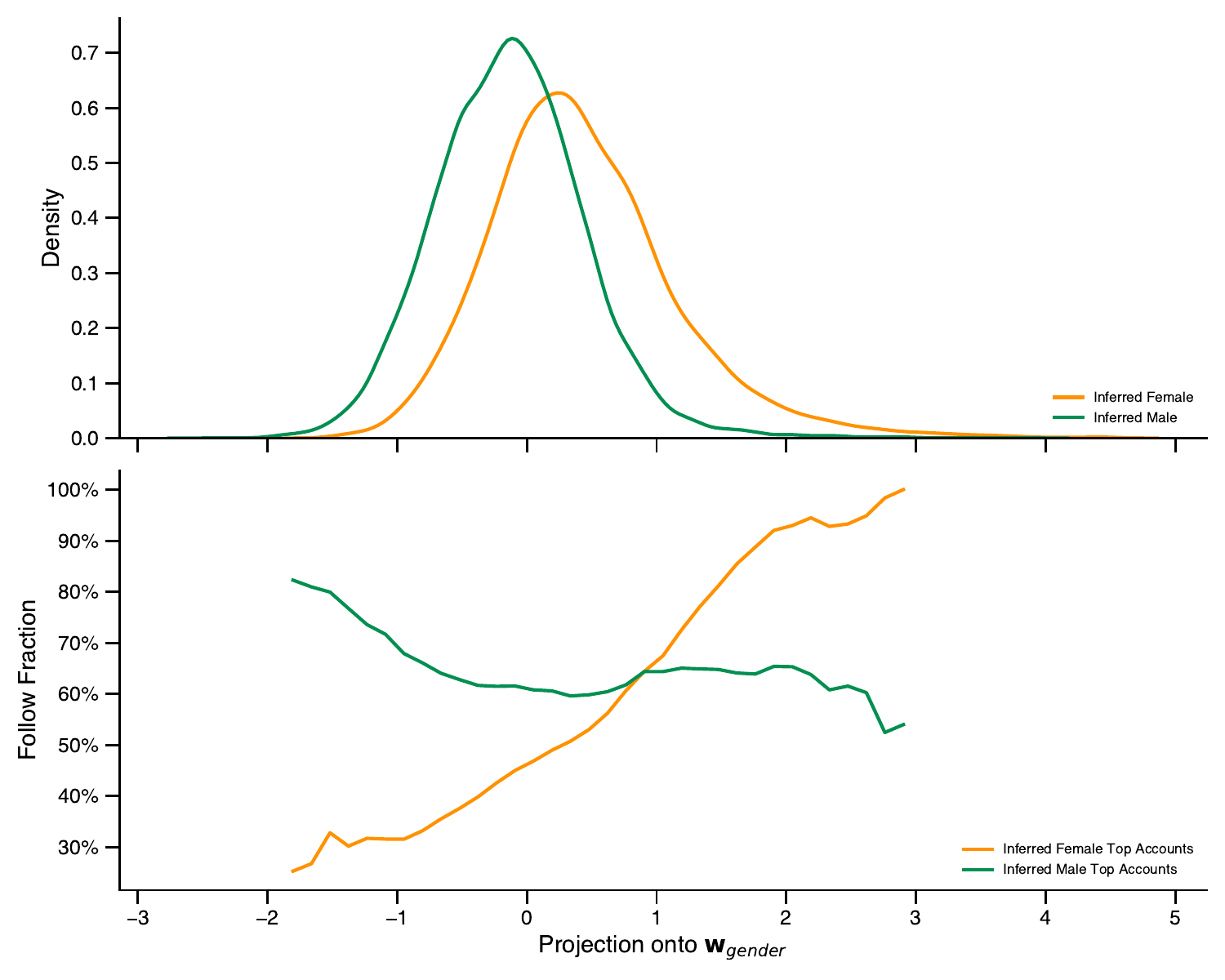}
\caption{Positioning of users by the recommender along direction $\mathbf{w}_{gender}$ and validation. (Top figure) Distribution of inferred-male/female users according to their projection onto $\mathbf{w}_{gender}$. (Bottom figure) Fraction of users following at least one of the popular accounts , as function of their projection onto  $\mathbf{w}_{gender}$.}
\label{fig:valid_gender}
\end{figure}

The direction $\mathbf{w}_{gender}$ in the embedding space induced by user embeddings $\Phi$ that maximally correlates with the estimated probability of users being ``female" (binary classification, with complementary probability of being ``male"), achieves a Pearson correlation of $\rho = .384$, $p < 10^{-4}$. When framed as a classification task for these two groups, projections onto $\mathbf{w}_{gender}$ predict user gender with an AUC-ROC of $.731\pm.044$ (standard deviation, 10-fold cross validation). We display in Extended Figure~\ref{fig:valid_gender} the distribution of projections onto $\mathbf{w}_{gender}$ for users inferred as ``female" (\texttt{M3} prediction above 0.5) and ``male".

A Mann–Whitney U test reveals that the positions of users along $\mathbf{w}_{gender}$,  classified as ``female" are greater than the positions of users accounts classified as ``male", with $p\ll10^{-4}$.

Following the previously outlined methodology, we report the accounts, words and hashtags overrepresented in users' following networks and content production at opposite ends of the $\mathbf{w}_{gender}$ spectrum further validate this interpretation.

Extended Figure~\ref{fig:valid_gender} shows the fraction of users following at least one of the 15 popular accounts, reported below, as a function of their projection onto $\mathbf{w}_{gender}$. The fraction of users following inferred female-popular accounts increases along $\mathbf{w}_{gender}$, while those following inferred male-popular accounts decreases to a lesser extent along this direction. As before, we report the accounts followed, and words and hashtags used by users at the extremes of direction $\mathbf{w}_{gender}$ in Extended Lists~\ref{list:embedding_male_words}~\&{}~\ref{list:embedding_female_words}. \newline

\begin{customlist}
\label{list:embedding_male_words}
Accounts followed, words used, and hashtag used by 10\%{} with the highest probability of being male according to the embedding space of recommender system (i.e., having low values along $\mathbf{w}_{gender}$).

\begin{itemize}
\item \textbf{Popular followed accounts:} \textit{@RMCsport, @beinsports\_FR, @Romain\_Molina, @PSG\_inside, @lequipe, @AntoGriezmann, @equipedefrance, @KMbappe, @X, @NetflixFR, @AAstierOff, @NASA, @Michel\_Goya, @realDonaldTrump, @vincentglad}
\item \textbf{Popular words:} \textit{genesio, mourinho, zerbi, borussia, lacazette, bundesliga, kolo muani, rabiot, pogacar, clasico, away section, juventus, scorers, muani, goalkeeper,}
\item \textbf{Popular hashtags:} \textit{\#{}uel (Europa League), \#{}ligue1 (French Football League), \#{}tdf2023 (Tour de France 2023), \#{}om (Olympique Marseille), \#{}ucl (Champions League), \#{}teamom (Team Olympique Marseille), \#{}cycling, \#{}ompsg (Olympique Marseille vs Paris Saint-Germain), \#{}top14 (French Rugby League), \#{}defense, \#{}frenchxv (French Rugby Team), \#{}bitcoin, \#{}pgw (Paris Games Week), \#{}tdf2024 (Tour de France 2024)}
\end{itemize}
\end{customlist}

Users projecting low onto $\mathbf{w}_{gender}$, i.e., having high probability of being male, demonstrate strong engagement with sports content across their following patterns and published content, frequently mentioning football players and associated hashtags.\newline

\begin{customlist}
\label{list:embedding_female_words}
Accounts followed, words used, and hashtag used by 10\%{} with the highest probability of being female according to the embedding space of recommender system (i.e., having high values along $\mathbf{w}_{gender}$).

\begin{itemize}
\item \textbf{Popular followed accounts:} \textit{@ONUFemmes, @Ovidieofficiel, @osezlefeminisme, @NousToutesOrg, @BeaulieuBap, @PepiteSexiste, @PenelopeB, @alicecoffin, @MaiaMazaurette, @monachollet, @Klaire, @LyesLouffok, @faizazrl, @misspress, @ATrapenard}
\item \textbf{Popular words:} \textit{Stéphane Plaza (TV host under investigation), masculinists, PPDA (Patrick Poivre d'Arvor - former news anchor accused of SGBV), Godrèche (Judith - actress speaking against SGBV), Haenel (Adèle - actress speaking against SGBV), plaintiffs (multiple accusers in various cases), ableism, intersex, male domination, pimping, Judith Godrèche (leading voice against SGBV), incest, femicides, Jacquot (Benoît - film director accused of SGBV)}
\item \textbf{Popular hashtags:} \textit{\#{}rape, \#{}ppda (Accused of SGBV), \#{}rapes, \#{}metoo, \#{}pelicot (Survivor of SGBV), \#{}sexism, \#{}mazan, \#{}depardieu (Accused of SGBV), \#{}feminism, \#{}insociety, \#{}feminism, \#{}sexualviolence, \#{}giselepelicot (Survivor of SGBV), \#{}bedos (Accused of SGBV) , \#{}womensrights}
\end{itemize}
\end{customlist}

Users projecting high onto $\mathbf{w}_{gender}$, i.e., having high probability of being female, demonstrate significant engagement in French public discourse surrounding sexual violence, industry abuse, and feminist movements. Their content frequently references individuals involved in Sexual and Gender Based Violence cases, including both alleged perpetrators and survivors, alongside associated advocacy hashtags.

\subsection{Inferred age direction $\mathbf{w}_{age}$}
\label{subsec:inferred_wage}

We denote by $\mathbf{w}_{age}$ the direction in the embedding space that achieves the strongest correlation with users' estimated age.
Because \texttt{M3} provides categorical values ($\leq$18, 19-29, 30-39, and $\geq$40), we computed, for each user, estimated age as a continuous variable by considering the middle points of each category (15.5 years for the $\leq 18$ bin, 24.5 years old for the 18-29 bin, 34.5 years old for the 30-39, and 45 years for the $\geq40$ bin), weighted by the probability of belonging to the age category as given by the \texttt{M3} model.
We then compared these continuous age values with positions along $\mathbf{w}_{age}$, achieving a Pearson correlation of $\rho = .562$, $p < 10^{-4}$.

\begin{figure}
\centering
\includegraphics[width=\linewidth]{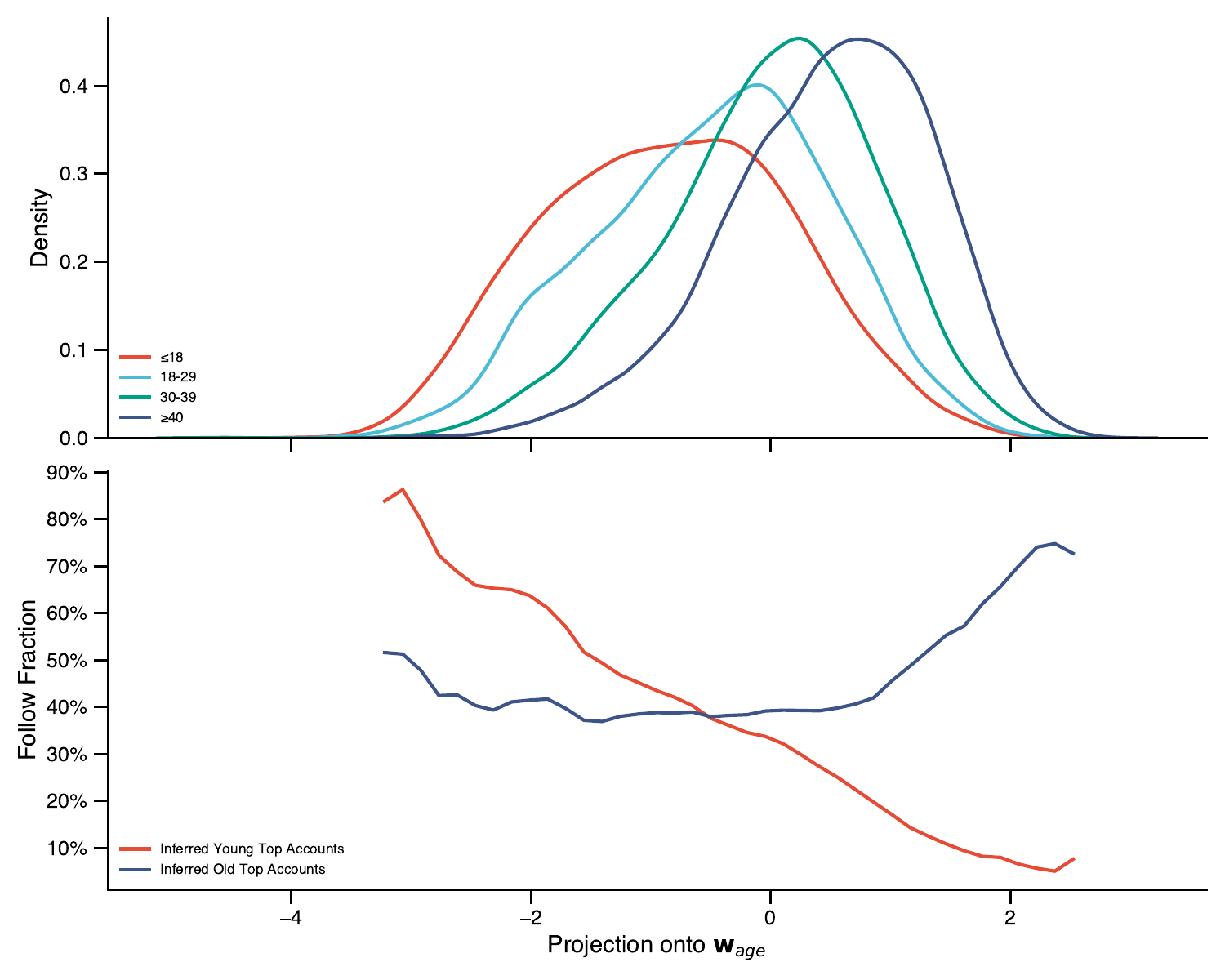}
\caption{Positioning of users by the recommender along direction $\mathbf{w}_{age}$ and validation. (Top figure) Distribution of users projection onto $\mathbf{w}_{age}$. (Bottom figure) Fraction of users following at least one of the age-popular accounts, as function of their projection onto  $\mathbf{w}_{age}$.}
\label{fig:valid_age}
\end{figure}

When framed as a multi-categorical classification task, projections onto $\mathbf{w}_{age}$ predict user \texttt{M3} age bins with an AUC-ROC of $.692 \pm .030$ (standard deviation, 10-fold cross validation). Extended Figure~\ref{fig:valid_age} displays the distribution of projections onto $\mathbf{w}_{age}$ for each age category inferred by \texttt{M3}.

A Mann–Whitney U test reveals that the distribution of projection onto $\mathbf{w}_{age}$ of accounts classified as $\leq 18$ is  stochastically lesser than the distribution underlying accounts classified as $19-29$, similarly the distribution associated with $19-29$ accounts is stochastically lesser than the one associated with $30-39$, finally the distribution of projection of account classified as $\geq 40$ is stochastically greater than the one of $30-39$; with $p\ll10^{-4}$.

{For validation purposes, we identify the direction in the embedding that achieves the strongest correlation with volunteers' self-declared age. The direction identified for this narrow set of users aligns with the direction previously identified considering the M3-inferred population with a cosine similarity of 0.63 and as observed with the M3-inferred direction poorly aligns with the left-right political direction with a cosine similarity of 0.16.}

Following the previously outlined methodology, we report the accounts, words and hashtags overrepresented in users' following networks and content production at opposite ends of the $\mathbf{w}_{age}$ spectrum further validate this interpretation.

Extended Figure~\ref{fig:valid_age} shows the fraction of users following at least one of the 15 popular accounts reported below, as a function of their projection onto $\mathbf{w}_{age}$. The fraction of users following inferred young-popular accounts significantly decreases along $\mathbf{w}_{age}$, from being followed by 86\% of users projecting low onto $\mathbf{w}_{age}$ to being followed by only 5\% of those scoring high along this same direction. Conversely, the fraction of users following inferred old-popular accounts increases, yet to a lesser extent, along the direction. As before, we report the accounts followed, and words and hashtags used by users at the extremes of direction $\mathbf{w}_{age}$ in Extended Lists~\ref{list:embedding_young_words}~\&{}~\ref{list:embedding_old_words}.\newline

\begin{customlist}
\label{list:embedding_young_words}
Accounts followed, words used, and hashtag used by 10\%{} with the highest probability of being young according to the embedding space of recommender system (i.e., having low values along $\mathbf{w}_{age}$).

\begin{itemize}
\item \textbf{Popular followed accounts:} \textit{@Theorus\_, @idkbvtimleila, @KoteiZousa, @EcrisioFX, @KennyStream, @JoelPostbad, @Grimkujow\_, @LRB\_LoL, @DephaseuR, @Chap\_GG, @mastuu\_, @Doigby, @Inoxtag, @KarmineCorp, @Kammeto}
\item \textbf{Popular words:} \textit{karmine corp, jcrois, faker, mdrrrrrrr, jpense, ptdrrrr, sdm, jveux, ninho, oomf, tqt, wsh, zevent, ptdrrrrr}
\item \textbf{Popular hashtags:} \textit{\#{}azurlane (Mobile game), \#{}bluearchive (Mobile game), \#{}worlds2024 (League of Legends Championship), \#{}lec (European League Championship), \#{}kcorp (French esports team), \#{}zevent2024 (French charity stream event), \#{}nintendoswitch (gaming console), \#{}pgw (Paris Games Week), \#{}onepiece (manga/anime), \#{}advertising (ad disclosure), \#{}rt (retweet), \#{}ad (advertisement), \#{}illustration (art content), \#{}staracademy (French TV show)}
\end{itemize}
\end{customlist}

Users projecting low onto $\mathbf{w}_{age}$, i.e., inferred young, demonstrate strong engagement with gaming and entertainment content in their following patterns and published content, frequently referencing video games, e-sport events, and their associated hashtags.\newline

\begin{customlist}
\label{list:embedding_old_words}
Accounts followed, words used, and hashtag used by 10\%{} with the highest probability of being old according to the embedding space of recommender system (i.e., having high values along $\mathbf{w}_{age}$).

\begin{itemize}
\item \textbf{Popular followed accounts:} \textit{@DarmonMichael, @frebsamen, @medef, @JLBorloo, @LaurentFabius, @chevenement, @pierremoscovici, @ericwoerth, @MarisolTouraine, @jwaintraub, @JBLemoyne, @danycohnbendit, @VenturaAlba, @jccambadelis, @jeanmarcayrault}
\item \textbf{Popular words:} \textit{competitiveness, Jacques Delors (former European Commission President), industrial sovereignty, decentralization, social cohesion, Mario Draghi (former ECB head), ministerial, Michel Barnier (EU Brexit negotiator, Former French Prime Minister), de-escalation, Slava Ukraini}
\item \textbf{Popular hashtags:} \textit{\#{}zan (France's ``zero net artificialization" environmental law), \#{}decarbonization, \#{}unga79 (UN General Assembly), \#{}cop16, \#{}bonjourchezvous (morning TV show), \#{}SMEs (Small/Medium Enterprises), \#{}sustainabledevelopment, \#{}localauthorities, \#{}carbon, \#{}france2030 (national investment plan), \#{}cop29, \#{}proximity, \#{}choosefrance (investment attraction program)}
\end{itemize}
\end{customlist}

Users projecting high onto $\mathbf{w}_{age}$, i.e., inferred old, demonstrate significant engagement with French national and international political discourse.

\subsection{Inferred directions in the embedding space of the recommender system and that capture interest in different topics as defined by \cite{cardiffNLP}}
\label{subsec:inferred_wtopics}

\subsubsection{News $\mathbf{w}_{News}$}

\begin{figure}
\centering
\includegraphics[width=\linewidth]{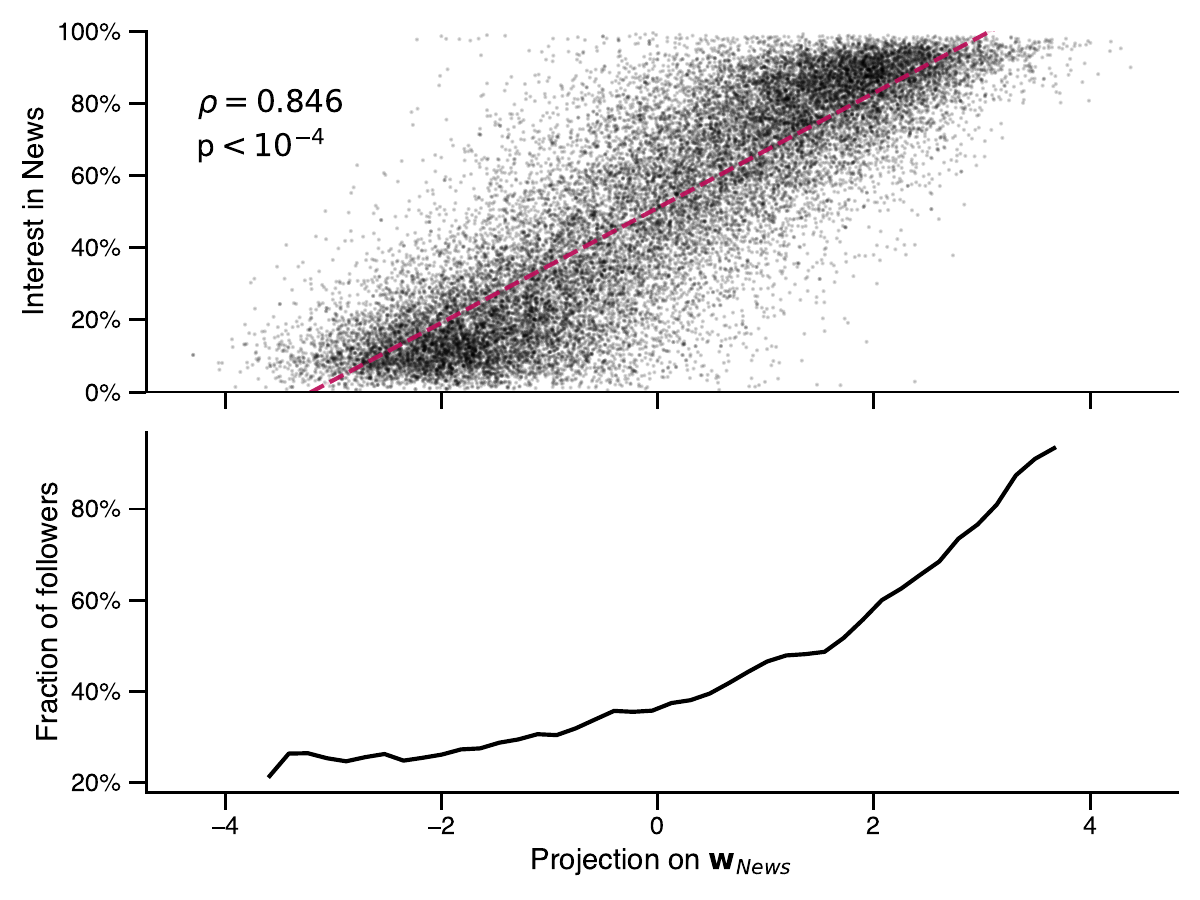}
\caption{Positioning of users by the recommender along direction $\mathbf{w}_{News}$ and validation. (Top figure) Scatter plot of user projections onto embedding direction $\mathbf{w}_{News}$ versus their fraction of news-related tweets. (Bottom figure) Fraction of users following at least one account popular among news enthusiasts.}
\label{fig:valid_topic_news}
\end{figure}

We plot in Extended Figure~\ref{fig:valid_topic_news} the projection of users onto $\mathbf{w}_{News}$ versus the fraction of tweets published or relayed by users related to ``News \& Social Concern'', showing a strong Pearson correlation; $\rho = .851$, $p < 10^{-4}$.

We also show, as function of users projection onto $\mathbf{w}_{News}$, the fraction of users following at least one account from the 15 selected accounts that are popular among news-enthusiast users, according to our previously described methodology. We report the accounts followed by users at the extreme positions in direction $\mathbf{w}_{News}$ in Extended List~\ref{list:embedding_accounts_news}.\newline

\begin{customlist}
\label{list:embedding_accounts_news}
15 selected accounts that are popular among news-enthusiast users according to their positing along the direction of the embedding that induces the highest correlation with interest in news.\\
    \textit{@JosepBorrellF, @KenRoth, @AgnesCallamard, @haaretzcom, @EP\_President, @SecBlinken, @amnesty, @UNHumanRights, @maxseddon, @coe, @L\_insoumission, @hrw, @Europarl\_EN, @Refugees, @EUCouncil}
\end{customlist}

\subsubsection{Sports $\mathbf{w}_{Sports}$}

\begin{figure}
\centering
\includegraphics[width=\linewidth]{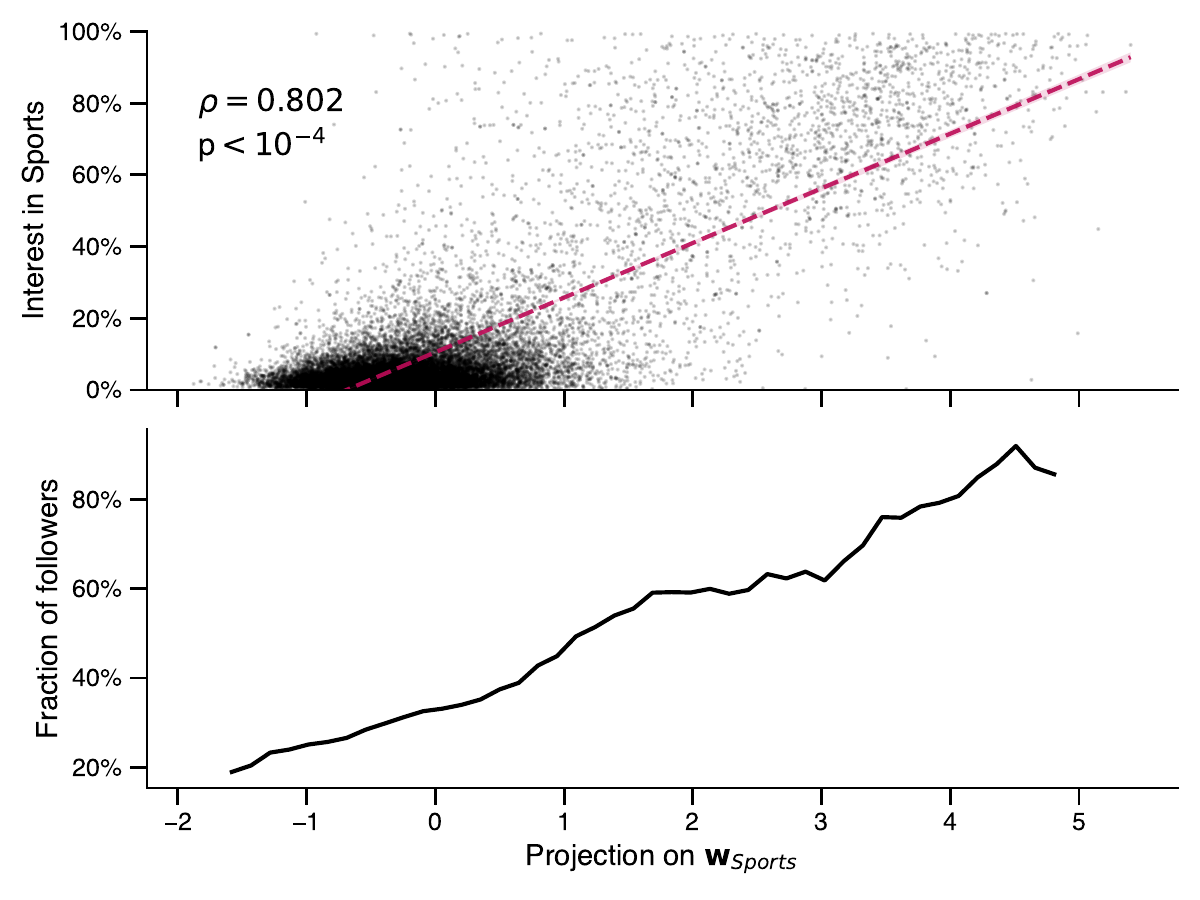}
\caption{Positioning of users by the recommender along direction $\mathbf{w}_{Sports}$ and validation. (Top figure) Scatter plot of user projections onto embedding direction $\mathbf{w}_{Sports}$ versus their fraction of sports-related tweets. (Bottom figure) Fraction of users following at least one account popular among sports enthusiasts.}
\label{fig:valid_topic_sports}
\end{figure}

We plot in Extended Figure~\ref{fig:valid_topic_sports} the projection of users onto $\mathbf{w}_{Sports}$ versus the fraction of tweets published or relayed by users related to ``Sports'', showing a strong Pearson correlation; $\rho = .809$, $p < 10^{-4}$.

Additionally, we show, as function of users projection onto $\mathbf{w}_{Sports}$, the fraction of users following at least one account from the 15 selected accounts that are popular among sports-enthusiast users, according to our previously described methodology. We report the accounts followed by users at the extreme of direction $\mathbf{w}_{Sports}$ in Extended List~\ref{list:embedding_accounts_sports}.\newline

\begin{customlist}
\label{list:embedding_accounts_sports}
15 selected accounts that are popular among sports-enthusiast users according to their positing along the direction of the embedding that induces the highest correlation with interest in sports.\\
    
    \textit{@UNFP, @EuropaLeague, @AS\_Monaco, @FIFAcom, @LFPfr, @premierleague, @ChampionsLeague, @PVSportFR, @losclive, @ballondor, @Eurosport\_FR, @Ligue1, @CanalplusFoot, @Omar\_daFonseca, @lachainelequipe}
\end{customlist}

\subsubsection{Science \&{} Technology $\mathbf{w}_{SciencesTech}$}

We plot in Extended Figure~\ref{fig:valid_topic_sports} the projection of users onto $\mathbf{w}_{SciencesTech}$ versus the fraction of tweets published or relayed by users related to ``Science \& Technology'', showing a strong Pearson correlation; $\rho = .758$, $p < 10^{-4}$.

Additionally, we show, as function of users projection onto $\mathbf{w}_{SciencesTech}$, the fraction of users following at least one account from the 15 selected accounts that are popular among science and technology-enthusiast users, according to our previously described methodology. We report the accounts followed by users at the extreme of direction $\mathbf{w}_{SciencesTech}$ in Extended List~\ref{list:embedding_accounts_science}.\newline

\begin{customlist}
\label{list:embedding_accounts_science}
15 selected accounts that are popular among sciences and technology-enthusiast users according to their positing along the direction of the embedding that induces the highest correlation with interest in sciences and technology.\\
    
    \textit{@karpathy, @ylecun, @techreview, @OpenAI, @sama, @NaturePortfolio, @Nature, @newscientist, @ScienceMagazine, @TechCrunch, @esa, @UNFCCC, @tim\_cook, @IPCC\_CH, @Google}
\end{customlist}

\begin{figure}
\centering
\includegraphics[width=\linewidth]{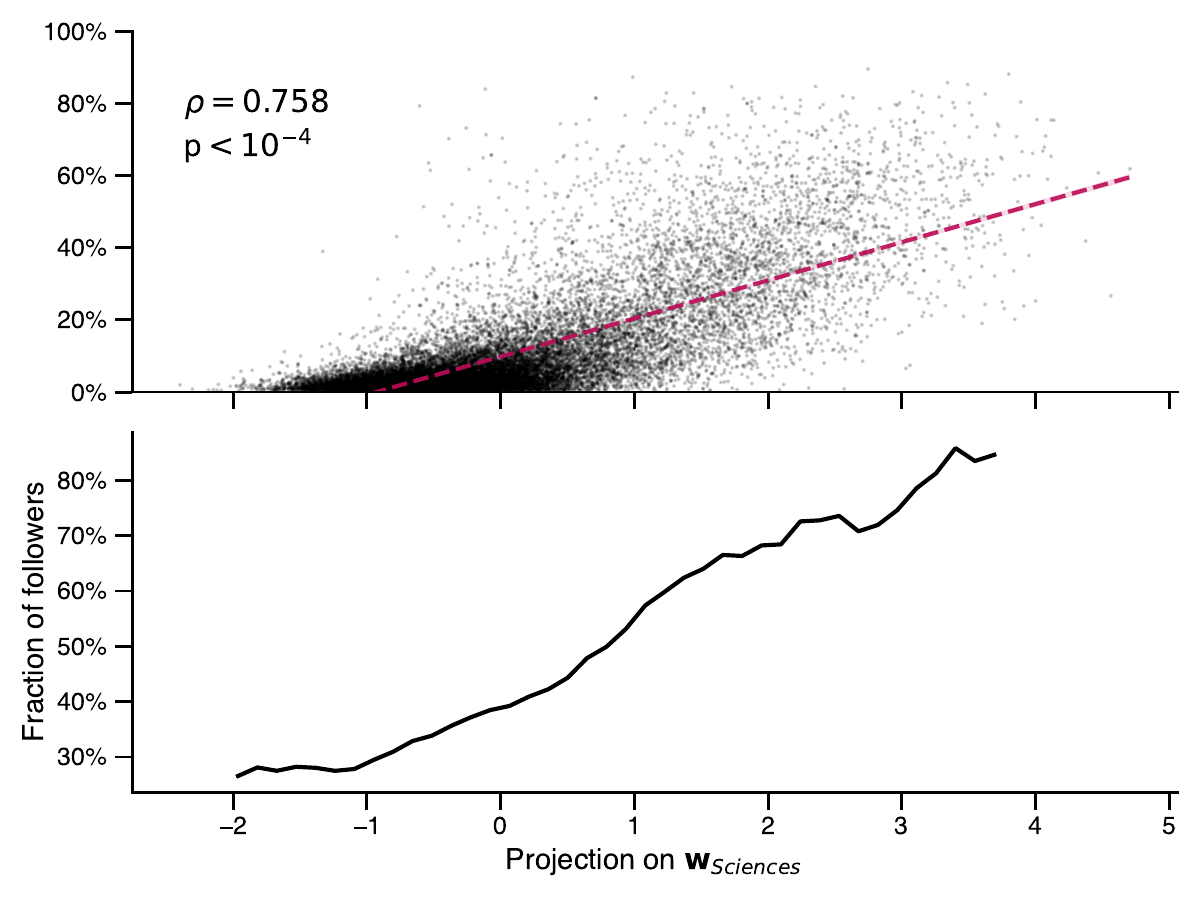}
\caption{Positioning of users by the recommender along direction $\mathbf{w}_{SciencesTech}$ and validation. (Top figure) Scatter plot of user projections onto embedding direction $\mathbf{w}_{SciencesTech}$ versus their fraction of sciences-and-technology-related tweets. (Bottom figure) Fraction of users following at least one account popular among sciences and technology enthusiasts.}
\label{fig:valid_topic_sciences}
\end{figure}

\subsection{Inferred directions in the embedding space of the recommender system and that the lexical complexity $\mathbf{w}_{complexity}$}

\begin{figure}
\centering
\includegraphics[width=\linewidth]{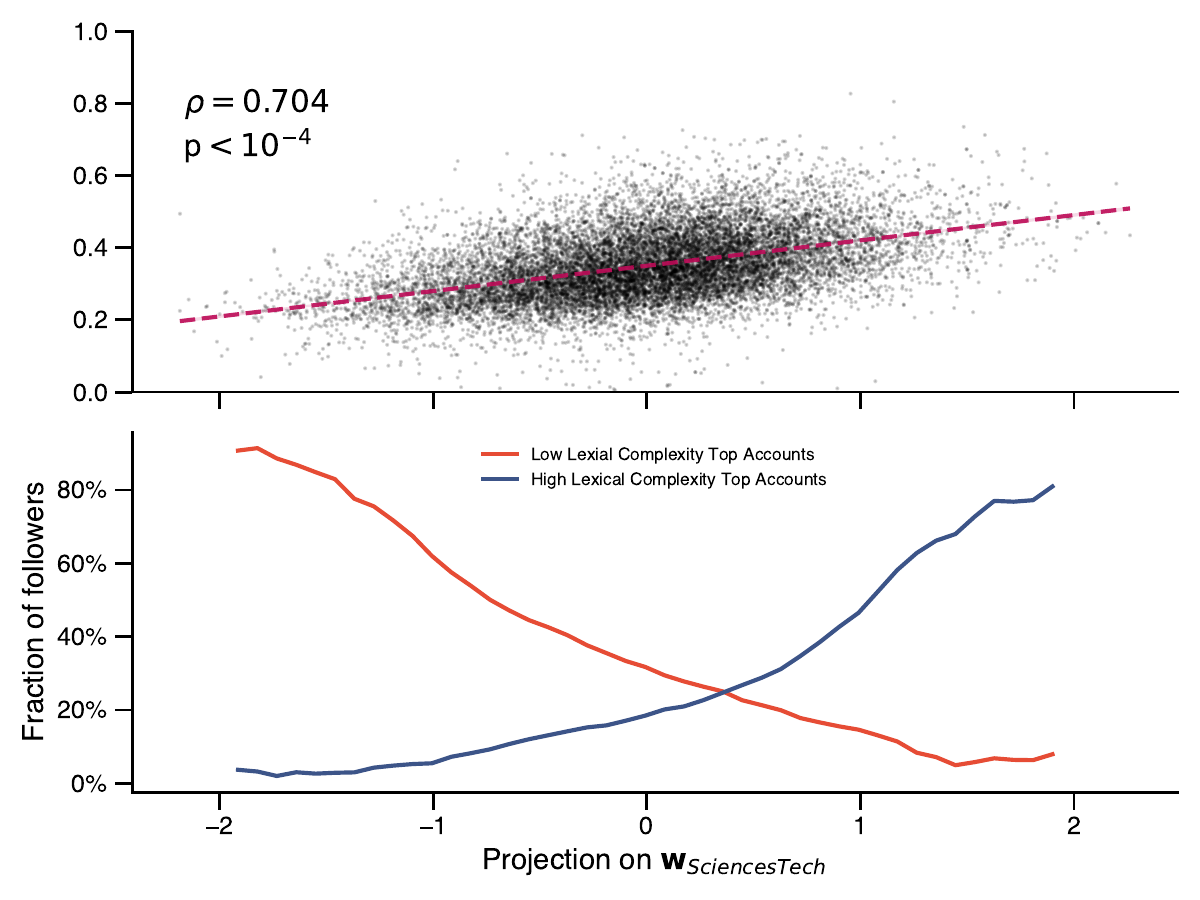}
\caption{Positioning of users by the recommender along direction $\mathbf{w}_{complexity}$ and validation. (Top figure) Scatter plot showing users' Lexical Complexity against their projection onto $\mathbf{w}_{complexity}$, the embedding direction maximizing feature correlation; $\rho = .704$, $p < 10^{-4}$. (Bottom figure) Fraction of users following at least one of the low-complexity/high-complexity -popular accounts, as function of their projection onto  $\mathbf{w}_{complexity}$.}
\label{fig:valid_lexical_complexity}
\end{figure}

We plot in Extended Figure~\ref{fig:valid_lexical_complexity} users' Lexical Complexity ---measured as the average word length---- against their projection onto $\mathbf{w}_{complexity}$. This projection exhibits a Pearson correlation, $\rho = 0.704$, $p < 10^{-4}$, with users' Lexical Complexity.

Following the previously outlined methodology, we report in Extended Lists~\ref{list:embedding_low_lexical_complexity}~\&{}~\ref{list:embedding_high_lexical_complexity} the accounts, words and hashtags overrepresented in users' following networks and content production at opposite ends of the $\mathbf{w}_{complexity}$ spectrum further validate this interpretation.\newline

\begin{customlist}
\label{list:embedding_low_lexical_complexity}
Accounts followed, words used, and hashtag used by 10\%{} with the highest probability of displaying low linguistic complexity according to the embedding space of recommender system (i.e., having low values along $\mathbf{w}_{Complexity}$).
    
    \begin{itemize}
\item \textbf{Popular accounts:} \textit{@Theorus\_, @\_ronare, @Grimkujow\_, @EcrisioFX, @Pandrezz, @maximebiaggi, @DephaseuR, @le\_egar, @Feldup\_official, @Mousstery, @KoteiZousa, @DylanKujo2, @kronomuzik, @Chap\_GG, @KarmineCorp}
\item \textbf{Popular words:} \textit{jcrois, frr, oomf, nn, stv, jss, jpense, jme, jveux, oe, wsh, jte, karmine corp, bz, jvais,}
\item \textbf{Popular hashtags:} \textit{\#{}lec, \#{}kcorp, \#{}worlds2024, \#{}j2s, \#{}zevent2024, \#{}businesscollaboration, \#{}staracademy, \#{}pgw, \#{}newprofilephoto, \#{}rt, \#{}kohlanta, \#{}twitch, \#{}competition, \#{}rapes, \#{}medalalert}
\end{itemize}
\end{customlist}

\begin{customlist}
\label{list:embedding_high_lexical_complexity}
    Accounts followed, words used, and hashtag used by 10\%{} with the highest probability of displaying high linguistic complexity according to the embedding space of recommender system (i.e., having high values along $\mathbf{w}_{Complexity}$).
    
\begin{itemize}
\item \textbf{Popular accounts:} \textit{@INRAE\_France, @Edu\_Num, @LUsineDigitale, @sup\_recherche, @Polytechnique, @usinenouvelle, @Inserm, @palaisdecouvert, @SorbonneParis1, @UNESCO\_fr, @citedessciences, @Travail\_Gouv, @iledefrance, @AFD\_France, @Bpifrance}
\item \textbf{Popular words:} \textit{european week, activity report, apply, pollinators, aap, labeled, entrepreneurship, guided tours, artificial intelligence, european heritage, adopt, rse, call for applications, signing agreement, apply}
\item \textbf{Popular hashtags:} \textit{\#{}france2030, \#{}sustainabledevelopment, \#{}orientation, \#{}fds2024, \#{}sme, \#{}jobs, \#{}ecologicaltransition, \#{}carbon, \#{}event, \#{}waste, \#{}innovations, \#{}decarbonation, \#{}ecologicaltransition, \#{}sustainable, \#{}territory}
\end{itemize}
\end{customlist}

\subsection{Summary of correlations of user attributes in the embedding spaces of the recommender system}
\label{subsec:inferred_w_summary}

\begin{table}
\caption{Pearson correlations induced by CCA directions in the embedding space of the recommender for different attributes of users. CCA directions are those that maximize the correlation between the attributes and the projection of users in the direction. All p-value are $<10^{-4}$.}
\label{tab:embedding_cca_correlations}
\begin{tabular}{|l | r | l | r|}
\hline
Attributes & Correlation & Attributes & Correlation \\
\hline
Left-Right Leaning & 0.89 & Interest in Diaries \& Daily Life & 0.63 \\
Anti-Elite Rhetoric & 0.86 & Verified Status & 0.62 \\
Interest in News \& Social Concern & 0.85 & Interest in Music & 0.60 \\
Interest in Sports & 0.80 & Age & 0.56 \\
Interest in Learning \& Education & 0.76 & Interest in Travel \& Adventure & 0.54 \\
Interest in Science \& Technology & 0.75 & Interest in Other Hobbies & 0.53 \\
Followers Count (Percentile Rank) & 0.73 & Interest in Relationships & 0.50 \\
Followers Count (Log) & 0.73 & Interest in Family & 0.41 \\
Average Word Length & 0.70 & Daily Likes & 0.40 \\
Youth \& Student Life & 0.70 & Male-Female & 0.38 \\
Arts \& Culture & 0.67 & Followers Count (Raw value) & 0.37  \\
Interest in Film, TV \& Video & 0.67 & Friend/Follower Ratio & 0.35 \\
Interest in Gaming & 0.64 & Fashion \& Style & 0.34 \\
Interest in Fitness \& Health & 0.64 & Interest in Food \& Dining & 0.32 \\
Interest in Business \& Entrepreneurs & 0.63 & Daily Posts & 0.26 \\
Interest in Celebrity \& Pop Culture & 0.63 & & \\
\hline
\end{tabular}
\end{table}

\begin{figure}
\centering
\includegraphics[width=\linewidth]{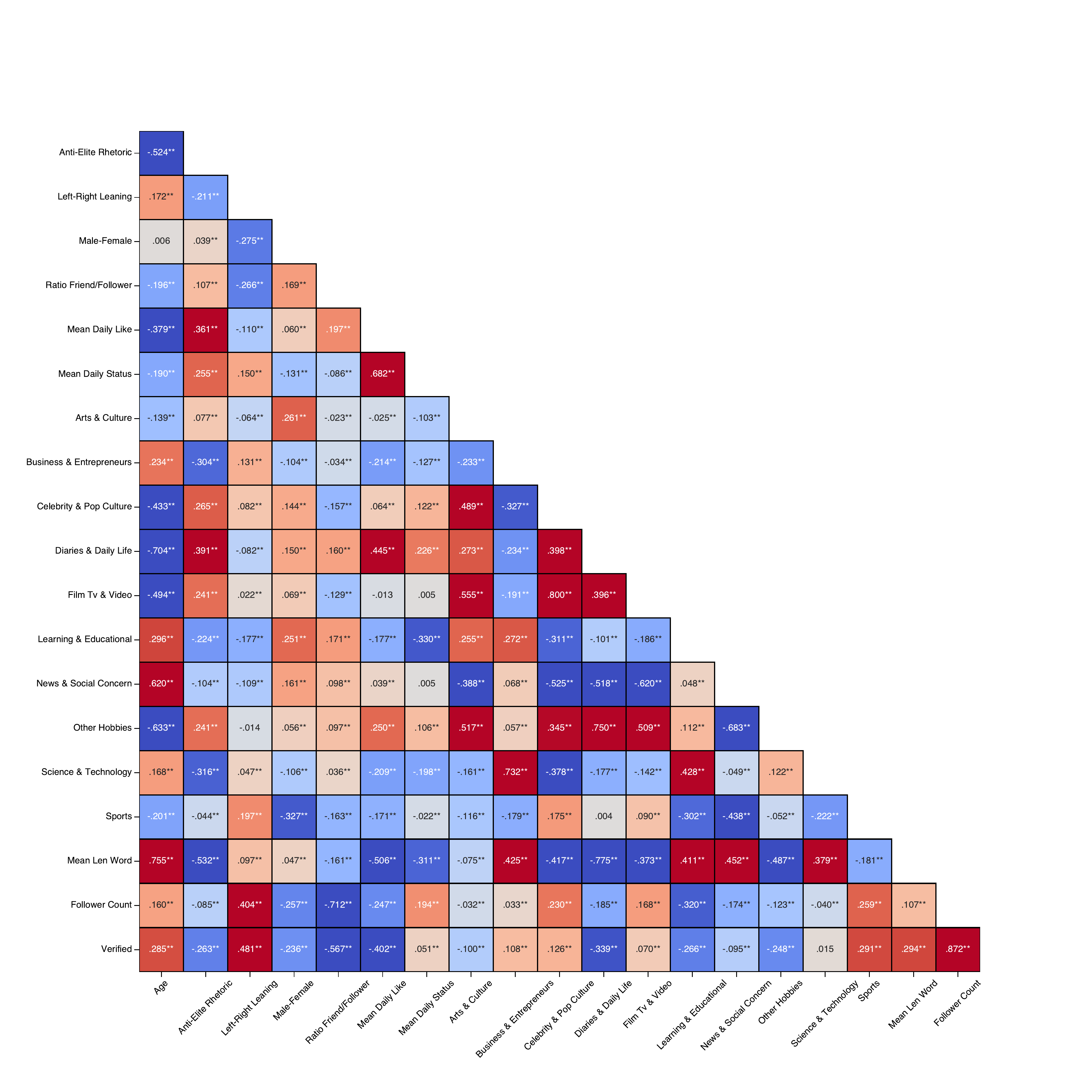}
\caption{Spearman correlation between projection of users in the study onto the CCA directions of the embedding corresponding to different attributes of users. (* indicates p-value $<0.01$, ** $<0.001$)}
\label{fig:correlation_projection_full}
\end{figure}

\begin{figure}
\centering
\includegraphics[width=\linewidth]{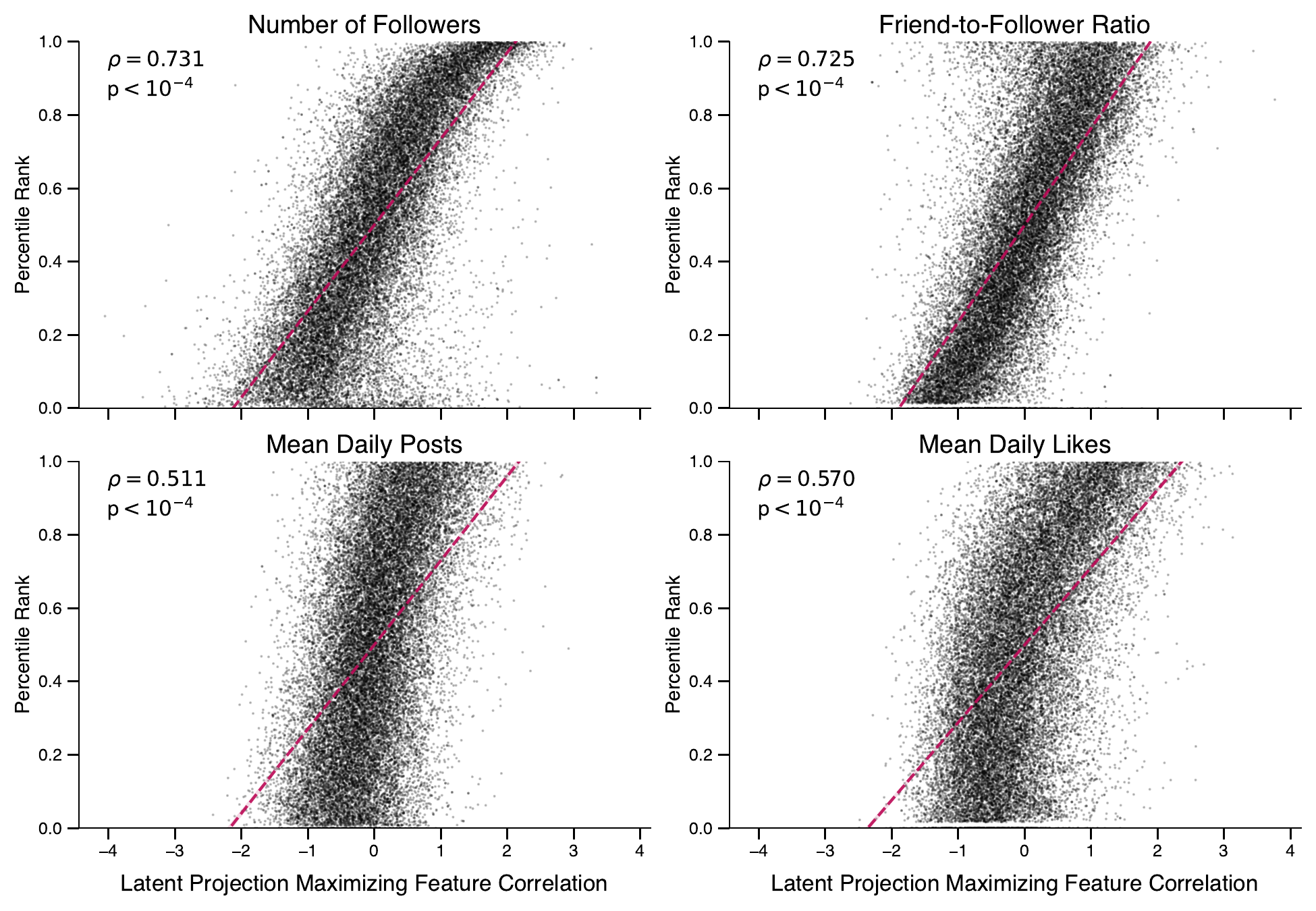}
\caption{Scatter plot of user features versus projections onto embedding directions that maximally correlate with user characteristics: number of followers, friend-to-follower ratio, mean daily posts, and mean daily likes (percentile rank).}
\label{fig:result_activity}
\end{figure}

\begin{figure}
\centering
\includegraphics[width=\linewidth]{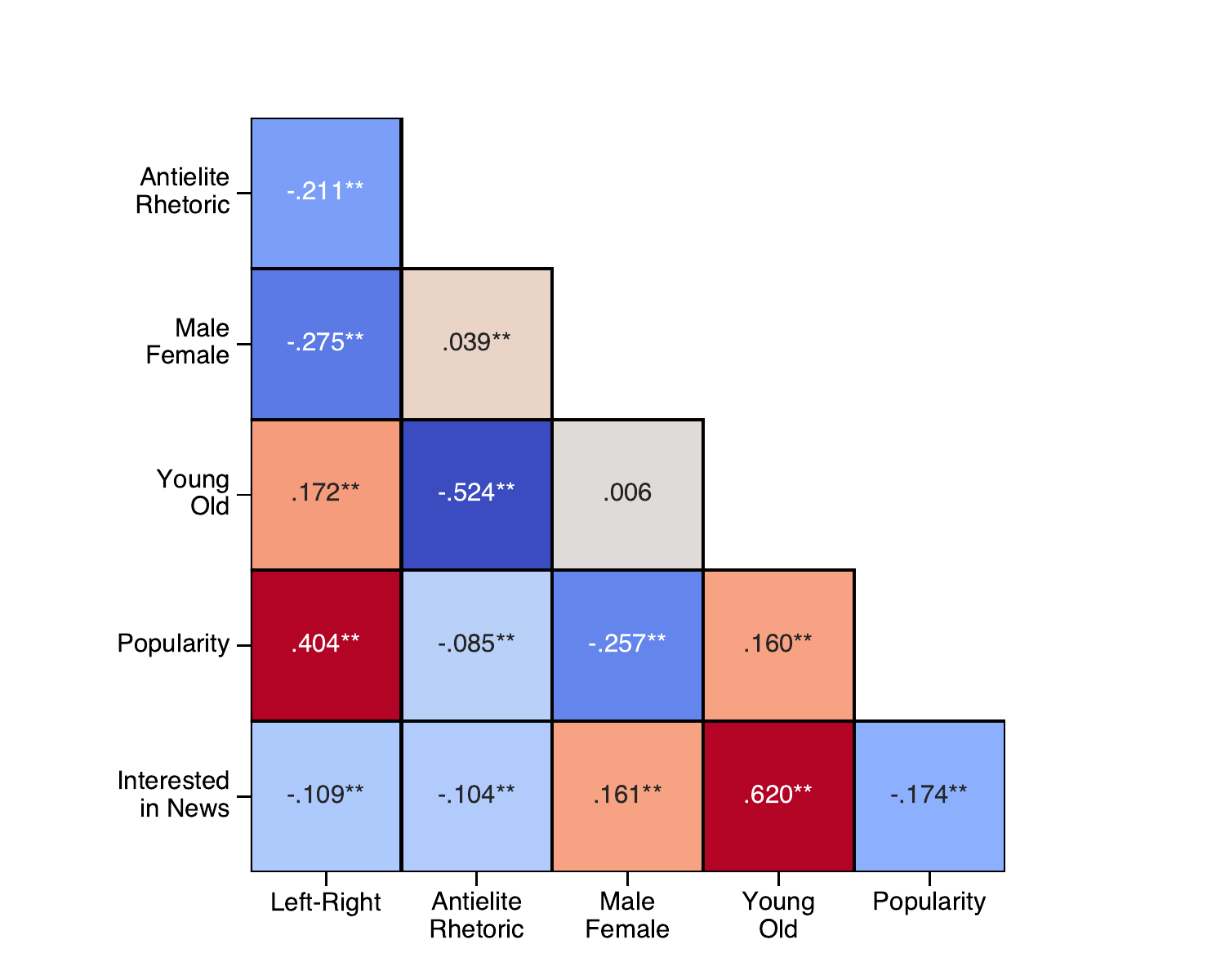}
\caption{Spearman correlation between projection of users  in the study onto the CCA directions of the embedding corresponding to different attributes of users. (* indicates p-value $<0.01$, ** $<0.001$)}
\label{fig:correlation_projection}
\end{figure}

\begin{figure}
\centering
\includegraphics[width=\linewidth]{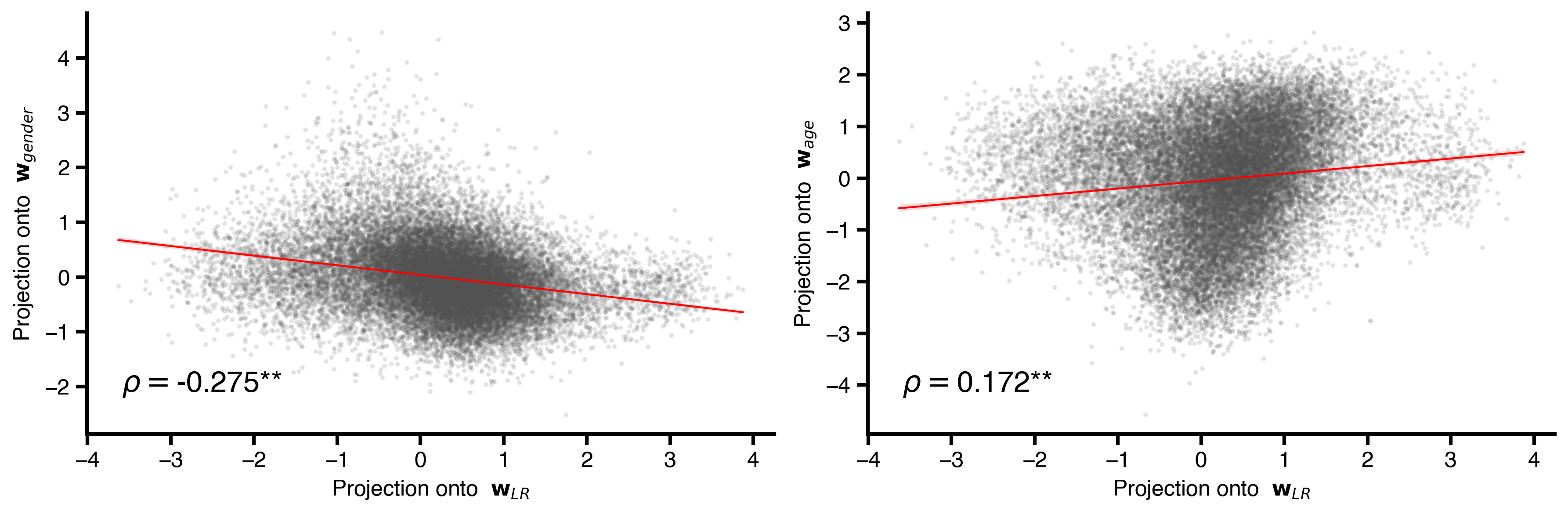}
\caption{Scatter plots between projection of users  in the study onto the CCA directions of the embedding corresponding to Left-Right leaning $\mathbf{w}_{LR}$, gender, $\mathbf{w}_{gender}$ and age $\mathbf{w}_{age}$. Low correlation values show the degree of independence of these attributes as learned by the recommender system.}
\label{fig:proj_gender_age_pol}
\end{figure}

We display in Extended Figure~\ref{fig:correlation_projection} the Spearman correlation between the projection of population $\mathcal{E}$ onto direction $\mathbf{w}_a$ and its projection onto $\mathbf{w}_b$. We observe that, contrary to the feature space where users' age negatively correlates with the probability of being ``female" (i.e., younger women and older men), the projections onto the recommender system's learned directions $\mathbf{w}_{age}$ and $\mathbf{w}_{gender}$ show no significant correlation. However, the correlations between age and anti-elite rhetoric, as well as interest in news, are amplified in the projected space compared to the feature space. We display in Extended Figure~\ref{fig:proj_gender_age_pol} the projections of users embedding onto latent directions $\mathbf{w}_{LR}$, $\mathbf{w}_{gender}$ and $\mathbf{w}_{age}$, showing low correlations.

\section{Constraining political position information encoded in the recommender system}\label{sec:constrained_recommendation}

In this section we propose a method for limiting the information encoded in identifiable directions $\mathbf{w}$ associated with user features, amounting to constrained learning in that the information encoded in this directions is rendered unavailable for the downstream recommendation tasks under the hypothesis of linear encoding.
In Section~\ref{subsec:iterative_projection} we first specified the proposed procedure for constraining learned attributes.
In Section~\ref{subsec:protocol_evaluation_constrained_learning} we then specify a protocol to assess the loss in accuracy and changes in diversity of recommendations resulting from this constraints.
Finally, in Section~\ref{subsec:results_constrained_learning} we show the results for different strategies and at different levels of aggregation on users based on their features.

\subsection{Iterative projections onto subspace orthogonal to feature directions $\mathbf{w}_{\mathbf{a}}$ in the embedding space $\Phi$}
\label{subsec:iterative_projection}

Having identified directions in the user embedding space inducing the highest correlations with user attributes, we now seek to constrain the use of such attributes when formulating recommendations. This task methodologically relates to ``debiasing" language models, similar to Bolukbasi et al.'s identification of the ``gender subspace" in word embeddings \cite{Bolukbasi}, which allows zeroing out the components of neutral words in the direction of the ``gender subspace" to mitigate gender stereotypes in word embeddings.

We adapt this methodology to protect user attributes in recommender systems. Consider a user attribute $\mathbf{a}$. Using Canonical Correlation Analysis, we first identify the direction $\mathbf{w}_{\mathbf{a}}$ where the user embeddings projection $\mathbf{\Phi}\mathbf{w}_{\mathbf{a}}$ maximally correlates with $\mathbf{a}$. We then project embedded positions of users $\mathbf{\Phi}$ onto $\mathbf{w}_{\mathbf{a}}^\perp$ the subspace orthogonal to $\mathbf{w}_{\mathbf{a}}$, effectively removing information about $\mathbf{a}$ along this direction.

However, as shown by Gonen and Goldberg \cite{gonen_goldberg} and Ravfogel et al \cite{ravfogel}, sensitive attributes may be encoded along multiple orthogonal directions in the embedding space. A single projection therefore only masks the most prominent direction $\mathbf{w}_{\mathbf{a}}$ identified by the CCA, leaving the attribute recoverable through other directions. To address this limitation, we adopt, as Ravfogel et al \cite{ravfogel}, an iterative approach. At each step, we identify and project out the direction most correlated with attribute $\mathbf{a}$.

Relying on the null distribution of correlations obtained through permutation tests in Section~\ref{section:cca}, we continue the projections until we can no longer find a statistically significant, with a p-value at 1\%, correlation between directions in the embedding space and $\mathbf{a}$.

Importantly, this procedure only removes linearly-encoded information, meaning that attribute $\mathbf{a}$ might still be recovered from the protected embeddings $\mathbf{\Phi}'$ using non-linear regression models. However, since profile recommendations are formulated through inner products between (translated) user embeddings, our procedure effectively constrains recommendations from relating to attribute $\mathbf{a}$.

In principle, any feature direction can be targeted to be constrained in the embeddings space through the described procedure. In this section, we illustrate its consequences in accuracy and diversity of recommendations using the Left-Right political dimensions. We choose this dimension because it is the leading political dimension structuring national politics, and because it is the one that more clearly falls into the definition of regulation pertaining to data privacy and profiling on sensitive categories such as political opinion and preferences.

\subsection{Protocol for the evaluation of accuracy loss of protected embeddings $\Phi'$}
\label{subsec:protocol_evaluation_constrained_learning}

We evaluate the protected embeddings $\mathbf{\Phi}'$ through both attribute correlation and recommendation accuracy. First, we assess the ``Who to Follow" predictive accuracy of the protected embeddings $\mathbf{\Phi}'$ on a 10\% held-out test dataset. Second, we analyze the political diversity of recommendations by measuring homophily in the top-k recommended profiles.

For each user $s \in \mathcal{E}$, we score every non-friend user $t$ using the TransE scoring function from Eq.~\eqref{eq:transe}: $f(s, wtf, t) = (\phi_s + \varphi_{wtf})^T \phi_t$, where $\varphi_{wtf}$ represents the ``Who to Follow" relation embedding. We then select the k highest-scoring users as recommendations. The political diversity of these recommendations is measured as the standard deviation of the Left-Right political leaning users in the top-k recommendations.

Additionally, we seek to assess whether recommendations align with users' interest in ``News \& Social Concern" from the topic ontology \cite{cardiffNLP} used in our study, while being constrained to not reflect their political leanings. 
We measure this alignment as the difference between user $s$'s interest in ``News \& Social Concern" and that of recommended users, where interest is quantified as the fraction of tweets published or shared that were classified as related to ``News \& Social Concern" by Antypas et al \cite{cardiffNLP} \texttt{cardiffnlp/tweet-topic-21-multi} model.
We choose to evaluate the impact on the topic ``News \& Social Concern" because of its potential alignment with politics. By erasing information about political preferences encoded in the embedding, the risk is that we also erase information about the interest users have in news, thus limiting one the major lever of relevance for contents served to users. 
In other words, if interest in news and salient Left-Right positions were highly aligned, constraining recommendations to remove the effect of political preferences might also erase information about interests that users have in news, and thus serve recommendations that do not rely on politics, but that are less relevant.
In our subsequent analysis we show that this is not the case, and that it is possible to erase information about political preferences for recommendations without affecting the relevance and the weight that interest in news (on the part of users) may have in these recommendations.

To estimate the French Twitter population-level effect size of the intervention, we bootstrapped our embedded population $\mathcal{E}$ to match the Left-Right Leaning distribution observed in the French Twitter population from the dataset of politically-positioned users taken from the dataset \cite{ramaciottimorales:hal-04807916} (as described in Section~\ref{subsubsec:epo_dataset}); displayed in Extended Figure~\ref{fig:distrib_epo_vs_sample}. We randomly sampled 250 times while maintaining the sample size to derive 95\% confidence intervals for our estimates.

\subsection{Results: assessing the effects on accuracy and diversity of recommendations}
\label{subsec:results_constrained_learning}

For the top-$k=50$ recommended users in the French Twitter population, maximal removal of linearly encoded political information from embedding (intervention strength 1) produced an effect size of $d = 0.477,~ 95\%~ CI ~[0.465, 0.490]$, relative to the original embedding $\Phi$.
In other words, completely removing linearly-encoded Left-Right information increased the Left-Right diversity of recommended users by $.477$ standard deviations.

Notably, removal of linearly-encoded political information from embeddings demonstrated no significant alteration in the alignment between users' interest in the news and recommended users, relative to intervention-free recommendations.
With an effect size of $d = 1.11,~ 95\%~ CI ~[1.06, 1.16]$, we observe the intervention produces significantly greater increases in political diversity for profile recommendations served to Left-leaning users compared to those served to Center- and Right-leaning users.

\begin{figure}
\centering
\includegraphics[width=\linewidth]{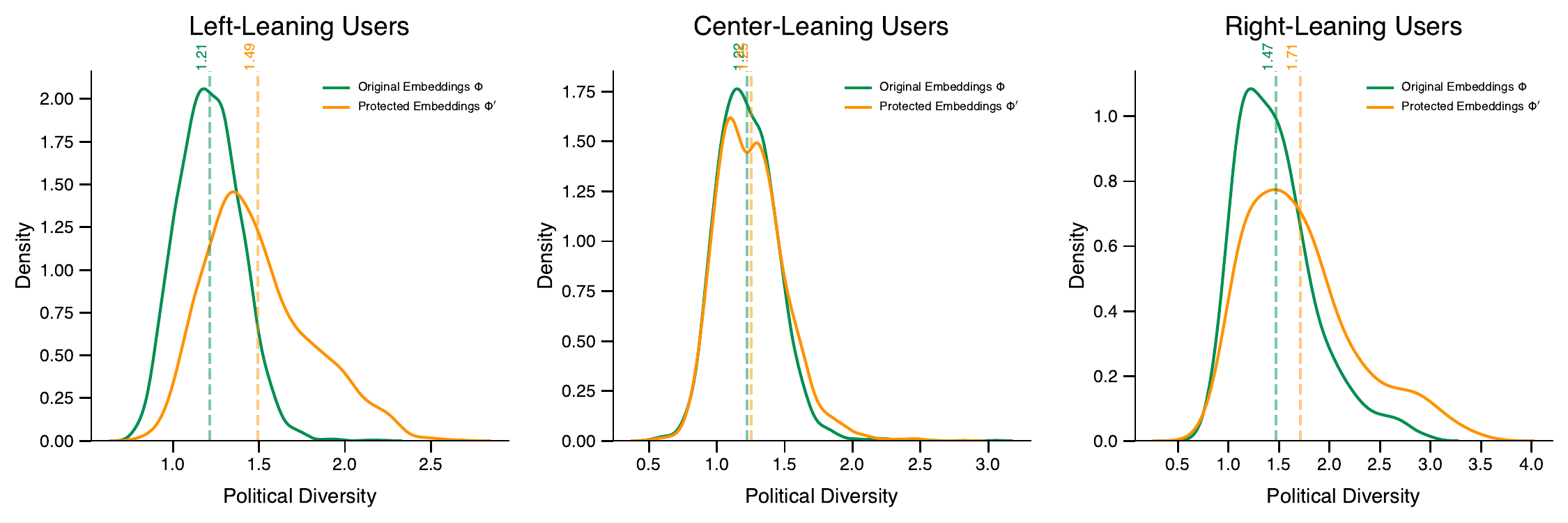}
\caption{Distribution of political diversity in profile recommendations induced by original unprotected embeddings $\Phi$ (in green) and protected embeddings $\Phi'$ (in orange), segmented by users' Left-Right leaning. Average values are displayed.}
\label{fig:diversity_rec}
\end{figure}

This situation mirrors the original distribution of political diversity, where, as displayed in Extended Figure~\ref{fig:diversity_rec}, profile recommendations served to Left-leaning users are significantly less diverse than those served to Center- and Right-leaning users. The distribution of political diversity in recommendations induced by the protected embeddings $\Phi'$ (intervention strength 1) highlights that, irrespective of users leaning, political diversity in recommendations increases after removing the political information linearly encoded in user embeddings.

We emphasize that the above procedure solely removes linearly encoded Left-Right political leanings from the embeddings $\Phi$. 
{Additionally, we display in Figure, the distribution of inferred gender and age in the audience served with the recommendations of accounts listed in Extended List \ref{list:embedding_female_words} and Extended List \ref{list:embedding_old_words}, considering the original and political protected embeddings.}

\begin{figure}
\centering
\includegraphics[width=\linewidth]{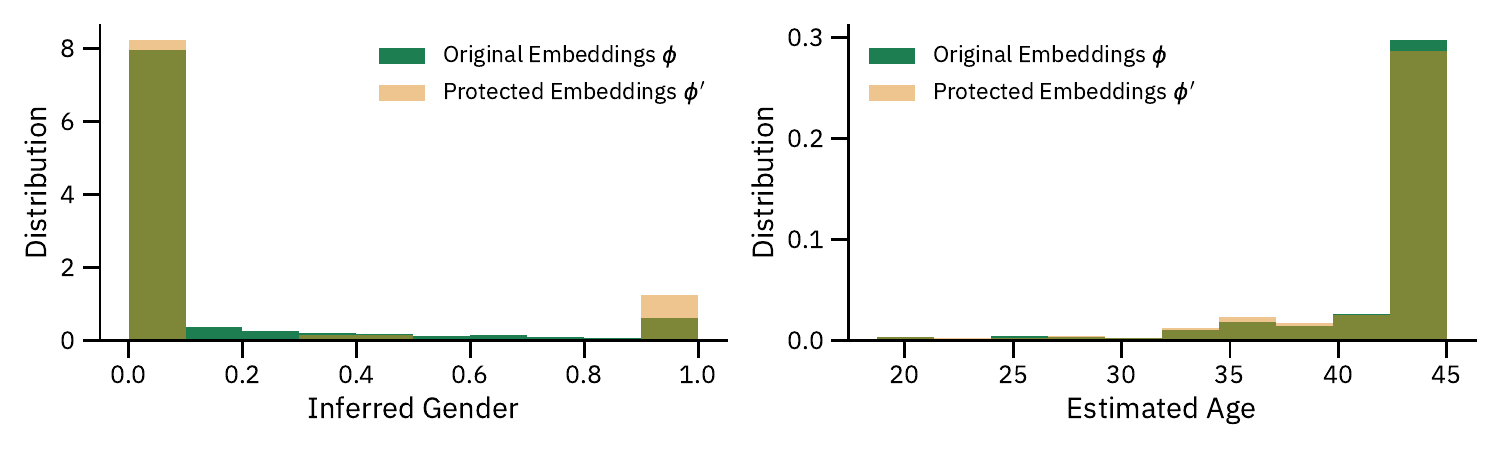}
\caption{Distribution of gender and age in the population served with the recommendations of listed accounts under the original and political-protected embeddings.}
\label{fig:gender_age_after_polprotect}
\end{figure}

\subsection{Constraining other attributes information encoded in the recommender system}
\label{subsec:contraining_other_attributes}

{
To assess the robustness and generalizability of the protection protocol, we extend it beyond political information to gender and age attributes. Following the same procedure, we generate two additional sets of embeddings: one where gender cannot be linearly retrieved beyond chance levels, and another for age.}

{We evaluate these protections by examining the distribution of user attributes among those receiving stereotypical recommendations. Specifically, we identify accounts over-represented in the follows of users inferred as female (Extended List \ref{list:embedding_female_words}) and determine which users receive recommendations for these accounts under different embeddings. We measure disparity using the Wasserstein distance between the gender distribution in the overall population and in the population served these recommendations.}

{The Wasserstein distance decreases by a factor of 10.4 when using gender-protected embeddings compared to unprotected embeddings, indicating significantly reduced gender disparity in recommendations. Similarly, the Wasserstein distance between the age distribution in the overall population and in the audience served with recommendations for the accounts listed in Extended List \ref{list:embedding_old_words} is 26.2\% smaller for age-protected embeddings compared to original ones.}

\begin{figure}
\centering
\includegraphics[width=\linewidth]{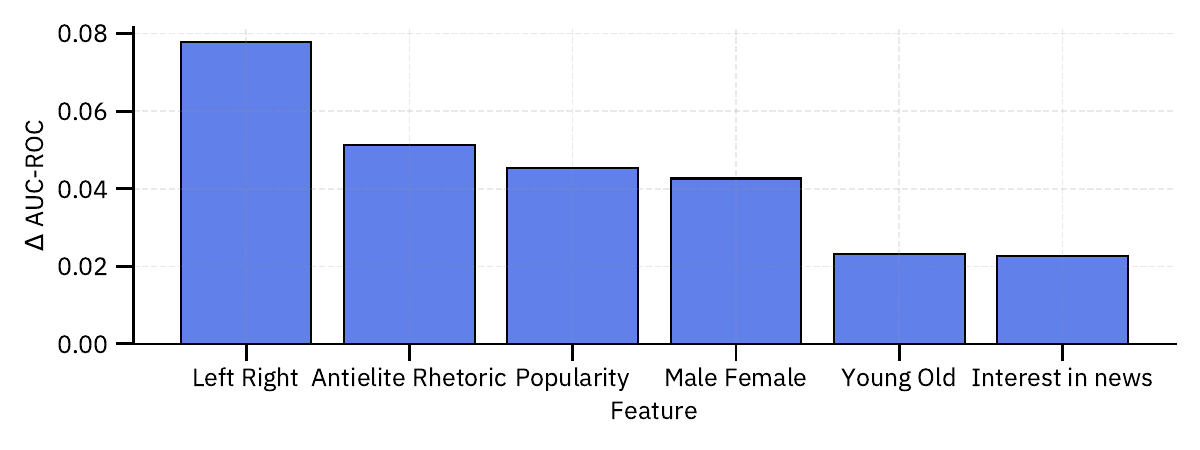}
\caption{Drop in AUC-ROC in the prediction of ``Who To Follow'' edges via the protected embedding as a function of the protected user attributes.}
\label{fig:weight_feature_delta_auc}
\end{figure}

{Finally, we report in Extended Figure \ref{fig:weight_feature_delta_auc} the drop in predictive performance of the embedding as a function of which attribute is protected. We observe that Left-Right political leaning emerges as the most important attribute among those tested, with constraining information encoding it leading to the highest performance drop.}

\end{document}